\documentclass{aa}  
\usepackage{graphicx}
\usepackage{siunitx}
\DeclareSIUnit \parsec {pc}
\DeclareSIUnit\year {yr}
\usepackage{txfonts}
\usepackage{multirow}
\usepackage{booktabs}
\usepackage{amsmath,amssymb}
\usepackage{tikz}
\usetikzlibrary{positioning,arrows.meta,calc}
\usepackage[utf8]{inputenc}
\usepackage{newunicodechar} % To write the definition on the next line
\usepackage{amsmath}
\usepackage{bm}
\newunicodechar{⌀}{\ensuremath{\diameter}}
\DeclareUnicodeCharacter{00A0}{~}	
\usepackage{natbib}
\bibpunct{(}{)}{;}{a}{}{,} % to follow the A&A style
\usepackage{booktabs}
\usepackage{tabularx}
\usepackage{array}
\usepackage{marvosym}

\usepackage{tikz}
\usetikzlibrary{positioning,arrows.meta,calc}
\usepackage[colorlinks=true, linkcolor = blue, urlcolor=blue, citecolor = blue]{hyperref}
\usepackage{txfonts}

\newcommand*{\rom}[1]{\expandafter\@slowromancap\romannumeral #1@}

\usepackage{xcolor}
\AtBeginDocument{\let\linenumbers\nolinenumbers}

\authorrunning{Donevski et al.}
\titlerunning{Dust and PAHs in late-stage galaxy evolution}

\begin{document}

   %\title{Direct evidence for diverse dust-to-gas mass ratios in old quiescent galaxies with ALMA}

   \title{Dust and PAHs in late-stage galaxy evolution:}
   \subtitle{Imprints of TP-AGB dust injection, grain growth and AGN feedback in high-$z$ quiescent galaxies with JWST and ALMA}

\author{
D. Donevski\inst{1,2,3}\corrauth{darko.donevski@ncbj.gov.pl},
A. Nanni\inst{1,4},
K. E. Whitaker\inst{5,6},
A. W. S. Man\inst{7},
A. Faisst\inst{8},
A. Lapi\inst{2,12},
I. Garc\'ia-Bernete\inst{9},
M. Romano\inst{10},
T. Petrushevska\inst{11},
G. Gururajan\inst{2,12},
G. Lorenzon\inst{1},
D. Narayanan\inst{13,6}
%S. Belli\inst{12},
%A. Vijayan\inst{11}
%A. Long\inst{10,11},
%M. M. Lee\inst{12},
%Junais\inst{13},
%I. Damjanov\inst{14},
%R. Davé\inst{15,16},
%G. Rodighiero\inst{17,4},
%D. Liu\inst{18},
%C. Lovell\inst{20},
%K. Lisiecki\inst{1},
%P. Sawant\inst{1},
%C. Pappalardo\inst{21},
%M. Hamed\inst{22}
}

	%\fnmsep\thanks{Fill other institutions later...}

\institute{
    National Center for Nuclear Research, Pasteura 7, 02-093 Warsaw, Poland
   % \email{darko.donevski@ncbj.gov.pl}
    \and
    SISSA, Via Bonomea 265, 34136 Trieste, Italy
    \and
    INAF, Osservatorio Astronomico di Trieste, via Tiepolo 11, I-34131, Trieste, Italy
    \and
    INAF - Osservatorio Astronomico d'Abruzzo, Via Maggini SNC, 64100, Teramo, Italy
    \and
    Department of Astronomy, University of Massachusetts, Amherst, MA 01003, USA
    \and
    Cosmic Dawn Center (DAWN), Copenhagen, Denmark
    \and
    Department of Physics \& Astronomy, University of British Columbia, 6224 Agricultural Road, Vancouver, BC V6T 1Z1, Canada
    \and
    Caltech/IPAC, 1200 E. California Blvd. Pasadena, CA 91125, USA
    \and
    Centro de Astrobiología (CAB), CSIC-INTA, Camino Bajo del Castillo s/n, E-28692 Villanueva de la Cañada, Madrid, Spain
    \and
    Max-Planck-Institut für Radioastronomie, Auf dem Hügel 69, 53121, Bonn, Germany
    \and
    %INAF, OAPD, Vicolo dell'Osservatorio, 5, 35122 Padova, Italy
    %\and
    Center for Astrophysics and Cosmology, University of Nova Gorica, Vipavska 11c, 5270, Ajdovščina, Slovenia
    \and
    IFPU – Institute for Fundamental Physics of the Universe, Via Beirut 2, I-34014 Trieste, Italy
    \and 
    Department of Astronomy, University of Florida, 211 Bryant Space Sciences Center, Gainesville, FL, USA
   }
\date{}

% \abstract{}{}{}{}{} 
% 5 {} token are mandatory
 
  %\abstract

\abstract
{A major unknown in late-stage galaxy evolution is what regulates the cold interstellar medium (ISM) after quenching, a question central to interpreting molecular gas, dust, and stellar content in quiescent galaxies (QGs) now probed by ALMA and JWST to $z\sim7$. We present the first semi-analytic model that follows the coupled post-quenching evolution of dust, cold gas, and polycyclic aromatic hydrocarbons (PAHs), using flexible star-formation histories and a framework tracking small and large carbonaceous and silicate grains. At $z\sim1$, we find that QGs of similar mass ($M_\star\sim8\times10^{10} M_\odot$), stellar-population age ($\sim2$ Gyr), and cold gas fractions ($f_{\rm gas}\sim1$-$10\%$), span $2$-$3$ dex in $M_{\rm dust}/M_\star$ and $M_{\rm dust}/M_{\rm gas}$, ranging from star-forming-like ratios to highly depleted dust states. The diversity arises from delayed dust injection by thermally pulsing asymptotic giant branch (TP-AGB) stars and ISM grain growth, which sustain dust enrichment for up to $\sim2.5$ Gyr after quenching. Without these channels, the pre-quenching $M_{\rm dust}$ falls below $10\%$ of its initial value within $\lesssim0.5$-$1$ Gyr, and twice as fast when AGN feedback is active. The imprint of post-quenching dust processing persists in substantial reservoirs of small carbonaceous grains, with PAH fractions of $\sim2$-$3\%$, even if the cold-dust budget falls below typical ALMA continuum detection limits ($M_{\rm dust}/M_\star\lesssim10^{-4}$). Such signatures may remain detectable with JWST/MIRI at $\mu$Jy depths, probing chemically enriched dust phases in otherwise ALMA-faint galaxies. Altogether, dust and PAHs provide independent probes of distinct stages of ISM evolution in QGs, rather than simply tracing the residual cold ISM of the preceding star-forming phase.
}

% \abstract{}{}{}{}{} 
% 5 {} token are mandatory
 
\keywords{galaxies: evolution -- galaxies: ISM -- dust}
\authorrunning{DD} \titlerunning{UNDUST} 
\maketitle
%\linenumbers
%
%-------------------------------------------------------------------
\section{Introduction}

Quiescent galaxies (QGs) are traditionally described as systems in which star formation has largely ceased and cold interstellar medium (ISM) content is expected to be low \citep[see e.g.][for a recent review]{whitaker26}. In this picture, molecular gas and dust reservoirs should fade soon after galaxies leave the star-forming main sequence (MS). However, how and under what conditions QGs process their leftover ISM fuel, and which physical mechanisms govern this processing, remain a major puzzle in galaxy evolution. This puzzle has become pressing in the James Webb Space Telescope (JWST) era: although the stellar populations of QGs are now well characterised even at early cosmic times \citep[e.g.,][]{setton24, weibel25}, the fate of their cold ISM remains considerably more uncertain. 

Recent observations challenge the view of QGs as uniformly dust- and gas-poor systems. Stacking analyses and infrared studies with low-resolution instruments reveal non-negligible average cold dust and gas budget in massive QGs \citep{man16, gobat2018unexpectedly, magdis2021interstellar, blanquez2023gas, donevski23}. In parallel,  Atacama Large Millimeter/submillimeter Array (ALMA) studies of individual QGs via CO \citep[e.g.,][]{belli21, lorenzon25b, spilker25, Umehata_2025}, [C\,II] \citep{deugenio26,valentino26}, and/or dust continuum \citep[e.g.,][]{morishita22, lee2023high} show that they can host substantial dust and molecular gas to $z\!\sim\!7$.

These inhomogeneous samples with different tracers and selections reveal a potentially very complex ISM in QGs that is challenging to interpret. At present, there is no consensus on how gas and dust evolve after quenching. Both rapid decline ($\lesssim$0.5 Gyr) and prolonged survival ($\gtrsim$1 Gyr) of the cold ISM are reported, with broad ranges of gas fractions ($f_{\rm H_2}\!=\!M_{\rm H_2}/M_\star$) of 1--15\%; \citep[e.g.,][]{hayashi18, whitaker21b, wu23, Umehata_2025} and dust fractions ($10^{-2}\!\lesssim f_{\rm dust}\!=\!M_{\rm dust}/M_\star\!\lesssim10^{-4}$; \citealt{donevski23, lee2023high, michalowski23}), often accompanied by non-detections in ultra-massive ($\log (M_\star/M_{\odot})>11.2$) QGs (\citealt{williams21,chang26}). 
The two ALMA studies that simultaneously investigate both CO and dust emission at $z\!\sim\!0.3-1$ \citep{spilker25, lorenzon25b} sharpen this tension, revealing dust-to-${\rm H_2}$ gas ratios ($\delta_{\rm DGR}$) spanning two orders of magnitude ($\sim\!1/1200$ to $1/50$) and deviating from the values typical for star-forming galaxies (SFGs; $\delta_{\rm DGR}\sim1/100$; \citealt{magdis2021interstellar}). The emerging observational picture is therefore heterogeneous, and may suggest a surprisingly diverse evolution of gas and dust after the onset of quenching. 

From the theoretical point of view, cosmological simulations with dust naturally produce broad scatter in $\delta_{\rm DGR}$ among QGs across redshift \citep{whitaker21a,lorenzon25a,chandro26}, suggesting that this diversity reflects galaxy-to-galaxy differences in dust lifecycle. However, as many channels act simultaneously, such simulations make it difficult to isolate which processes dominate dust replenishment or suppression after quenching. A further challenge is the lack of physically motivated diagnostics that can test these channels observationally. With JWST, such diagnostics can now be extended beyond the ALMA sub-mm regime to shorter wavelengths, i.e., the mid-infrared (MIR), where polycyclic aromatic hydrocarbon (PAH) emission traces
gas-phase aromatic species linked to the ultrasmall carbonaceous
grains \citep[e.g.][]{draine01}. 

This motivates us to re-examine the conditions under which dust is destroyed, survives, or is replenished across post-quenching phases. To explore this, we develop \texttt{UNDUST}, to our knowledge the first semi-analytic framework that couples cold gas, dust, and PAHs specifically in the post-quenching regime of QGs. \texttt{UNDUST} treats delayed stellar dust injection, grain growth and reprocessing, destructive processes, and AGN-driven gas outflows and heating as separable channels. In QGs, where SF-driven processes are weak, this allows the dominant dust-processing channels to be isolated more cleanly and their impact on the observable ISM evolution \textit{after} quenching to be identified. We use \texttt{UNDUST} to quantify how these channels shape quantities such as $f_{\rm dust}$, $\delta_{\rm DGR}$, and the PAH fraction ($q_{\rm PAH}=M_{\rm PAH}/M_{\rm dust}$).

We organise the work as follows: in Section~\ref{sec:undust} we present the \texttt{UNDUST} framework. We show the post-quenching evolution of $f_{\rm dust}$ and $\delta_{\rm DGR}$ in Section~\ref{sec:results}, grain-sizes and PAHs in Section~\ref{sec:section4}. In Section~\ref{sec:section5} we study the imprints of these processes with JWST and ALMA, and discuss their effects on late stages of galaxy evolution. We outline our conclusions in Section~\ref{sec:conclusions}.

%-------------------------------------------------------------
\section{An overview of the \texttt{UNDUST} model}
\label{sec:undust}
%-------------------------------------------------------------
\texttt{UNDUST} (UNified model of DUST evolution across galaxy stages) is a one-zone semi-analytic framework that follows the coupled evolution of stars, gas, metals, dust, and PAHs. Its full-lifecycle implementation, in which these reservoirs evolve self-consistently along the star-formation history (SFH), is described in Donevski \& Nanni (in prep.). Here, we focus on the post-quenching regime and initialise the residual ISM at the onset of quenching using empirically motivated cold gas ($M_{\rm gas}$, hereafter) and dust masses. We then decompose the dust budget into three channels: (A) the \textit{remnant} reservoir inherited from the SFG progenitor; (B) \textit{delayed dust injection} by thermally pulsing asymptotic giant branch (TP-AGB) stars; and (C) \textit{ISM grain growth} on gas-phase metals. This allows us to isolate the processes regulating dust survival, destruction, and replenishment after quenching without reconstructing the full pre-quenching baryon cycle.

\begin{figure*}[t]
\centering
\includegraphics[width=0.61\textwidth]{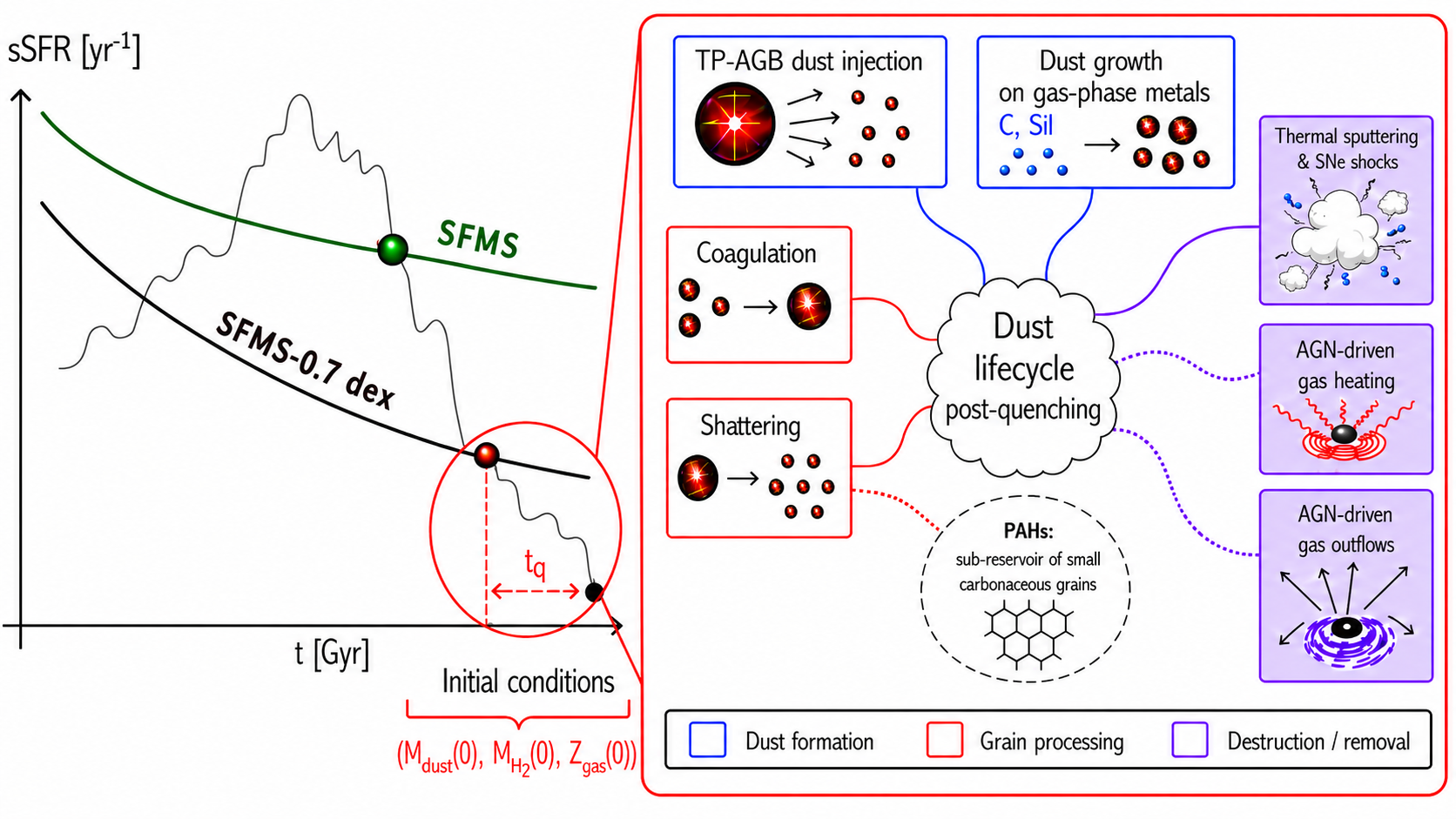}
\vspace{-0.3cm}
\caption{Illustration of the \texttt{UNDUST} post-quenching framework. The left-hand panel defines $t_{\rm q}$, which begins when the galaxy falls $5\times$ (0.7 dex) below the star-forming MS (SFMS). The cold ISM is then initialised with empirically motivated $M_{\rm dust,0}$, $M_{\rm H_2,0}$, and $Z_{\rm gas,0}$. The right-hand panel shows the dust lifecycle: delayed TP-AGB injection, ISM growth on gas-phase metals, coagulation, shattering, sputtering, SN shocks, and AGN-driven gas heating and outflows. PAHs are treated as a sub-reservoir of small carbonaceous grains. Symbols in the legend distinguish dust production, destruction, and grain-processing channels, while solid and dotted lines connect processes that affect dust directly and indirectly, respectively. Azure blue particles describe gas-phase metals, to which sputtering and SN shocks return the mass of fully destroyed dust grains (red particles). }
\label{fig:Fig1}
\end{figure*}

Before introducing the individual model ingredients, we summarise the global physical framework in Fig.~\ref{fig:Fig1}. After quenching of star formation, the inherited dust reservoir is reduced by thermal sputtering, residual supernova (SN) shocks, and astration. At the same time, TP-AGB stars inject fresh dust seeds that can grow further in residual cold, dense, metal-enriched gas. Grain--grain processing redistributes the reservoir: shattering of large grains replenishes small grains, including PAHs as its smallest carbonaceous component, whereas coagulation transfers small grains into larger ones. The cold gas reservoir evolves through adopted SFH, residual consumption, stellar mass return, and gas removal. Our one-zone framework does not explicitly resolve a multiphase ISM: dense and diffuse components are treated as fractions of the cold gas, while the hot phase is described by characteristic density and temperature. AGN feedback acts separately through cold-gas outflows and heating, thereby modifying the conditions for dust processing.

%-----------------------------------%
\subsection{Star-formation histories and quenching timescales}
%-----------------------------------%
We evolve the model along an adopted star-formation history, $\rm SFH={\rm SFR}(t)$, determining when the system departs from the star-forming MS (SFMS). Our fiducial SFH combines a delayed component with an exponentially declining post-quenching tail\footnote{In Donevski \& Nanni (in prep.) we also test results from stochastic-regulator SFH with time-variable inflow and feedback, and find that it does not significantly alter the main trends presented in this work.}:
\begin{equation}
	{\rm SFR}(t)=
	\begin{cases}
		{\rm SFR}_{\rm del}(t), 
		& t < t_{\rm q}^{0}, \\[0.05em]
		{\rm SFR}_{\rm q}^{0}
		\exp\left[-\dfrac{t-t_{\rm q}^{0}}{\tau_{\rm quench}}\right],
		%\exp(t-t_{\rm q}^{0}/\tau_{\rm quench})
		& t \geq t_{\rm q}^{0},
	\end{cases}
\end{equation}
where
\begin{equation}
	{\rm SFR}_{\rm del}(t)
	=
	A\,t\,
	\exp\left(-{t}/{\tau_{\rm main}}\right);
	\qquad
	{\rm SFR}_{\rm q}^{0}
	\equiv
	{\rm SFR}_{\rm del}(t_{\rm q}^{0}).
\end{equation}
Here, $t_{\rm q}^{0}$ marks the onset of quenching and $\tau_{\rm quench}$ controls the exponential decline of the residual SFR. From a chosen formation redshift, $z_{\rm form}$, we evolve the SFH forward in cosmic time and identify $t_{\rm q}^{0}$ as the first epoch at which the galaxy departs at least $5\times$ below \citet{speagle2014highly} SFMS, and transition to quiescent regime. We quantify the departure as
\begin{equation}
	\Delta_{\rm MS}(t)
	=
	\log_{10}
	\left[
	\frac{{\rm SFR}(t)}
	{{\rm SFR}_{\rm MS}(M_\star,z)}
	\right].
\end{equation}
Once the galaxy is $5\times$ below the MS ($\Delta_{\rm MS}<-0.7$), we express its subsequent evolution with time since quenching,
$t_{\rm q}=t_{\rm obs}-t_{\rm q}^{0}$.% allowing us to place systems with different formation histories and quenching epochs on a common post-quenching timescale

%------------------------------------------------
%------------------------------------------------

%-----------------------------------%
\subsection{Gas and metals}
\label{sec:gas_metals}
%-----------------------------------%

We evolve the gas and metal reservoirs from the initialised quenching state described in Section~\ref{sec:initial_grid}. The residual cold-gas reservoir changes as: \begin{equation} \dot M_{\rm gas}(t) = -{\rm SFR}(t) +\dot M_{\rm ret}(t) +\dot M_{\rm in}(t) -\dot M_{\rm out}(t). \label{eq:gas_evolution} \end{equation} 
The first term is gas consumption by residual star formation, while $\dot M_{\rm ret}$ accounts for stellar mass return, computed from SFH and stellar-population ages. Gas inflow, $\dot M_{\rm in}$, is set to zero to isolate internal post-quenching dust channels. We write the gas-removal rate as $\dot M_{\rm out}=\dot M_{\rm out,0}+\dot M_{\rm out,AGN}$, where $\dot M_{\rm out,0}=M_{\rm gas}/\tau_{\rm gas}$ is gradual depletion of cold gas. We adopt $\tau_{\rm gas}=1.5$ Gyr, as a reference mid-range value of broad depletion times measured in post-SB and QGs \citep[e.g.][]{magdis2021interstellar,whitaker21b, belli21, woodrum22}. The AGN-term is described in Section~\ref{sec:feedback}.

The metal reservoir ($M_Z=Z_{\rm gas}M_{\rm gas}$) evolves with grain depletion, astration, gas inflows and outflows, metal return from dust destruction, and delayed TP-AGB metal return. %We include the metal-rich gas released by AGB stars,while subtracting the fraction already condensed into AGB dust to avoid double counting. 
Starting from $M_{Z,0}=Z_{\rm gas,0}M_{\rm gas,0}$, the metal reservoir evolves self-consistently with the gas and dust. We adopt initial solar metallicity, $Z_{\rm gas,0}=Z_\odot=0.02$. Throughout this work, $M_{\rm gas}$ denotes the effective cold gas mass, initialised from observed $\rm H_2$ masses and taken as $M_{\rm gas}\!\sim\!M_{\rm H_2}$ when comparing with ALMA data. 
%-----------------------------------%
\subsection{Dust evolution}
\label{sec:dust_ev}
%-----------------------------------%
We model the post-quenching dust evolution as a well-mixed, one-zone reservoir coupled to stellar sources and gas--dust processing. We adopt
a two-size model in which grains are divided at radius 
$a=0.03\,\mu{\rm m}$ \citep[e.g.][]{hirashita15a,aoyama20}: grains
below and above this radius belong to the small- and large-grain
reservoirs, $M_s$ and $M_l$, respectively. Small grains dominate
accretion, whereas large grains initially contain most of the dust
mass. We set their initial partition with small-grain mass fraction ($f_{\rm small,0}=M_{s,0}/M_{\rm dust,0}=0.4$) after which shattering and coagulation redistribute mass between the grain reservoirs. We evolve $\mathrm{C}-$ and $\mathrm{Sil}-$grains separately, with PAHs treated as a subset of the small carbonaceous reservoir. 

We describe the two grain-size dust evolution as:
\begin{equation}
\begin{aligned}
\dot M_{i}(t)
={}&
\underbrace{
f_{{\rm inj},i}\dot M_{d,{\rm AGB}}
}_{\text{AGB dust injection}}
+
\underbrace{
\delta_{i,s}G_{\rm acc}
}_{\text{ISM grain growth}}
\\
&+
\kappa_i
\left[
\underbrace{
\frac{M_l}{\tau_{\rm sh}}
}_{\text{shattering}}
-
\underbrace{
\frac{M_s}{\tau_{\rm co}}
}_{\text{coagulation}}
\right]
-
\underbrace{
\frac{M_i}{\tau_{{\rm dest},i}}
}_{\text{sputtering /SN shocks}}
\\
&-
\underbrace{
\frac{M_i}{M_{\rm gas}}{\rm SFR}
}_{\text{astration}}
+
\underbrace{
I_i
}_{\text{dust inflow}}
-
\underbrace{
O_i
}_{\text{dust outflow}},
\qquad i\in\{s,l\}.
\end{aligned}
\label{eq:dust_ev}
\end{equation}
Here, we consider dust production from TP-AGB stars, where $\dot M_{d,{\rm AGB}}$ is dust-injection rate, and $f_{\rm inj,i}$ is dust fraction assigned to the small- and large-grains as $f_{{\rm inj},s}$ and
$f_{{\rm inj},l}=1-f_{\rm inj,s}$, respectively. The term $G_{\rm acc}$ describes ISM grain growth via accretion of gas-phase metals, and is applied only to the
small-grain reservoir ($\delta_{i,s}=1$ for $i=s$, and $\delta_{i,s}=0$ for $i=l$). Grain–grain interactions redistribute material between bins while conserving the total $M_{\rm dust}$. Shattering transfer mass from large to small grains, and coagulation drives reverse transfer. We encode this direction with $\kappa_s=+1$ and $\kappa_l=-1$, and use $\tau_{\rm sh}$ and $\tau_{\rm co}$ for the corresponding processing timescales. The destruction term $\tau_{{\rm dest},i}$ represents grain-size-dependent dust destruction via thermal sputtering and unresolved SN-driven shocks. Astration removes dust in proportion to the residual SFR, while $I_i$ and $O_i$ are optional dust transport via inflows and entrainment in outflows. %\footnote{This is defined as $O_i =
%\epsilon_{d,{\rm out}}
%\frac{M_i}{M_{\rm gas}}
%\dot M_{\rm out},
%\qquad i=s,l .
%\label{eq:dust_outflow}$}

We note that, aside from processes outlined in Eq.~\ref{eq:dust_ev}, that affect dust directly, we also introduce AGN feedback (Sect.~\ref{sec:feedback}) which enters the dust evolution indirectly, by changing the cold gas conditions that regulate dust growth and destruction. We do not include SNe~Ia, but note that their effect on $M_{\rm dust}$ in QGs is likely minor ($\lesssim0.15$ dex; Donevski and Nanni, in prep.)

%-----------------------------------%
\subsubsection{TP-AGB dust injection}
\label{sec:agb}
%-----------------------------------%

We compute dust injection by evolved stars from TP-AGB yield tables of \cite{nanni13,nanni14}, which tabulate the total dust ejecta as a function of initial $M_{\star}$ and $Z_{\rm gas}$, which extends to super-solar values. We use these tables to split the stellar dust into $\mathrm{C}-$ and $\mathrm{Sil}-$based component. %The grids span low to solar metallicity in \citet{nanni13} and extend to super-solar metallicities in \citet{nanni14}, providing species-dependent yields as functions of initial stellar mass and metallicity. Oxygen-rich stars primarily produce silicates and other oxygen-bearing species, whereas carbon-rich stars mainly form amorphous carbon and silicon carbide.
Following standard chemical-evolution models, we convert tabulated yields into a time-dependent source term by convolving them with the adopted SFH. For each stellar generation formed at time $t-\tau$, we identify the stars that reach the TP-AGB phase at time $t$ through the lifetime--mass relation. We then weight their interpolated dust yield by the IMF and the SFR at the formation time.

We distribute the total AGB dust-injection rate 
$\dot M_{d,{\rm AGB}} =
\dot M^{\rm C}_{d,{\rm AGB}}
+
\dot M^{\rm sil}_{d,{\rm AGB}}$,
between the small and large grain bins as
$\dot M_{s,{\rm AGB}}=f_{\rm inj,s}\dot M_{d,{\rm AGB}}$ and $\dot M_{l,{\rm AGB}}=(1-f_{\rm inj,s})\dot M_{d,{\rm AGB}}$. Motivated by AGB wind calculations and theoretical studies in which freshly condensed stellar grains typically have radii above our adopted division at $a=0.03\,\mu{\rm m}$, we assume $f_{{\rm inj},s}=0.2$, so that the injected AGB dust is predominantly associated with the large-grain bin \citep{yasuda12, nanni18, matsumoto24}. For the TP-AGB dust yields, we adopt the solar-models ($Z=0.02$) as our reference and compare them with super-solar
in Appendix~\ref{AppendixB}. The dependence of adopted yields on $Z_{\rm gas}$ is particularly relevant for massive QG progenitors, which are expected to be chemically evolved before quenching (e.g., \citealt{kriek19}). The \citet{nanni13,nanni14} models show that metallicity affects the composition of TP-AGB dust more strongly than its total dust-to-gas yield: near solar metallicity, C-rich TP-AGB stars produce carbonaceous dust, whereas super-solar populations are increasingly dominated by oxygen-rich TP-AGB stars and
$\mathrm{Sil}$-dust production. We discuss the consequences for late-time carbonaceous and PAH reservoirs in Section~\ref{sec:section4}.

%-----------------------------------%
\subsubsection{ISM dust growth}
\label{sec:growth}
%-----------------------------------%

We model ISM grain growth as the accretion of gas-phase metals onto available grains in the cold, dense ISM. We follow standard prescriptions \citep[e.g.][]{hirashita15b, zhukovska16, toyouchi26}, where accretion becomes more efficient with higher cold gas densities, but weakens when little gas remains in dense, shielded phases or when heating reduces sticking efficiency. We describe the dust growth rate as:
\begin{equation}
\tau_{\rm grow,eff}
=
\tau_{\rm acc,0}
\left(\frac{Z_{\rm gas}}{Z_\odot}\right)^{-1}
\left(\frac{n_{\rm H}}{n_{\rm H,0}}\right)^{-1}
f_{\rm dense}^{-1}
f_T(T_{\rm cold}).
\end{equation}
Here, $\tau_{\rm acc,0}$ is the initial accretion timescale, $n_{\rm H}$ is the effective cold gas density, and $f_{\rm dense}$ is the fraction of cold gas phase used for accretion (e.g. \citealt{matsumoto24}). The factor $f_T(T_{\rm cold})$ depicts temperature-dependent suppression of sticking: $f_T\approx1$ in cold gas, and $f_T>1$ %lengthens the effective growth time%
in warm or feedback-heated conditions.

We apply the corresponding growth rate to the small-grain reservoir as
$G_{\rm acc}=\frac{M_s}{\tau_{\rm grow,eff}}S_{\rm sat}(\delta_{\rm DTM})$.
%because small grains provide the largest surface area per unit mass. 
We add the saturation factor $S_{\rm sat}(\delta_{\rm DTM})$, where $\delta_{\rm DTM}$ is an instantaneous dust-to-metal ratio. %$\xi\equiv\frac{M_{\rm dust}}{Z_gM_{\rm gas}}$,
Therefore, accretion stops as the dust reservoir approaches the maximum allowed value which we set to $\delta_{\rm DTM}=0.5$ based on the maximal values inferred in nearby galaxies with low sSFR \citep[e.g.][]{delooze20, casasola2022resolved} and in QGs from \textsc{simba} simulation \citep{donevski23}.

We normalise the accretion timescale to
$\tau_{\rm acc,0}=50\,{\rm Myr}$ at $Z_{\rm gas}=0.02$, $n_{\rm H,0}=100\,{\rm cm^{-3}}$, and $T_{\rm cold,0}=25$ K. 
Adopted $\tau_{\rm acc,0}$ is deliberately longer than commonly assumed for SFGs ($\tau_{\rm acc,0}\sim1$--$25$ Myr; \citealt{dwek1998evolution,asano2013dust,popping17,liqi19, toyouchi26}), or local ellipticals ($\tau_{\rm acc,0}\sim30$ Myr; \citealt{hirashita17}), and reflects the less favourable conditions for ISM growth expected in QGs. We set $f_{\rm dense,0}=0.25$, within the range used in dust models \citep[e.g.][]{hirashita21}, and let it decline with age to match the lower values observed in local post-SBs \citep{french2023state}.

%-----------------------------------%
\subsubsection{Shattering and coagulation}
\label{sec:destruction}
%-----------------------------------%
%\paragraph{Shattering and coagulation.}
Shattering transfers mass from large to small grains,
$(dM_s/dt)_{\rm sh}=M_l/\tau_{\rm sh}$, while coagulation transfers small grains into the large-grain reservoir,
$(dM_l/dt)_{\rm co}=M_s/\tau_{\rm co}$. %We treat both mechanisms as phase-dependent sub-grid processes within the one-zone ISM.
Because we do not evolve separate diffuse and dense gas reservoirs, we calculate the intrinsic timescale under representative local conditions and weight the rate by the fraction of gas occupying the relevant unresolved phase. This follows the physical picture of two-size models, in which shattering occurs mainly in diffuse turbulent gas and coagulation in cold, very dense clouds \citep[e.g.][]{asano2013dust,hirashita15a,aoyama18, hirashita19, narayanan23}.

For shattering, we quantify the mass fraction in diffuse phase via
$f_{\rm diff}=1-f_{\rm dense}$, and adopt
\begin{equation}
\tau_{\rm sh,eff}
=
\tau_{\rm sh,0}
\left(
\frac{n_{\rm H,diff}}
{1\,{\rm cm}^{-3}}
\right)^{-1}
\left(
\frac{v_{\rm turb}}
{10\,{\rm km\,s^{-1}}}
\right)^{-1}
f_{\rm diff}^{-1}.
\label{eq:tau_sh_eff}
\end{equation}
Here, $\tau_{\rm sh,0}=0.5$ Gyr is the reference timescale at
$n_{\rm H,diff}=1\,{\rm cm}^{-3}$ and
$v_{\rm turb}\!=\!10\,{\rm km\,s^{-1}}$. We adopt
$n_{\rm H,diff}=0.3\,{\rm cm}^{-3}$, consistent with effective conditions in two-size dust models \citep{hou19,romano22}, and probe $1\lesssim v_{\rm turb}\lesssim 50\,{\rm km\,s^{-1}}$, encompassing the broad range inferred for QGs and AGN-dominated local galaxies \citep{li_SNIa_20,haidar26}. We set $\tau_{\rm sh,eff}\rightarrow\infty$ when $f_{\rm diff}=0$ or $n_{\rm H,diff}\geq1\,{\rm cm}^{-3}$, thereby switching off shattering outside the adopted environment.

Coagulation operates through low-velocity sticking collisions in unresolved dense cold structures:
\begin{equation}
\tau_{\rm co,eff}
=
\tau_{\rm co,0}
\left(
\frac{n_{\rm H,co}}
{100\,{\rm cm}^{-3}}
\right)^{-1/2}
f_{\rm coag}^{-1},
\label{eq:tau_co_eff}
\end{equation}
where 
$f_{\rm coag}$ is the effective fraction of gas residing in dense, coagulation-active structures. In the fiducial model we adopt $f_{\rm coag,0}=0.05$, $n_{\rm H,co}=10^{3}\,{\rm cm}^{-3}$, $T_{\rm co}=50$ K, and $\tau_{\rm co,0}=0.05$ Gyr. These parameters correspond to local coagulation times of a few to several tens of Myr under dense-cloud conditions ($n_{\rm H,co}\gtrsim 10^{3}\,{\rm cm}^{-3}$), broadly consistent with turbulent dense-cloud calculations \citep{ormel09,hirashita21}. Coagulation is switched off when $f_{\rm coag}=0$, $n_{\rm H,co}<100\,{\rm cm}^{-3}$.

%-----------------------------------%
\subsubsection{Dust destruction channels}
\label{sec:destruction}
%-----------------------------------%
Dust destruction by SN shocks and thermal sputtering acts separately in small- and large-grain reservoirs as
$1/{\tau_{\rm dest,i}}= 1/({\tau_{\rm sput,i}}+\tau_{\rm SN,i})$, where $i=s,l$. 
\vspace{-0.3cm}
\paragraph{Thermal sputtering} For thermal sputtering of grains in hot plasma, we adopt an effective timescale based on the \citealt{1995Tsai}, in the form commonly used in models without explicit grain-size distribution tracking (e.g., \citealt{McKinnon2017, osman25}):
\begin{equation}
\tau_{\rm sput}
=
\tau_{\rm sput,0}
\left(
\frac{n_{\rm hot}}
     {10^{-3}\,{\rm cm}^{-3}}
\right)^{-1}
\left[
1+
\left(
\frac{2\times10^{6}\,{\rm K}}
     {T_{\rm hot}}
\right)^{2.5}
\right].
\label{eq:tau_sput_hot}
\end{equation}
As a reference sputtering timescale we use $\tau_{\rm sput,0}=0.2$ Gyr. 
We adopt $T_{\rm hot}=10^{7}$ K as the fiducial hot-phase temperature and probe range $T_{\rm hot}=10^{6}$--$10^{8}$ K. 
Motivated by theoretical works that allow large grains to persist longer than small grains \citep{biscaro16,slavin20}, we apply $\tau_{{\rm sput},j}^{\rm eff}=s_j\tau_{\rm sput}/f_{\rm unshield}$, where $j\in\{s,l\}$, $s_s=1$, and $s_l=2$. We also account for dust protected in surviving cold, dense gas pockets, as suggested by resolved simulations and JWST studies of multiphase outflows \citep{richie2024dust,richie26,veilleux25,cronin26}. We parametrise the exposed dust fraction as $f_{\rm unshield}=\exp(-M_{\rm gas}/M_{\rm shield})$, adopting $f_{\rm unshield,0}=0.90$, which corresponds to conservative $10\%$ effective shielding at quenching. As the residual gas declines, $f_{\rm unshield}$ approaches unity and sputtering becomes more efficient.

\vspace{-0.3cm}
\paragraph{SN-driven shocks} To model the destruction of dust by unresolved SN-driven shocks, we adopt the standard effective scaling \citep[e.g.][]{slavin15}
$\tau_{{\rm SN},i} = M_{\rm gas} /{R_{\rm SN}M_{{\rm cl},i}}$,
where SN rate is $R_{\rm SN}=\nu_{\rm SN}{\rm SFR}$ and $M_{{\rm cl},i}$ is the effective $M_{\rm gas}$ cleared of dust per SN for grain bin $i$. The later absorbs the destruction efficiency and swept-up mass, which in detailed shock models depend on density, metallicity, and shock structure \citep[e.g.][]{dwek07,slavin15}. We keep $M_{{\rm cl},i}$ fixed, probing $M_{{\rm cl},s}=1000\,{\rm M_\odot}$ and $M_{{\rm cl},l}=650\,{\rm M_\odot}$, so that small grains are destroyed more efficiently than large grains \citep[e.g.][]{hirashita15a, hu19}.

As $R_{\rm SN}$ rapidly decreases after quenching, SN shock destruction is generally subdominant in QGs. For ${\rm sSFR}\sim10^{-12}\,{\rm yr^{-1}}$, the resulting timescales are several to $\gtrsim10$ Gyr, while recently quenched systems with ${\rm sSFR}\gtrsim10^{-10}-10^{-11}\,{\rm yr^{-1}}$ can reach $\sim0.5$--$1$ Gyr for small grains and $\sim\!1$--$3$ Gyr for large grains, depending on the remaining gas fraction. Our reference values are set to reflect this range. Finally, we also include an astration that removes dust at a rate $(M_i/M_{\rm gas})\times{\rm SFR}$, while we shut off dust advection by inflows ($I_i$) and outflows ($O_i$) in fiducial runs.

%-----------------------------------%
\subsection{PAHs}
\label{sec:pah}
%-----------------------------------%

We describe PAHs as a sub-component of the small carbonaceous reservoir \citep{narayanan23}. Since our two-size framework does not explicitly resolve PAH-sized grains, we track them within the small-C reservoir using a subgrid prescription. We model their evolution as:

\begin{equation}
\dot{M}_{\rm PAH}
=
f_{\rm sh\rightarrow PAH}
%\dot{M}_{s,{\rm C}}
%f_{\rm PAH,size}
\frac{M_{l,{\rm C}}}{\tau_{\rm sh, C}}
+
\dot{M}_{\rm PAH,AGB}
-
\frac{M_{\rm PAH}}{\tau_{\rm co,PAH}} 
-
\frac{M_{\rm PAH}}{\tau_{\rm dest, PAH}}.
\end{equation}

The first term represents PAH production from shattering: $M_{l,{\rm C}}/\tau_{\rm sh,C}$ is the mass flux
transferred from large- to small-$\mathrm{C}$ grains, and $f_{\rm sh\rightarrow PAH}$ is the fraction of this flux assigned to PAHs. To determine which part of the shattered material reaches PAH sizes, we follow \citet{jones96} and adopt a fragment spectrum $n_{\rm frag}(a)\propto a^{-3.3}$ over $3\leq a\leq300\,\text{\AA}$. We count as PAHs only the aromatic fragments in the smallest part of this distribution, $a\leq20\:\text{\AA}$ \citep{rau19, narayanan26}, while larger fragments remain in the non-PAH small-$\mathrm{C}$ reservoir.

The second term describes the direct TP-AGB contribution to PAH as $\dot{M}_{\rm PAH,AGB} =\epsilon_{\rm PAH,AGB}\dot{M}_{\rm C,AGB}$, where $\dot{M}_{\rm C,AGB}$ is the $\textrm{C-}$-dust injection rate predicted from the adopted TP-AGB dust yield tables, and$\epsilon_{\rm PAH,AGB}$ is the fraction assigned to PAHs. This contribution is motivated by evidence for evolved-star contributions to PAH emission in early-type galaxies \citep{galliano08,vega10,perez25}. As conversion of circumstellar carbon dust into PAH remains uncertain, we follow \citet{letter91} whose Eq.~(2) implies $\epsilon_{\rm AGB}\sim0.15$--$0.23$ relative to the combined PAH and amorphous-carbon ejecta. We adopt the conservative reference value $\epsilon_{\rm AGB}=0.15$ and treat it as a minor contribution. Therefore, most PAH production remains associated with the shattering of large $\textrm{C-}$grains.

PAHs are transferred out of the PAH reservoir via coagulation to larger $\textrm{C-}$grains on a timescale $\tau_{\rm co,PAH}$, and are destroyed by hard radiation fields and shocks \citep{montillaud13,bernete24}. We apply an effective processing timescale $\tau_{\rm dest,PAH}$ that, for a timestep $\Delta t$, removes a fraction $1-\exp(-\Delta t/\tau_{\rm dest,PAH})$ of the current $M_{\rm PAH}$. Guided by PAH lifetimes of a few $10^8$ yr inferred for interstellar shocks \citep{micelota13}, we adopt $\tau_{\rm dest,PAH}=0.3$ and 0.6 Gyr in runs with and without AGN feedback, respectively.

%--------------------------------%
\subsection{AGN feedback}
\label{sec:feedback}
%--------------------------------%

We include AGN feedback phenomenologically to test how its impact on the cold gas affects post-quenching dust evolution. We do not model a self-consistent AGN luminosity, and AGN activity does not directly destroy grains. Instead, once activated, AGN mode acts by enhancing cold ISM gas removal and heating. % thereby suppressing the dense, shielded phase required for efficient grain growth and PAH replenishment.

First, AGN feedback induces gas-outflow as
\begin{equation}
\dot M_{\rm out,AGN}
=
\eta_{\rm out}\,{\rm SFR},
\label{eq:agn_outflow}
\end{equation}
where $\eta_{\rm out}$ is the effective mass-loading factor. For the fiducial massive QGs at $z\approx1$, we adopt constant values $\eta_{\rm out}=1.25$ when feedback is enabled and $\eta_{\rm out}=0$ otherwise. This choice is motivated by molecular outflows in local galaxies \citep{fluetsch19} and neutral-outflow constraints in post-starbursts, where
$\dot M_{\rm out}\sim0.3$--$1\,{\rm M_\odot\,yr^{-1}}$ implies
$\eta_{\rm out}\sim0.4$--$1.5$ for a massive QG with
${\rm sSFR}\sim10^{-11}\,{\rm yr^{-1}}$ \citep{baron22}. It is known that some 
rapidly quenching systems exhibit much larger mass-loading factors ($\eta_{\rm out}\gtrsim5-50$, i.e., \citealt{davies24,valentino25}) and we briefly explore this regime in the Appendix~\ref{AppendixA}, while a self-consistent analysis of AGN-driven quenching will be presented in Lorenzon et al. (in prep.).

Second, AGN feedback heats or disrupts the cold dense ISM \citep{gaspari12,vogelsberger19,viaene20,haidar26}. This suppresses ISM grain growth by raising the effective cold-gas temperature
$T_{\rm cold,eff}=T_{\rm cold,0}(1+k_{\rm heat})$, where $T_{\rm cold,0}=25$ K and $k_{\rm heat}$ sets the heating amplitude. The grain accretion timescale then increases as
\begin{equation}
\tau_{{\rm acc},i}^{\rm AGN}
=
\tau_{{\rm acc},i}^{0}
\left(\frac{T_{\rm cold,eff}}{T_{\rm cold,0}}\right)^p
=
\tau_{{\rm acc},i}^{0}(1+k_{\rm heat})^p,
\end{equation}
where exponent $p$ captures the reduced fraction of cold, dense gas able to sustain grain growth. In the fiducial AGN-on runs we fix $p=1$ and test arbitrary values $k_{\rm heat}=[2,3,4]$.

%--------------------------------%
\subsection{Initial conditions and model grid}
\label{sec:initial_grid}
%--------------------------------%
We initialise each model at the onset of quenching with
$\left\{
M_{\star,0},\,
M_{\rm gas,0},\,
Z_{\rm gas,0},\,
M_{\rm dust,0},\,
f_{\rm s,0}
\right\}_{t=t_{\rm q}^{0}}$.
These quantities describe the chosen pre-quenching ISM state of the star-forming progenitor. The fiducial SFH represents a massive QG observed at $z_{\rm obs}\approx1$ with a mass-weighted stellar age of $\sim2$ Gyr. For the reference starting value (inherited from the SF progenitor) we use $M_{\star,0}=8\times10^{10}\,{\rm M_\odot}$, $f_{\rm gas,0}=0.08$, $Z_{\rm gas,0}=0.02$, and $\delta_{\rm DGR,0}=1/250$. These values are empirically motivated and deliberately conservative. The initial $\delta_{\rm DGR}$ lies near the middle of the ALMA-observed range for QGs (see e.g. \citealt{lorenzon25b}), while $f_{\rm gas}$ and $f_{\rm dust}$ are $\sim7$--$10$ times below the $\rm H_2$--gas and dust scaling of MS SFGs at comparable mass and redshift \citep[e.g.,][]{liu19b,donevski20}. We therefore approximate a general population of quenched descendants of a normal, dusty SFG at $z\sim2$ (e.g. \citealt{casey26}) to test the extent of cold dust, gas, and PAH evolution in a gas-poor system, rather than tuning the model to rare dust- or gas-rich outliers. 

We set initial small- and large-grain fractions as $f_{\rm s,0}=0.4$ and $f_{\rm l,0}=0.6$, i.e., an ISM-processed grain population from the preceding star-forming phase, as expected after $\sim$Gyr of grain processing \citep{aoyama20}. We neglect dust transport via inflows/outflows to isolate AGN effects on the cold gas, avoiding degeneracy with direct dust entrainment \citep{nanni20}.

We explore three dust-destruction regimes: mild, moderate, and harsh. These represent progressively less favourable conditions for dust survival, combining lower $n_{\rm H}$, shorter grain-survival times in SNe shocks, and stronger thermal sputtering in hot gas. In Table~\ref{tab:undust_fiducial_modes} we list parameters that define these modes.

We solve the coupled gas, dust, and PAH equations numerically on a uniform time grid and propagate parameter uncertainties via Monte Carlo sampling. For each configuration, we show the median tracks and the corresponding 16th--84th percentile range. For each destruction regime, we compare a remnant-only model with one including excess dust from TP-AGB injeciton and ISM grain growth, thereby separating the contributions of inherited dust, stellar replenishment, and grain regrowth. 
%----------------------------------
\begin{table}[h]
\centering
\caption{Fiducial \textsc{UNDUST} parameters defining the mild, moderate, and harsh destruction modes, explored with and without AGN feedback.}
\label{tab:undust_fiducial_modes}

\footnotesize
\setlength{\tabcolsep}{3pt}
\renewcommand{\arraystretch}{1.1}

\begin{tabular}{lccc}
\hline
\hline
Parameter && Destruction mode \\
& Mild & Moderate & Harsh \\
\hline

\multicolumn{4}{l}{
\textit{for both AGN=OFF and AGN=ON runs}
} \\
\hline

$n_{\rm H}$ [cm$^{-3}$]
& $40$ & $30$ & $20$ \\

$\tau_{\rm SN,s}$ [Gyr]
& $1.6$ & $1.2$ & $0.8$ \\

$\tau_{\rm SN,l}$ [Gyr]
& $2.5$ & $2.0$ & $1.5$ \\

$T_{\rm hot}$ [$10^{6}$ K]
& $6.0$ & $8.0$ & $10.0$ \\

\hline
\multicolumn{4}{l}{
\textit{only in AGN=ON runs}
} \\
\hline
$\eta_{\rm out}$
& $1.25$ & $1.25$ & $1.25$ \\

$k_{\rm heat}$
& $2.0$ & $3.0$ & $4.0$ \\

\hline
\end{tabular}

\tablefoot{
$n_{\rm H}$ is the cold gas density,
$\tau_{\rm SN,s}$ and $\tau_{\rm SN,l}$ are the small (large) grain
SN-destruction timescales, $\eta_{\rm out}$ is the mass outflow rate, and $k_{\rm heat}$ is AGN heating boost. Initial reference parameters common to all runs are $M_{\star,0}=8\times10^{10}\,{\rm M_\odot}$, $f_{\rm gas,0}=0.08$, $\delta_{\rm DGR,0}=1/250$, $Z_{\rm gas,0}=0.02$, $\tau_{\rm acc,0}=50\,{\rm Myr}$, $f_{\rm s,0}=0.4$ and $f_{\rm l,0}=0.6$.
}

\end{table}
%----------------------------------

%-------------------------------------------------------------

\section{Evolution of dust-to-gas mass ratio and dust-to-stellar mass ratio in quenched galaxies}
\label{sec:results}
%-------------------------------------------------------------

\begin{figure*}
\centering
\includegraphics[width=0.76\textwidth]{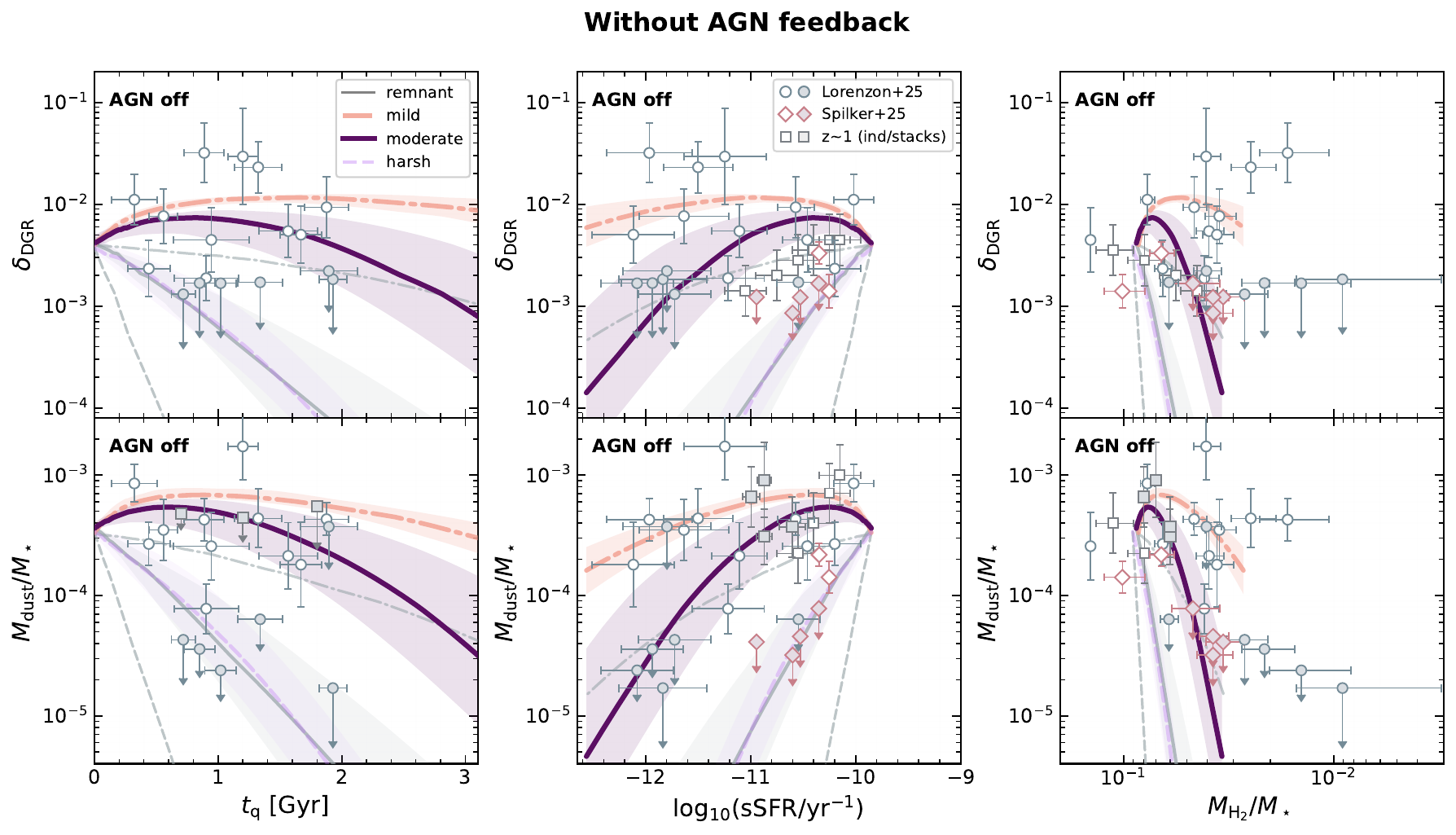}
\includegraphics[width=0.76\textwidth]{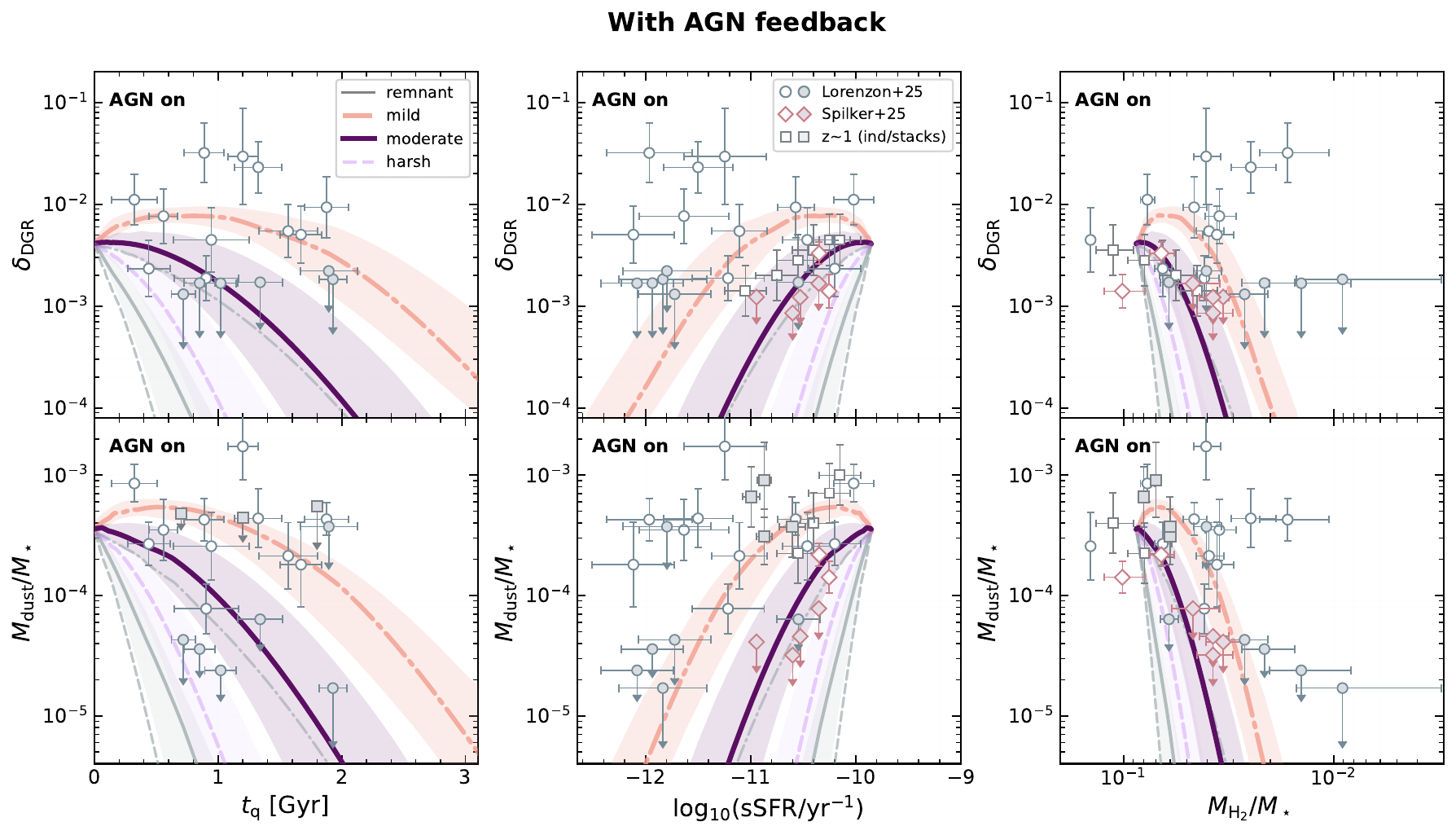}
\caption{Post-quenching evolution of $\delta_{\rm DGR}$ and $M_{\rm dust}/M_\star$ with $t_q$, sSFR, and $f_{\rm gas}\approx M_{\rm H_2}/M_{\star}$. Upper (lower) panels show runs with (without) AGN feedback. Grey curves show the evolution of the remnant dust reservoir, while colored tracks represent total dust reservoir including cumulative contribution of remnant dust, ISM grain growth and TP-AGB dust injection. Identical linestyles map to the same destruction mode in both sets of tracks: mild (dot-dashed), moderate (solid), and harsh (dotted), as indicated in the legend. Shaded regions are model uncertainty range. Literature results include individual QGs at $0.3<z<1.5$ studied in CO and/or dust continuum (circles: \citealt{lorenzon25b}; diamonds: \citealt{spilker25}; squares: \citealt{hayashi18,whitaker21b,gobat2022uncertain}) and stacked samples (grey squares: \citealt{magdis2021interstellar,blanquez2023gas,adscheid25}). Downward arrows indicate upper limits.}
\label{fig:Fig2}
\end{figure*}

We track the evolution of $f_{\rm dust}$ and $\delta_{\rm DGR}$ within multi-parameter space describing the post-quenching ISM, namely $t_{\rm q}$, sSFR and $f_{\rm H_2}$. In Fig.~\ref{fig:Fig2} we show the model tracks and literature results on ALMA studies of individual QGs up to $z\sim2$ \citep{lorenzon25a, spilker25, whitaker21b}, and QG stacking samples \citep{magdis2021interstellar, adscheid25}. 

%-------------------------------------------------------------
 
\subsection{Comparison to ALMA results}

As seen in Fig.~\ref{fig:Fig2}, our reference model broadly spans the literature constraints of $f_{\rm dust}$ and $\delta_{\rm DGR}$ in the multiparameter space. Passively fading \textit{remnant} tracks cannot fully reproduce the observed ISM diversity of QGs. Even for modest destruction, the decline of the pre-quenching $f_{\rm dust}$ with sSFR and $f_{\rm gas}$ is too steep with respect to data. Instead, post-quenching dust processing is required to reach the upper distribution of the data and reproduce the diversity in $f_{\rm dust}$ and $\delta_{\rm DGR}$. We see that even a modest decline in about one dex in gas fraction, $f_{\rm gas}=M_{\rm gas}/M_\star\sim0.01$--$0.1$, produces more than two dex of variation in $\delta_{\rm DGR}$, from nearly Milky-Way-like values typical of SFGs, $\delta_{\rm DGR}\sim10^{-2}$ \citep{magdis2012evolving}, to strongly depleted systems ($\delta_{\rm DGR}\lesssim10^{-3}-10^{-4}$). Such behaviour is qualitatively consistent with the broad range of $\delta_{\rm DGR}$ reported for observed QGs across cosmic epochs \citep{lorenzon25a,spilker25,valentino26}. This implies that scatter among ALMA data is primarily driven by dust grains responding abruptly to cold gas changes, and suggests that molecular-gas detections do not guarantee detectable dust continuum, and vice versa, in line with recent observational claims \citep{whitaker21a, morishita22, lorenzon25b, gangula26}.

%------------------------------------------------------------
From Fig.~\ref{fig:Fig2} we also see that stacking measurements \citep{magdis2021interstellar, blanquez2023gas, adscheid25} tend to lie near slow-to-moderate destruction tracks, possibly explaining why ALMA stacks often infer higher $f_{\rm dust}$ than targeted observations of individual QGs. Similar tracks overlap with gas-richer ($f_{\rm H_2}\gtrsim10\%$) QG from \citet{hayashi18}, caught in a transitioning phase toward red sequence. By contrast, dust-rich QGs at lower $f_{\rm gas}$ (\citealt{lorenzon25b}; see also \citealt{morishita22}) require efficient dust reformation in gas-poor environments. These dustiest outliers remain difficult to reproduce with our fiducial runs, and may require exceptionally efficient grain growth, potentially exceeding adopted dust-to-metal limit of $\delta_{\rm DTM}=0.5$, partial decoupling between dust and cold gas \citep{lorenzon25a}, or external dust supply through mergers \citep[e.g.][]{deugenio26}.\footnote{Our fiducial runs use single reference $f_{\rm gas}$ and $M_{\rm dust}$, although QGs may quench with higher or lower values. This mainly shifts the track normalisation, while preserving the relative trends and timescales; reproducing individual ALMA outliers is beyond the scope of this work.}

The lower panels of Fig.~\ref{fig:Fig2} show that AGN feedback further reduces the cold gas available for dust reformation, suppressing the upper end of the $f_{\rm dust}$ and $\delta_{\rm DGR}$ distributions even under mild destruction. In Appendix~\ref{AppendixA} we show that stronger outflows exacerbate this tension: higher $\eta_{\rm out}$ can temporarily enhance $\delta_{\rm DGR}\gtrsim1/100$ toward lower $f_{\rm H_2}$, while steeping the decline in $f_{\rm dust}$ at fixed sSFR. We interpret the broad overlap with the ALMA data as a support that these pathways as plausible, but not unique: similar locations in the dust--gas plane can arise from different routes; i.e., efficient destruction in no-AGN runs can resemble moderate-destruction tracks with AGN. Measurements of $f_{\rm dust}$ or $\delta_{\rm DGR}$ alone therefore cannot uniquely identify the dominant ISM-processing mechanism. Breaking this degeneracy requires joint constraints on $f_{\rm dust}$, $\delta_{\rm DGR}$, $t_q$, sSFR, and $f_{\rm H_2}$, motivating homogeneous observations that probe multiple dust, gas and stellar phases simultaneously.

Despite this ambiguity, the modelled tracks can identify the post-quenching epochs at which different dust-destruction modes are most diagnostic. The first $\lesssim0.5$ Gyr after quenching provides the strongest leverage on rapid destruction, because by $t_{\rm q}\sim0.3$--$0.5$ Gyr the mild-remnant tracks are already well separated from the harsh-destruction cases. Such diversity may already be present in recently quenched galaxies at $z\sim1$, where massive post-SBs are expected to be common (e.g., \citealt{wild16}). While systematic dust-to-gas studies of post-SBs beyond the local Universe are still limited, local samples show $>2$ dex scatter in $f_{\rm dust}$ at post-burst ages of $\sim0.2$--$0.6$ Gyr \citep{li19}. Many ALMA-studied post-SBs at $z\sim1$ retain gas reservoirs above those adopted in our fiducial QG model \citep[e.g.][]{belli21,zanella23}, but the subset already below the MS gas sequence is directly relevant for our tracks. Such systems span the transition from $f_{\rm gas}\sim0.1$ to $\sim0.01$ within $\sim0.5$ Gyr \citep{bezanson21}; ALMA observations reveal abundant CO reservoirs but a low ($\sim25\%$) 2 mm dust-continuum detection rate \citep{setton25}, making them useful targets for testing rapid decoupling of dust and gas.

In general, ALMA non-detections may not uniquely trace rapid destruction tracks of pre-quenching dust. Some QGs may continue to reform dust mass while remaining below typical ALMA continuum limits. We return to this point in Section~\ref{sec:section5}.

%------------------------------------------

\subsection{Comparison to simulations}

%------------------------------------------
We also compare the \texttt{UNDUST} tracks with QGs drawn from the 50 Mpc$h^{-1}$ box of \textsc{SIMBA} simulation (\citealt{simba2019, liqi19}). We analyse dust properties of QGs identified in \textsc{SIMBA} runs with and without AGN feedback physics. %This comparison should be interpreted as a phase-space consistency test, rather than as an object-by-object calibration, because \textsc{SIMBA} follows a cosmological distribution of SFHs, gas accretion, mergers, and feedback events, whereas \textsc{UNDUST} isolates controlled post-quenching dust evolution for idealized QGs.

\begin{figure}[h!]
	\centering
	\includegraphics[width=0.45\textwidth]{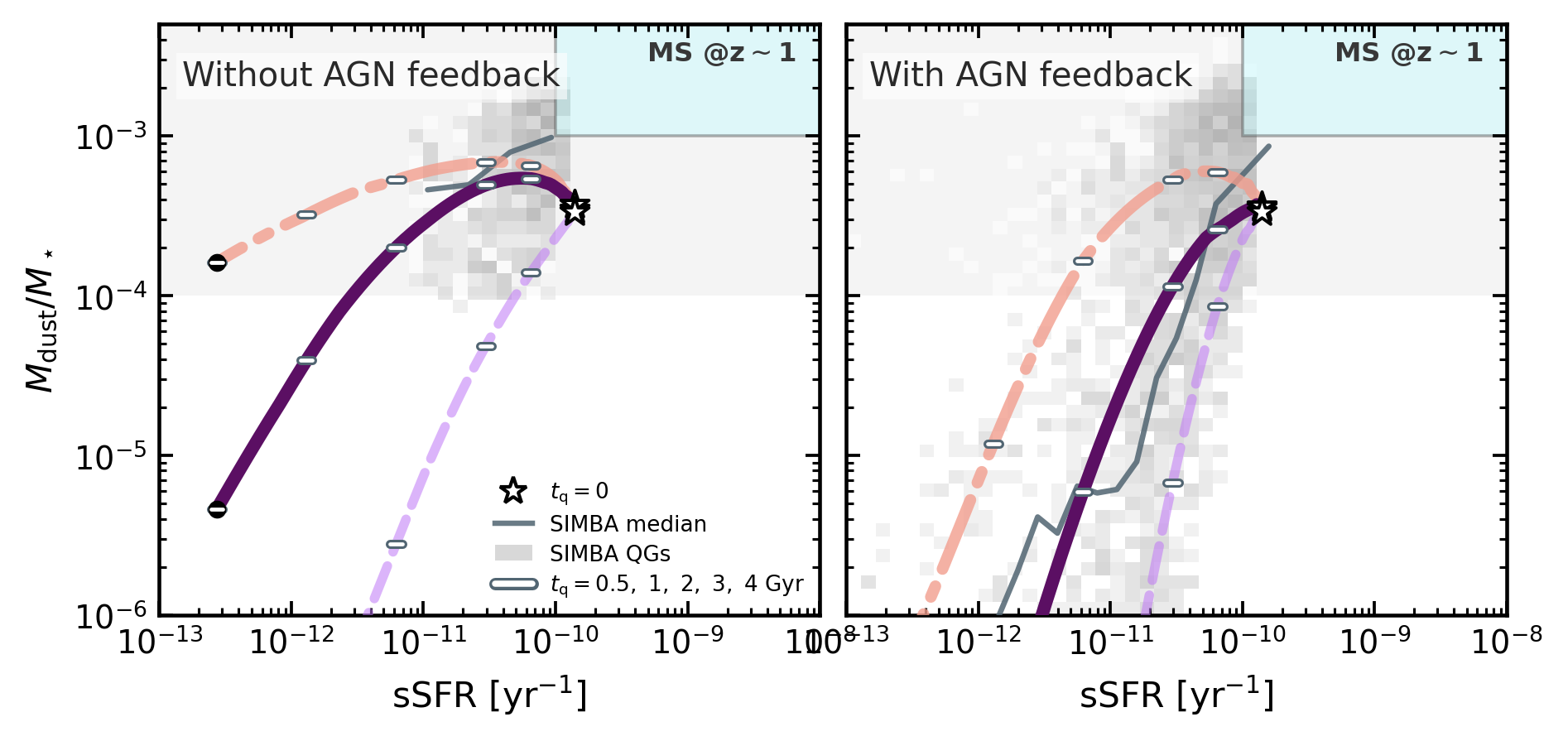}
	\caption{
		Post-quenching evolution in the $f_{\rm dust}$--sSFR plane for \textsc{UNDUST} models without and with AGN feedback, compared to $z\sim1$ \textsc{SIMBA} galaxies. Grey points and curves show the \textsc{SIMBA} QGs and median trends, while coloured tracks have the same meaning as in Fig.2. Cyan-blue region show position of $z\sim1$ MS SFGs \citep[e.g.][]{donevski20}.}
	\label{fig:Fig3}
\end{figure}

In Fig.~\ref{fig:Fig3} we see the sharp separation between the no-AGN and AGN \textsc{SIMBA} distributions in the $f_{\rm dust}$--sSFR plane. This reflects the known role of AGN feedback in regulating both quenching and the cold ISM in \textsc{SIMBA}. Without AGN feedback, massive galaxies remain insufficiently quenched and keep dense gas over tens of Gyr resulting in very weak decrease in sSFR and $f_{\rm dust}$. Instead, in the full \textsc{SIMBA} run, $\sim\!70\%$ of $z\sim1$ QGs are fast quenchers ($t_q<250$ Myr) that undergo AGN jet mode. In most of these QGs AGN feedback further injects energy via X-ray radiation, which is triggered when $f_{\rm gas}\lesssim15\%$ \citep{appleby2020impact}.
Such QGs exhibit relatively fast gas/dust removal, with a net decline of $\gtrsim\!2$ dex in $f_{\rm dust}$. \citealt{lorenzon25b} show that intrinsic dust replenishment can operate in harsh conditions, mostly among slow-quenchers and older QGs in \textsc{SIMBA}. Instead, in rapidly quenched galaxies at $z\sim\!1$, dust content is dominated by a preceding burst or merger \citep{faisst17, akins2022quenching} that can reform $M_{\rm dust}$ before quenching starts. %Their dust content at low sSFR is therefore expected to be dominated mainly by remnant dust from the pre-quenching phase, or by merger-supplied material, rather than by prolonged post-quenching replenishment, which occurs more efficiently in slowly quenched and older QGs. 
These processes jointly broaden a range in $f_{\rm dust}$ and $\delta_{\rm DGR}$ at fixed sSFR (\citealt{whitaker21a, donevski23, lorenzon25b}). Similar diversity has recently been found in $z\sim2$ QGs from COLIBRE simulations \citep{chandro26}. 

The \texttt{UNDUST} tracks qualitatively agree with this broad range from SIMBA, despite the different dust model and feedback implementation. In particular, for the case when  AGN feedback is included, moderate destruction track sits very close to the SIMBA median. This broad agreement supports the view that $f_{\rm dust}$--sSFR plane is not governed by a single gas depletion pathway, but encodes the complex interplay of SF suppression, gas removal, and the efficiency of dust regrowth post-quenching.

%-------------------------------------------------------------
\subsection{TP-AGB dust injection and ISM grain growth across post-quenching galaxy evolution}
\label{sec:3.3}

%-------------------------------------------------------------
\begin{figure*}
\centering
\includegraphics[width=0.8\textwidth]{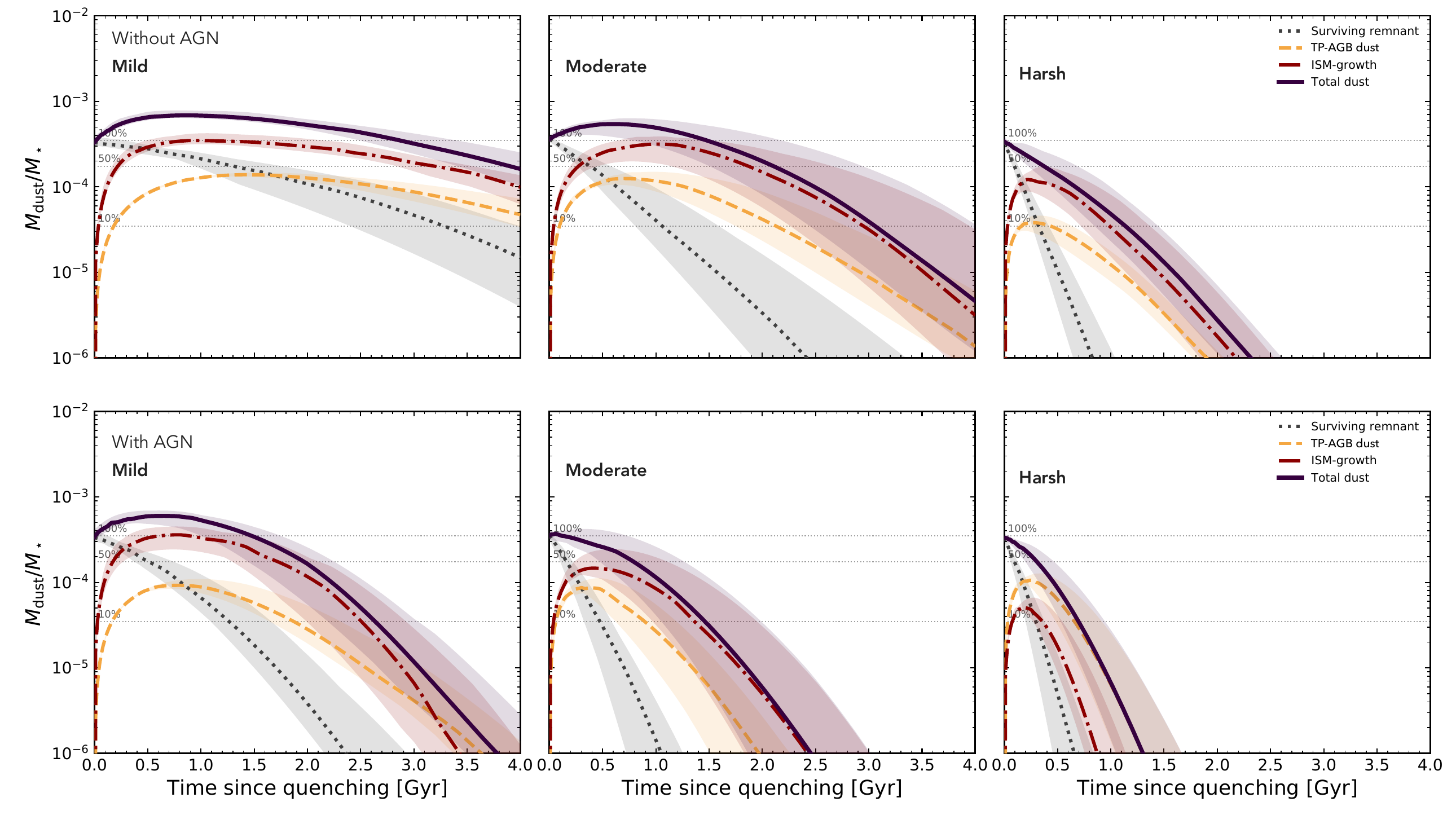}
\caption{Component-specific evolution of dust fraction after quenching. Panels compare mild, moderate, and harsh destruction regimes without AGN feedback (top row) and with AGN feedback runs (bottom row), as defined in Table~\ref{tab:undust_fiducial_modes}. The black dotted curves trace the pre-quenching remnant dust reservoir, the orange dashed curves show the excess dust supplied by delayed TP-AGB injection, the red dash-dotted curves show the excess produced by ISM grain growth, and the thick purple curves show the total $f_{\rm dust}$ including all three channels. Shaded regions mark 16th--84th percentile ranges. Horizontal reference lines mark different fractions ($100\%$, $50\%$, and $10\%$) of the initial dust reservoir.} 
\label{fig:Fig4}
\end{figure*}
%-----------------------------------------------------------
We further inspect how specific dust replenishment processes vary across post-quenching timescales. To do this, we decompose the dust budget into pre-quenching (\textit{"remnant"}) dust, TP-AGB dust channel, and the excess dust mass gained by ISM grain growth of seed grains. We illustrate these channels in Fig.~\ref{fig:Fig4} along with the total dust model that includes all three reservoirs.

TP-AGB stars are an important source of dust replenishment in QGs. Their contribution quickly exceeds the remnant dust reservoir, which declines below half of its initial mass within $\lesssim0.3$--$0.5$ Gyr under harsh (moderate) conditions. Although delayed TP-AGB injection cannot fully compensate for dust destruction, resulting in tracks that do not exceed $\sim 50\%$ of the initial $f_{\rm dust}$ regardless of ISM conditions, they enrich the ISM with fresh grains. For QGs of reference stellar population ages of $t_{\star}\!\sim\!2$ Gyr, cumulative TP-AGB dust yields are of $M_{\rm dust}^{\rm AGB}\!\sim\!3$--$8\times10^6\:M_\odot$ across the post-quenching phase. As discussed by \citet{nanni14}, total $M_{\rm dust}^{\rm AGB}$ depends only weakly on $Z_{\rm gas}$, with metallicity primarily affecting its composition: C-rich dust dominates at early times, while $\mathrm{Sil-}$rich production peaks later as lower-mass stars reach the TP-AGB phase. In Appendix~\ref{AppendixC} we illustrate this age dependence: AGB dust production peaks for $\sim1-2$ Gyr old stellar populations, but remains non-negligible for older systems up to $t_{\star}=3$ Gyr. This ages are typical of massive QGs at $z\sim1$ \citep{carnall19, hamadouche2023connection, beverage24}, making our predicted TP-AGB replenishment broadly representative of the bulk of this population.

ISM grain growth is the dominant dust source over much of the post-quenching evolution.  Under favourable residual ISM conditions, ISM grain growth can maintain more than $50\%$ of the initial $M_{\rm dust}$ for up to $\sim1.5$--$2$ Gyr, flattening the model tracks and helping to reproduce the $f_{\rm dust}$ range of ALMA-detected QGs in Fig.~\ref{fig:Fig2}. In the no-AGN models, typical growth timescales are $\tau_{\rm grow}\sim0.3$--$0.7$ Gyr during the first $\sim1.5$ Gyr after quenching, with the shortest timescales in the mild regime. At later times, $\tau_{\rm grow}$ slow to $\sim0.8$--$3$ Gyr as $n_{\rm H}$ and $f_{\rm dense}$ decline and the small-grain reservoir is depleted. Raising $Z_{\rm gas}$ to super-solar values slightly shortens $\tau_{\rm grow}$ by
$\sim0.1$--$0.2$ dex (Appendix~\ref{AppendixB}).

A key take-away from this result is that efficient ISM grain growth in QGs does not require very high cold gas densities typical of SFGs, but can remain viable for the modest initial range $n_{\rm H}\sim20$--$40~{\rm cm^{-3}}$ explored here. The most optimal conditions combine a joint presence of a modest (and declining) residual dense-gas fraction typical for adopted $n_{\rm H}$ ($f_{\rm dense}\lesssim0.1-0.2$; \citealt{neumann23, lin24}), and enough diffuse turbulent gas for shattering to convert large TP-AGB grains into small-grain seeds for accretion. This condition helps ISM growth counteract destruction for $\sim1$--$2$ Gyr in the mild and moderate regimes, but it fails under harsh conditions, where rapid sputtering ($\tau_{\rm sput}\lesssim0.2$ Gyr) quickly halts re-accretion. Sustained ISM growth ultimately requires $\tau_{\rm grow,eff}\lesssim\tau_{\rm sput}$; once this condition is lost, TP-AGB injection can become comparable to, or exceed, the ISM-growth contribution to dust budget. Interestingly, JWST/MIRI observations show that TP-AGB Sil dust can persist in the harsh radiation field within the central parsec of the Milky Way \citep{peissker26}, implying that AGN feedback need not suppress delayed dust production directly.

AGN feedback further shifts this balance against grain growth. In our model, AGN heating/outflows reduce the dense gas fraction and lengthen the growth time as $\tau_{\rm grow,AGN}=\tau_{\rm grow,0}(1+k_{\rm heat})^p,$ until growth becomes fully suppressed once the effective $T_{\rm cold}$ approaches the adopted cutoff $T_{\rm acc,cut}\simeq140$ K. For our fiducial $T_{\rm cold,0}=25$ K and $p=1$, the mild, moderate, and harsh destruction modes with AGN feedback lengthens $\tau_{\rm grow}$ by factors of 3-5. Therefore, AGN does not need to destroy all dust directly to make QGs dust-poor. By lowering the dense and shielded gas fraction and slowing re-accretion, it can prevent AGB-supplied grains from fully regrowing the dust reservoir.

Two consequences for late-stage galaxy evolution emerge from this analysis. First, in QGs that retain H$_2$ but remain quiescent because their dense-gas fraction has declined \citep[e.g.][]{french2023state}, ISM growth can stay active up to $\sim1.5$ Gyr after quenching. Second, the dust content in QGs cannot be fully attributed to passive aging or offsets from H$_2$-gas scaling relations \citep[e.g.][]{tacconi2018phibss,liu19b}. Only the remnant tracks in Fig.~\ref{fig:Fig2} and Fig.~\ref{fig:Fig4} resemble an age sequence driven by fading molecular gas, as proposed by \citet{gobat20} and \citet{michalowski23}. We show that TP-AGB injection and ISM grain processing substantially modify this evolution. Consequently, QGs observed at $t_{\rm q}\sim0.5$--$2$ Gyr are optimal for testing these channels and departure from the age sequence.

%-------------------------------------------------------------
\section{Grain sizes and PAH evolution in quenched galaxies}
\label{sec:section4}
%-----------------------------------------------------------
\subsection{Evolution of the grain-sizes}
\label{sec:grain_size_evolution}
% ---------------------------------------------------
The self-consistent two-size dust framework in \texttt{UNDUST} provides useful information on how the $\mathrm{C-}$ and $\mathrm{Sil-}$grains are distributed after quenching. We examine the changes in the grain-size balance to better understand the build-up and evolution of the bulk dust mass and PAH-bearing carbonaceous reservoirs. 

\begin{figure}[t]
\centering
\includegraphics[width=0.33\textwidth]{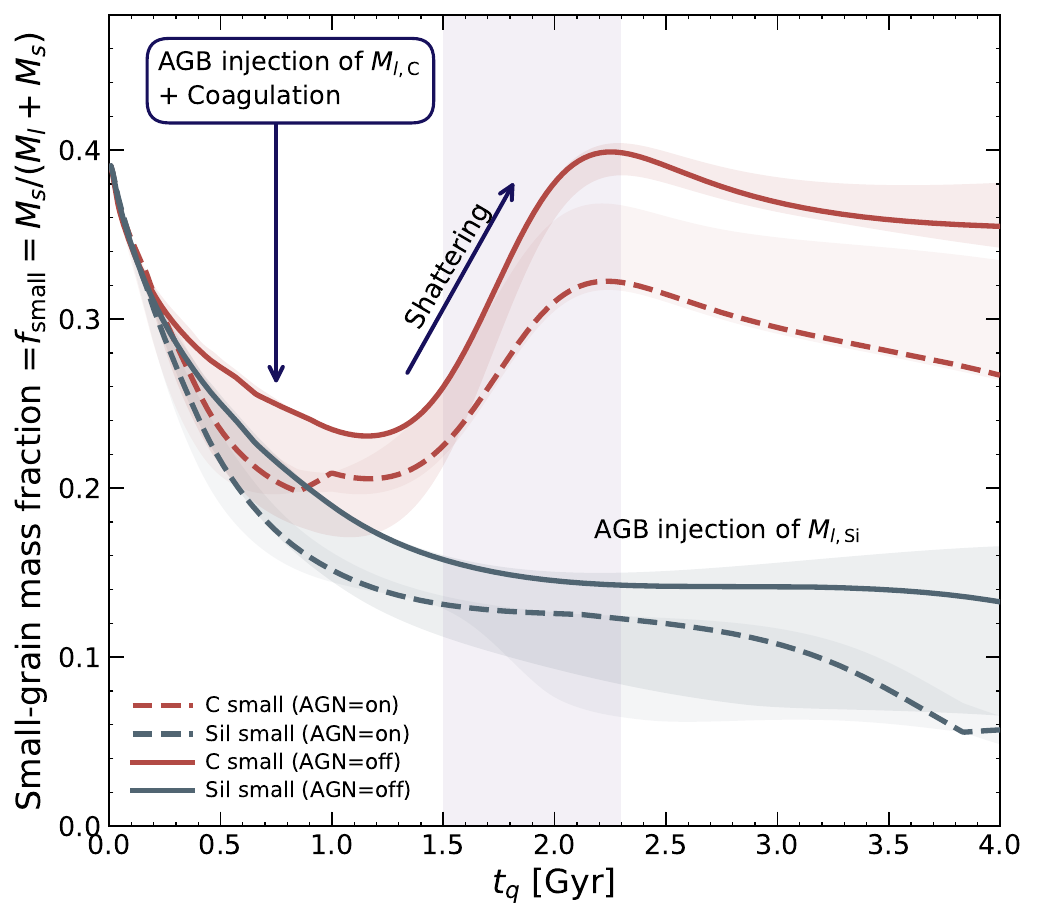}
\caption{
Median evolution of the small-grain mass fraction,
$f_{\rm small}=M_s/(M_l+M_s)$, for $\mathrm{C}$-grains
(red) and $\mathrm{Sil}-$grains (slate). Solid and dashed curves show
no-AGN and AGN-on models, respectively. The tracks show the median across
all three destruction modes, with 16th--84th percentile ranges.
Annotations mark the key processes driving the early decline, the
carbonaceous upturn (highlighted with the shaded area), and the late-time dominance of TP-AGB injected large $\mathrm{Sil}$ grains.}
\label{fig:Fig5b}
\end{figure}

Figure~\ref{fig:Fig5b} shows that the median post-quenching evolution of a small-grain mass fraction ($f_{\rm small}=M_{s}/(M_{l}+M_{s})$) is strongly composition-dependent. During the first $t_{\rm q}\lesssim1$ Gyr, both $f_{\rm small,{\rm C}}$ and $f_{\rm small,{\rm Sil}}$ exhibit decline which is attributed to two channels: coagulation, which transfers mass from small to large grains, and TP-AGB predominantly injecting large grains, which we initialised with a small-to-large grain mass ratio of 1:4. At $t_{\rm q}\sim0.5$--$0.6$ Gyr, ISM growth becomes effective, but is first mainly sustains the total dust reservoir rather than immediately shifting it toward small grains. Within $t_{\rm q}\sim1.4$--$2.3$ Gyr, the small $\mathrm{C-}$grains develop a temporary rise independently of AGN presence. The accumulated carbonaceous TP-AGB reservoir provides more large C grains for shattering, which overcomes coagulation and destruction. This creates delayed $M_{s,{\rm C}}$ upturn even though the strongest phase of ISM regrowth has already begun earlier. The silicates reservoir do not show the same trend at $t_{\rm q}\gtrsim1.5$ Gyr, as it remains more strongly weighted toward large-grain injection\footnote{We show underlying C- and $\mathrm{Sil-}$production rates in Appendix~\ref{AppendixC}.}. This increases $M_{l,{\rm Sil}}$, while coagulation and efficient small-grain destruction keep $f_{{\rm small},{\rm Sil}}$ flat ($f_{\rm small}\sim0.1$).

A qualitatively similar non-monotonic evolution has been found for composition-integrated $f_{\rm small}$ in the \textsc{COLIBRE} simulations \citep{vijayan26}. Their stellar dust is initialised with small-to-large fraction of 1:9, after which accretion rapidly rises the small-grain reservoir to $f_{\rm small}\sim0.4-0.5$. The range spanned by the \texttt{UNDUST} tracks broadly agrees with \textsc{COLIBRE} galaxies at $z\sim1$ over comparable masses ($M_\star\sim 8\times10^{10}$--$10^{11}\,M_\odot$). Importantly, our post-quenching evolution split for C- and $\mathrm{Sil-}$rich reservoirs enable revealing a temporary carbon enhancement that would otherwise be diluted in a composition-integrated ratio.

Local SFGs with sSFRs comparable to those probed by our model typically show $f_{\rm small}\lesssim0.2$ \citep{relano22}. These values are consistent with our recently quenched systems ($t_q<0.5$ Gyr), whereas at $t_q\gtrsim1$ Gyr our tracks exceed the range seen in local SFGs. This delayed enhancement of the $\mathrm{C}$-small reservoir, despite declining total $M_{\rm dust}$, provides additional physical context for the evolution of the PAHs discussed in Sect.~\ref{sec:pah_evolution}.

%=====================================================================
\subsection{Evolution of PAHs}
\label{sec:pah_evolution}
\begin{figure}[h!]
\centering
\includegraphics[width=0.5\textwidth]{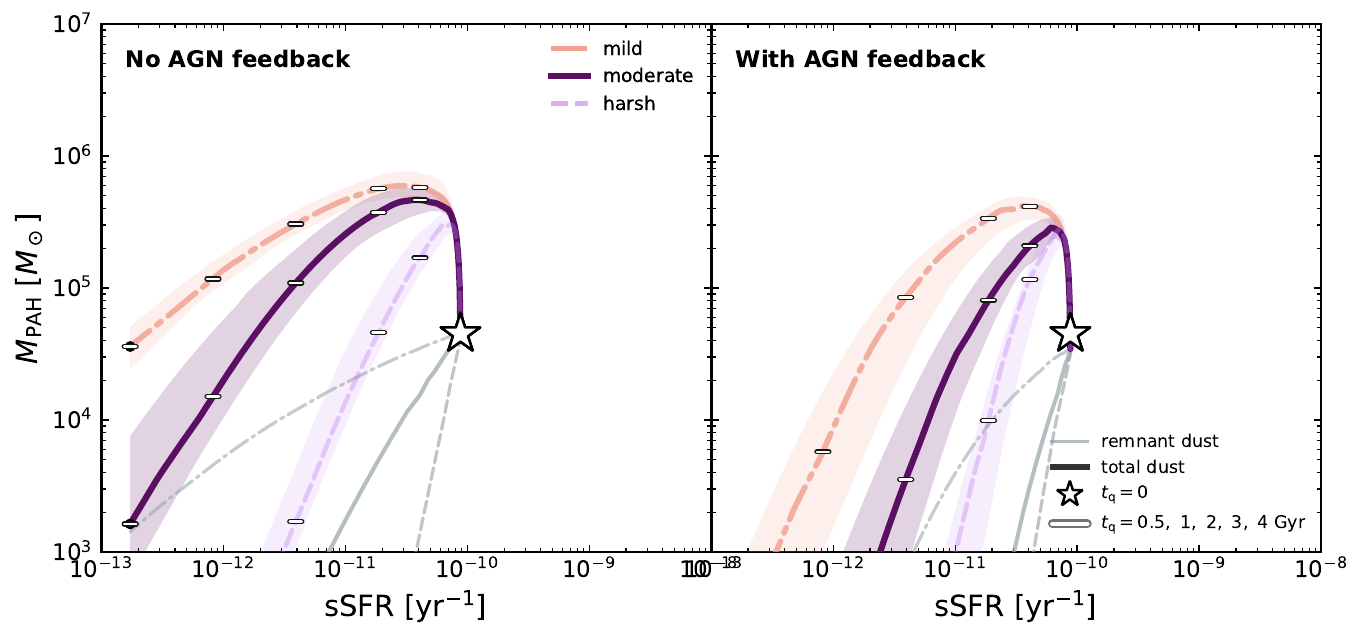}
\caption{Evolution of PAH mass, $M_{\rm PAH}$, with sSFR. Left and right panels correspond to models without and with moderate AGN outflows. The meaning of colored and grey tracks is the same as in Fig.~\ref{fig:Fig2}. Stars mark the onset of quenching, and horizontal white bars along the tracks indicate successive post-quenching times.}
\label{fig:Fig5}
\end{figure}

%=========================================================
%-------------------------------------------
The PAH mass reservoir follows diverse post-quenching pathways that broadly mirror the evolution of dust fraction (Fig.~\ref{fig:Fig5}). Models including TP-AGB injection and ISM grain growth maintain $M_{\rm PAH}\gtrsim10^5$--$10^6\,M_\odot$, whereas remnant tracks, especially with AGN feedback, decline rapidly because no fresh carbonaceous material is supplied after quenching. 

\begin{figure*}
\centering
\includegraphics[width=0.87\textwidth]{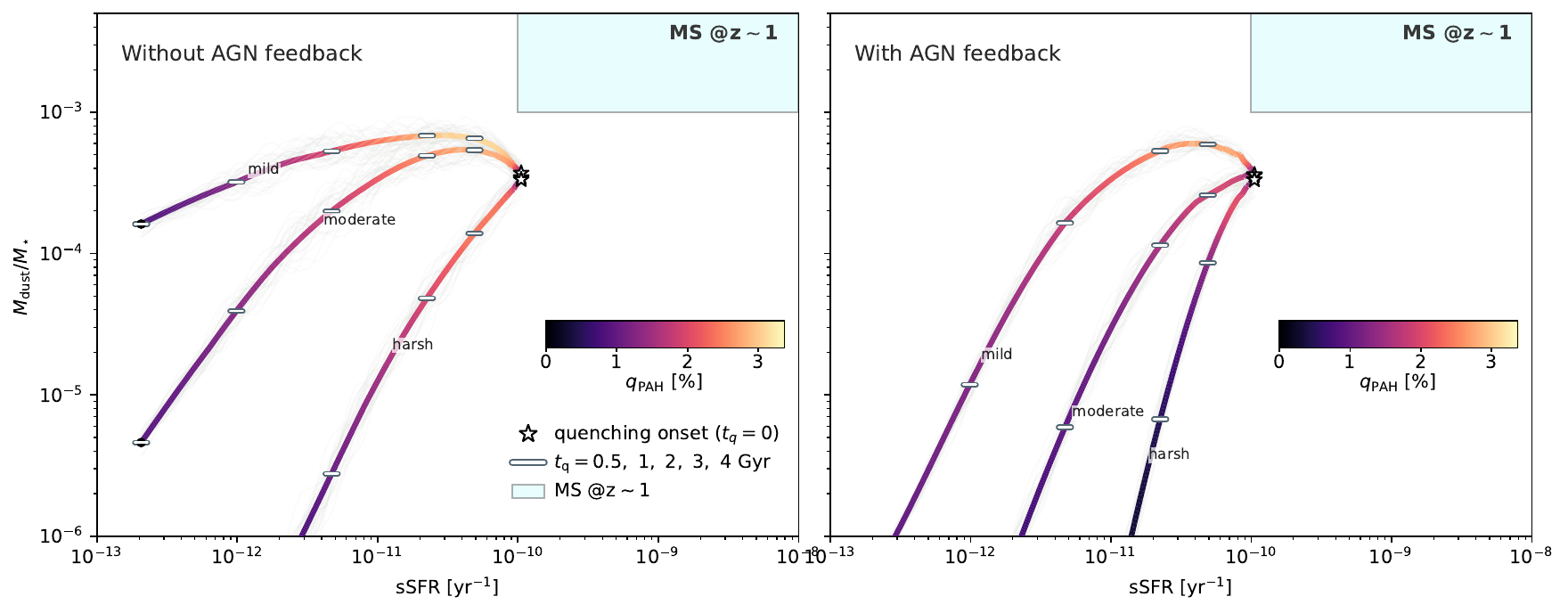}
\caption{PAH fraction in the $f_{\rm dust}$--sSFR plane. The left (right) panel shows \textit{total dust} models without (with) AGN feedback. Tracks relate to the mild, moderate, and harsh destruction regimes and are coloured by their PAH fraction, $q_{\rm PAH}$. Stars show the onset of quenching, and white bars mark successive post-quenching times (up to 4 Gyr) as indicated in the legend. Light cyan-blue region indicates the position in the plane for MS SFGs at $z\sim1$. Faded grey curves in the background show the individual model realisations underlying the coloured tracks.}
\label{fig:Fig6}
\end{figure*}

The temporary enhancement of $M_{s,{\rm C}}$ in Fig.~\ref{fig:Fig5b} provides an
additional reservoir for PAH formation, although it does not translate into a
comparable increase in $M_{\rm PAH}$. As we describe in Sec 2.5, in \texttt{UNDUST}, PAHs constitute an ultrasmall subset of the broader small-carbonaceous grain reservoir, rather than being identified with all small C-grains. This is motivated by models that associate PAHs with restricted C-grain intervals, typically $a\lesssim13$--$50\,\text{\AA}$, and, in some cases, require
the grains to be aromatic \citep{rau19, hirashita23, narayanan23}. We estimate that $\sim30\%$ of the total small carbon mass is transferred to PAHs. The remainder stays in the broader non-PAH small-C reservoir because it remains larger than the adopted PAH scale. This conversion bottleneck allows the small-C and PAH reservoirs to evolve independently after quenching. 
The resulting PAH response is moderate and is regulated by the balance between shattering, destruction, and the evolution of the total $M_{\rm dust}$. Consequently, if the real conversion of the shattered small C-grains into PAH-sized reservoir is more efficient, QGs could reach higher $M_{\rm PAH}$ particularly in AGN-off conditions.

In Fig.~\ref{fig:Fig6} we show the PAH fraction, (here defined as  $q_{\rm PAH}=(M_{\rm PAH}/M_{\rm dust})\times100$), as a function of sSFR. The models span from $q_{\rm PAH}\sim3.5\%$ during the early post-quenching phase to PAH-poor states ($q_{\rm PAH}\lesssim0.3\%$) at low sSFR, broadly consistent with the
diversity seen in resolved studies of local galaxies \citep{chestenet25}. We find that $q_{\rm PAH}$ generally rises during the first $\sim0.7$--$1$ Gyr, reaching $\sim3.5\%$ in the mild and moderate regimes and $\sim2.5\%$ under harsh processing with AGN feedback. This enhancement reflects effective shattering of large C-grains, together with delayed carbonaceous injection and survival of the residual grain reservoir, which enable QGs to match the medians of high-metalliticy and massive ($M_{\star}>5\times10^{10}\:M_{\odot}$) SFGs observed with JWST at $z\sim1-2$ (\citealt{shivaei24}). This suggests an important physical implication: $z\sim1$ galaxies with cold-dust fractions below typical MS values ($f_{\rm dust}\lesssim10^{-3}$) may still exhibit $q_{\rm PAH}$ comparable to those of coeval MS SFGs, especially if their evolution is unaffected by AGNs.

At later times, the PAH fraction is set by the balance between replenishment and destruction. Carbon-rich TP-AGB injection declines with time, while after $\sim2$ Gyr the delayed stellar contribution becomes increasingly dominated by $\mathrm{Sil}$-rich dust, broadly consistent with other AGB dust models in the literature \citep{ventura20}. For older stellar populations ($t_{\star}\gtrsim2$ Gyr), this transition occurs closer to the onset of quenching (Appendix~\ref{AppendixC}), shortening the interval over which enhanced $q_{\rm PAH}$ is expected. Once the $\mathrm{C-}$-rich TP-AGB contribution fades, PAH replenishment relies mainly on shattering of the remaining large C grains. Coagulation and radiative/shock destruction remove PAHs, while sputtering progressively reduces the broader small-C reservoir. Together, these processes drive most no-AGN tracks toward $q_{\rm PAH}\lesssim1.5\%$ after $\sim2$ Gyr.

AGN feedback changes this balance rather than destroying the PAH reservoir instantaneously. Depletion and heating of the cold ISM suppress grain growth and reduce the large-C reservoir available for shattering, while the lower dense-gas fraction also weakens coagulation. This partial compensation allows AGN-regulated models to retain $q_{\rm PAH}\gtrsim1.5\%$ during roughly the first $\sim$Gyr after quenching, despite their more rapid decline in total $M_{\rm dust}$. The
behaviour is qualitatively consistent with models in which shattering remains effective as the ISM becomes more diffuse and where declining $f_{\rm gas}$ induce temporarily elevated $q_{\rm PAH}$ \citep{narayanan26}. At later times, the same reduction in $f_{\rm gas}$ no longer supports enhanced PAH production: depletion of the large-C reservoir together with continued PAH destruction dominates over replenishment, driving most tracks toward low $q_{\rm PAH}$.  Overall, the diversity of $q_{\rm PAH}$ predicted by our model motivates systematic studies of PAHs in harsh conditions, for which early JWST observations already indicate diverse PAH survival in AGN environments \citep{bernete24}.

%--------------------------------------
\section{Observational implications for ISM evolution in quenched galaxies}

\label{sec:section5}
%--------------------------------------
%One of the main questions arising from the presented results is how the diverse fundamental dust physical processes can be observationally accessed given the depth of current instruments such as ALMA and JWST. In the following section, we explore the joint role of cold dust and PAHs in QGs. % with JWST beyond the threshold achievable with ALMA.
The above results show that, despite the many physical ingredients in \texttt{UNDUST}, the observable predictions are governed by a smaller set of dominant channels. The bulk dust budget is controlled mainly by delayed TP-AGB replenishment, ISM grain growth, thermal sputtering, and AGN-regulated suppression of the cold gas phase. The PAH reservoir is sensitive to the shattering of C-grains and, at the secondary level, grain distributions. Other processes (i.e., residual SN shocks, astration, shielding etc.) have secondary impact on the observed trends. %affect the normalisation and timing of the tracks, but have a more secondary impact on the main observable trends.

This has direct observational implications and naturally raises the question of whether the diverse post-quenching dust pathways can be distinguished with current facilities. In this section, we assess the joint diagnostic role of cold dust and PAHs in QGs using representative sensitivity limits for ALMA and JWST.

We link the \texttt{UNDUST} tracks to MIR observables by post-processing
the model outputs with the dust-emission library of \cite{dl07} (hereafter DL07). We assign the closest DL07 template to the modeled $q_{\rm PAH}$ and scale its luminosity by instantaneous $M_{\rm dust}$. Since we do not solve radiative transfer self-consistently, we follow models that compute dust and PAH SEDs
under an assumed radiation-field intensity \citep[e.g.,][]{hirashita20, matsumoto24}. We adopt a diffuse interstellar radiation field (ISRF) with constant Milky Way-normalised intensity $U=1$, representative of massive, metal-rich QGs at $z\sim1$ \citep{magdis2021interstellar}, and a minimal high-intensity component ($\gamma=0.001$), corresponding to $0.1\%$ of $M_{\rm dust}$ exposed to the DL07 power-law ISRF distribution. This mimics weak residual heating without introducing a star-forming/PDR component. We verify that the qualitative location of the tracks in the sSFR--$f_{\rm dust}$ plane is unchanged under this fiducial excitation. A stronger (weaker) ISRF would increase (decrease) the absolute MIR flux per unit $M_{\rm PAH}$, but would not erase the distinction between genuinely dust-poor systems and ALMA-faint systems retaining a non-negligible PAH component. The absolute MIRI fluxes should therefore be interpreted as estimates for a weakly heated ISM. The resulting rest-frame SEDs are redshifted and convolved with the relevant JWST/MIRI transmission curves. For our exemplary case, we focus on observed-frame F1500W, which at $z\sim1$ probes rest-frame $\simeq6.8$--$8.3\,\mu$m emission, and encompasses the charge-sensitive
$7.7\,\mu$m PAH complex together with the underlying dust continuum.

\subsection{Tracing cold dust and PAH pathways in QGs with ALMA and JWST}
\label{sec:sec5.1}
\begin{figure*}
\centering
\includegraphics[width=0.94\textwidth]{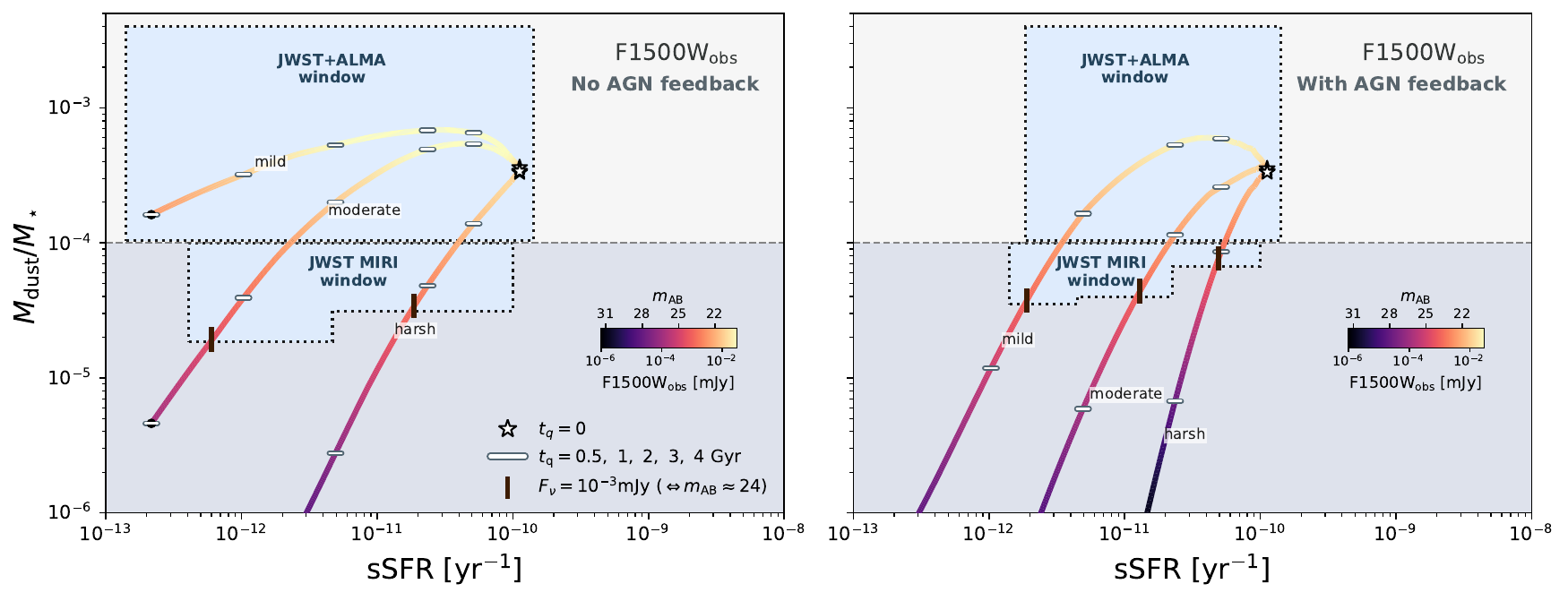}
\caption{
Modeled evolution of JWST/MIRI $\rm F1500W_{\rm obs}$ fluxes (corresponding to $\rm F770W_{\rm rest}$ for the QGs at $z\sim1$) within the $f_{\rm dust}$-sSFR plane. Panels, tracks and symbols (stars and horizontal bars) have the same meaning as in Fig.~\ref{fig:Fig6}. Vertical markers indicate where the modeled tracks reach a representative $5\sigma$ MIRI flux limit of $F_\nu\!\sim\!1\:\mu$Jy ($m_{\rm AB}\approx24$) typical for deep JWST MIRI surveys. The dotted regions denote observability windows: the upper "JWST+ALMA" window broadly corresponds to systems detectable both in deep MIRI emission and in deep ALMA dust continuum, while the lower "JWST MIRI" window highlights the regime where QGs may fall below the adopted ALMA threshold based on literature detection range ($M_{\rm dust}/M_\star\gtrsim10^{-4}$), but remain detectable with MIRI.} %AGN outflows shift the tracks toward lower dust fractions and shorter MIRI-detectable phases, but still allow a transient regime in which residual PAH emission traces small-grain ISM material missed by cold-dust continuum observations alone.}
\label{fig:Fig7}
\end{figure*}

Figure~\ref{fig:Fig7} shows the evolution of the predicted broad-band JWST/MIRI $\rm F1500W_{\rm obs}$ flux across the
$f_{\rm dust}$--sSFR plane. We define two schematic observability regimes. The upper "JWST+ALMA" window corresponds to objects with $M_{\rm dust}/M_\star \gtrsim 10^{-4}$, approximately the deepest cold-dust continuum regime accessible in individual ALMA Band~6/7 observations with $\textrm{rms}$ sensitivities of $\sim 8-15\,\mu$Jy (\citealt{lorenzon25b, chang26}). The lower "JWST MIRI" window marks the regime where galaxies fall below this cold-dust threshold but remain detectable in deep MIRI imaging, adopting a representative $\rm S/N\!\approx5$ limit of $F_\nu \!\sim \!1\,\mu$Jy ($m_{\rm AB}\approx24$), comparable to surveys such as SMILES \citep{alberts24}.\footnote{At $z\sim1$, $F1500W$ samples the redshifted $7.7\,\mu$m PAH complex; therefore, the colour-coding should be interpreted as a PAH-sensitive broad-band MIR flux predicted from the \texttt{UNDUST} dust and PAH masses, rather than as a pure PAH-line flux.} We find that the non-monotonic evolution of $f_{\rm dust}$ and $q_{\rm PAH}$ produces a broad range of F1500W at $z\!\sim\!1$. This is in line with the early studies of MIR emission in QGs at $z<0.7$ which find observed fluxes varying by $\gtrsim2$ dex at fixed $M_{\star}$ \citep{sase23b}. 

%The main result is that cold dust, PAHs, and MIR emission do not fade on the same timescale after quenching.
%In both the no-AGN and AGN-regulated modes, ongoing dust reformation keeps QGs in the JWST+ALMA window for$\gtrsim2$ Gyr in the mild destruction mode, $\gtrsim1.5-3$ Gyr in the moderate mode, and $\lesssim0.4-0.7$ Gyr in the harsh mode. 

%One of key predicitons of our model is illustrated in that cold dust and PAH-sensitive MIR emission do not fade on the same timescale after quenching. 
From Fig.~\ref{fig:Fig7} we see that in both the AGN-off and AGN-on models, ISM grain growth and TP-AGB dust can keep QGs within the JWST+ALMA window for $\gtrsim2-4$ Gyr in the mild destruction mode, $\gtrsim1.5$--$2.5$ Gyr in the moderate mode, and $\lesssim0.4$--$0.7$ Gyr in the harsh mode. Once the cold-dust reservoir falls below $M_{\rm dust}/M_\star\sim10^{-4}$, galaxies become ALMA faint. However, residual dust replenishment by TP-AGB stars and grain processing contribute to prolonged dust survival and PAH formation. As a consequence, %PAH-bearing component may remain observable with MIRI. This occurs because PAHs do not evolve strictly in lockstep with the total dust mass: shattering of surviving large carbonaceous grains, together with delayed AGB enrichment, can maintain $q_{\rm PAH}\sim1$--$3\%$ even when the galaxy becomes faint in ALMA continuum.
this produces a transient "JWST MIRI" phase in which dust in QGs remains detectable through PAH-sensitive MIR emission. For the mild and moderate tracks, this window remains accessible to $\log(\rm sSFR/yr)\gtrsim-12$ for the first $\sim2$ Gyr after quenching in the AGN-off case, and for $\sim1.2$--$1.5$ Gyr when AGN feedback is active. In the harsh mode, the corresponding phase shortens to $\sim1.2$ Gyr in AGN-off models, and to $\sim0.6$ Gyr in AGN-on models. We note that the AGN-on and AGN-off tracks represent alternative evolutionary channels rather than an observational classification of the sources. Our synthetic MIRI fluxes include both the PAH features and the
underlying C- and $\mathrm{Sil-}$ dust continuum provided by the DL07 templates. However, here we do not add a separate torus component to AGN-on models. In practice, at $z\sim1$, H$_2$ $S(5)$, [\ion{Ar}{ii}] $6.99\,\mu{\rm m}$, and, in AGN hosts, [\ion{Ne}{vi}] $7.65\,\mu{\rm m}$ also fall within F1500W. Hence, observed AGN hosts may show enhanced MIR continuum and reduced PAH equivalent widths relative to our predictions. Multi-band continuum subtraction or MIRI spectroscopy is thus required to isolate the PAH contribution \citep{donnelly25}.

This prolonged MIR-bright phase qualitatively agrees with \citealt{kelson10}, who argued that TP-AGB stars are important contributor to the MIR luminosity of post-SBs at $z\sim1-2$. As discussed in Section~\ref{sec:section4}, PAH replenishment becomes inefficient at later times, while destruction keeps eroding small C-grains. The tracks then enter a \textit{genuinely dust-poor} phase in which both dust continuum and MIR emission are weak.

We also test the sensitivity of our PAH predictions to the uncertain efficiency of AGB carbon dust into PAHs. We find that setting $\epsilon_{\rm PAH,AGB}=0$ does not erase the PAH reservoir: $q_{\rm PAH}$ remains close to the fiducial tracks, especially during the first $\sim1$ Gyr after quenching. This shows that PAHs are sustained mainly by the survival and shattering of C-grains. Increasing the efficiency to $\epsilon_{\rm PAH,AGB}=0.5$ raises $q_{\rm PAH}$ by $\sim0.1$--$0.3$ dex, depending on the mode, implying conservative MIRI fluxes if TP-AGB carbon dust is efficiently processed into PAH-sized grains. 

We therefore caution that ALMA non-detections should not be interpreted automatically as evidence for dust-free QGs. Figure~\ref{fig:Fig7} can help distinguish genuinely dust-poor systems from ALMA-faint QGs that retain a chemically enriched, PAH-bearing reservoir accessible to JWST/MIRI. It is important to note that although the figure predicts a relation between PAH-sensitive MIR emission and cold dust along individual evolutionary tracks, $q_{\rm PAH}$ alone does not uniquely determine $M_{\rm dust}$. PAHs may temporarily survive after the bulk dust reservoir declines, so similar $q_{\rm PAH}$ values can correspond to different cold $M_{\rm dust}$ and evolutionary pathways. Joint JWST measurements of PAH-sensitive emission and ALMA constraints on the cold-dust continuum, combined with $M_\star$, sSFR, and $t_{\rm q}$, can instead test the predicted loci and place model-dependent constraints on the remaining dust reservoir. 

\subsection{The role of dust reformation and processing in interpreting post-quenching galaxy evolution}
\label{sec:sec5.2}
The tracks presented in Section~\ref{sec:sec5.1} may translate into several key observational tests of dust processing accros the post-quenching phase. First, rapidly quenched galaxies or post-SBs (e.g., \citealt{greene20, whitaker21b, smercina22, davies24}) emerge as powerful probes of rapid dust destruction and AGN-regulated removal. AGN feedback shortens the MIRI-detectable phase, but does not necessarily erase $M_{\rm PAH}$ immediately unless destruction is maximally efficient. Recent JWST/MIRI observations of Cen A reveal PAH presence  in the AGN-regulated ISM despite strong processing and localised erosion \citep{pantoni26}. A deep MIRI non-detection of a young, ALMA-faint recently quenched QG would favour a harsh-destruction scenario, especially with AGN feedback. Conversely, a MIRI detection would indicate that the system is not simply dust-poor, but may retain a processed PAH-bearing component and residual dust below the ALMA detection limit. 

Second, QGs with residual dust fractions near $M_{\rm dust}/M_\star\sim10^{-4}$ may be especially important to disentangle ISM grain growth from the injection of TP-AGB dust. In the tracks shown in Fig.~\ref{fig:Fig4}, TP-AGB dust production can preserve up to $\sim50\%$ of the initial dust while still leaving the galaxy below the nominal ALMA detection threshold. %Since \texttt{UNDUST}is a one-zone model, we do not predict attenuation profiles directly. 
If centrally concentrated, this dust reservoir could produce significant reddening while remaining difficult to detect with ALMA in cold-dust continuum. Such systems may naturally explain reddened or highly attenuated cores recently reported in high-$z$ QGs \citep[e.g.][]{setton24,siegel2025, barrufet25, wang25}. %and in simulations \citep{akins2022quenching, matsumoto24}. 

Third, JWST/NIRSpec can test for intermediate-age TP-AGB activity in QGs. Recent detections of TP-AGB-sensitive spectral features in $z\sim1.5$ QGs and post-SBs demonstrate that this is now feasible \citep{lu25,bevacqua25}. Therefore, combining NIRSpec constraints on intermediate-age stellar populations with MIRI measurements of PAH-sensitive emission and ALMA limits on cold dust would directly test whether PAHs in quenched galaxies are maintained by shattering of carbonaceous grains, delayed AGB injection, or both. 

Finally, the presence of reformed dust in QGs can significantly affect interpretation of SED modeling results. %as dust-related quantities can remain close to the SF values ($f_{\rm dust}\sim10^{-3}, \delta_{\rm DGR}\sim1/100$), 
For example, energy-balance SED-fitting codes may misinterpret dust emission as obscured star formation rather than post-quenching ISM dust processing. Some recent studies demonstrate that inclusion of dust/IR constraints improves SED fits, but buried SF is generally not required to explain $\rm H_2$ depletion trend (\citealt{setton25}) or attenuation (\citealt{lisiecki26}). 

In general, QGs provide unique laboratories for fundamental dust physics, as many degeneracies induced by SF-driven ISM cycling (e.g., \citealt{nanni20}) are largely absent. Our model suggests that dust is also essential for interpreting post-quenching ISM conditions beyond the cold gas reservoir. Joint ALMA CO+dust continuum and JWST/MIRI observations can directly test whether QGs with similar $f_{\rm H_2}$ and $t_{\rm q}$ follow different dust-processing pathways. Deep MIRI surveys such as SMILES \citep{alberts24} and CEERS/MEGA \citep{backhaus25} are particularly suited to probe the MIRI-only window identified in Section~\ref{sec:sec5.1}. Looking ahead, PRIMA \citep{prima25} will provide MIR-to-FIR coverage and depth to constrain $q_{\rm PAH}$, $T_{\rm dust}$ and grain sizes up to $z\sim4$ \citep{long26}. This will help constrain the role of AGN feedback in QGs and distinguish among the dust-evolution scenarios proposed here.

%--------------------------------------
\subsection{Model caveats and future theoretical prospects}
%--------------------------------------

In this work, we explore plausible post-quenching dust pathways rather than provide unique fits to individual QGs. We adopt empirically motivated initial gas/dust ratios and grain populations, while the SFHs represent typical rather than the full range of progenitor histories. A complete treatment of population aging requires a redshift-dependent grid, since QGs observed at different epochs may enter quenching with different stellar ages, metallicities, and TP-AGB progenitor populations. For example, younger high-$z$ QGs may experience stronger delayed C-rich enrichment from TP-AGB progenitors of initial mass $\sim2$--$3,M_\odot$, whereas older, more metal-rich low-$z$ QGs may transition more rapidly toward $\mathrm{Sil}$-rich AGB injection. We also deliberately focus on the physical origin of the predicted $\delta_{\rm DGR}$ and $f_{\rm dust}$ at a fixed reference $M_{\star}$, leaving mass- and morphology-dependent pathways to future model extensions. Moreover, the total modeled $M_{\rm dust}$ need not correspond directly to that recoverable from ALMA continuum observations, since part of the reservoir may remain too cold to contribute appreciably at the observed wavelengths. Quantifying this requires radiative-transfer (RT) modeling of the dust-heating distribution. In Donevski \& Nanni (in prep.), we address these caveats by coupling redshift-dependent stellar evolution with RT predictions for QGs across key evolutionary phases, from cold-dust/PAH-rich to dust-poor states.

Our AGN implementation is phenomenological, based on prescribed outflow properties and heating, rather than on a self-consistent quenching model tied to an evolving AGN luminosity or coupling efficiency. A future extension will couple the time-dependent AGN luminosity and radiation field to gas and dust removal and heating, allowing the framework to predict both the dynamical impact of feedback and the corresponding IR observables. A dedicated analysis of AGN duty cycles and quenching will be presented in Lorenzon et al. (in prep.). %The fiducial $z<1$ models adopt $\eta_{\rm AGN}=1$, while the mild, moderate, and harsh modes represent effective levels of gas heating, sputtering, shielding loss, and dense-gas suppression, rather than specific AGN duty cycles, outflow geometries, or coupling efficiencies. Therefore, these tracks should be interpreted as controlled experiments on how AGN-regulated ISM conditions affect dust survival and re-growth, not as a self-consistent AGN-quenching model. 

Finally, in the present model we follow mass exchange among large/small grains, and PAHs using a simplified two-size description. Because the current model does not yet evolve a grain-size distribution, PAH charges or spatially resolved ISM phases, it cannot robustly predict the relative evolution of ionised and neutral PAH features. While implementing this is out of scope of the present study, future inclusion of self-consistent grain-size distribution in \texttt{UNDUST} model will provide a refined treatment of PAH production, destruction, and conversion between small and large grains, and will enable more realistic predictions for JWST/MIRI and future IR observations.

%-------------------------------------------------------------
\section{Conclusions}
\label{sec:conclusions}

%-------------------------------------------------------------

We presented \texttt{UNDUST}, to our knowledge the first 
semi-analytic model designed to follow the coupled post-quenching
one-zone evolution of cold gas, dust, metals, stars and PAHs in massive galaxies within a two-size framework for carbonaceous and silicate grains. By separating the roles of inherited remnant dust, delayed TP-AGB injection, ISM grain growth, and AGN feedback, we explored how dust and PAHs continue to evolve after star
formation has ceased. Our main results are:

\begin{itemize}

\item
Post-quenching dust evolution can decouple from cold gas depletion and
stellar-population aging. For QGs quenched at $z\sim1$ with stellar-population ages of $\sim2$ Gyr, grain replenishment and processing generate a wide range of cold dust states. At comparable $f_{\rm H_2}$ or $t_{\rm q}$, QGs can span SFG-like ($\delta_{\rm DGR}\sim10^{-2}$) to strongly depleted regimes
($\delta_{\rm DGR}\lesssim10^{-4}$--$10^{-5}$). Dust therefore provide an
independent probe of ISM conditions in QGs, as departures from
$\rm H_2$--gas scaling relations do not uniquely determine $f_{\rm dust}$ or
$\delta_{\rm DGR}$.

\item
Delayed TP-AGB injection and ISM grain growth can compete with dust destruction for up to $\sim2.5$ Gyr after quenching, and help rebuild more than half of the initial $M_{\rm dust}$. Crucially, in mild and moderate ISM destruction regimes, re-accretion can proceed within 0.4-0.7 Gyr, remaining effective even at limited dense-gas fractions ($f_{\rm dense}\lesssim0.2$), typical for QGs and post-SBs. This suggests that TP-AGB dust injection and ISM grain growth may be important and potentially common physical processes in late-stage galaxy evolution despite suppressed star formation.

\item
AGN feedback removes and heats the cold gas, suppressing the material available for grain re-accretion. At fixed sSFR, this accelerates the decline in $f_{\rm dust}$ by a factor of $\sim2$--$3$ relative to the AGN-off case, yielding $f_{\rm dust}<10^{-4}$ (> 1 dex below the dust MS) within $\lesssim1$ Gyr. Thus, AGN feedback can drive QGs into a dust-poor state primarily by preventing dust regrowth, without requiring direct grain destruction.

%\item 
%Similar $f_{\rm dust}$ and $\delta_{\rm DGR}$ can arise from different combinations of destruction, replenishment, and AGN-driven effects, requiring joint constraints on dust, cold gas, and stars in a multiparameter space that includes sSFR, $f_{\rm dust}$, $\delta_{\rm DGR}$ and $t_{\rm q}$ to break these degeneracies.

\item
Grain processing remain active long after quenching: silicate grains dominate the large grain budget required for cold dust, while shattering of TP-AGB supplied large C grains replenishes the small-C reservoir for up to $\sim2.5$ Gyr. This produces diverse PAH fractions ($q_{\rm PAH}\sim0.3$--$3.5\%$) that can temporarily match those of metal-rich SFGs at $z\!\sim\!1$ despite the declining cold dust reservoir.

\item
QGs can become ALMA-faint before becoming chemically depleted. Even after the bulk cold-dust mass falls below typical ALMA continuum limits ($M_{\rm dust}/M_\star\lesssim10^{-4}$), some systems retain $q_{\rm PAH}\sim2$--$3\%$ and PAH-sensitive MIR emission for $\sim1$--$2$ Gyr after quenching. ALMA non-detections therefore need not imply chemical depletion, but may trace a phase in which cold dust is faint while C-grain processing remains observable with JWST/MIRI.

\end{itemize}

Altogether, our results establish dust and PAHs as key time-sensitive diagnostics of the evolutionary stage and dominant ISM-processing channels after quenching. In a broader context, QGs may provide a clean laboratory for fundamental dust physics, while demonstrating why no single ISM tracer can fully describe their post-quenching state. Joint JWST, ALMA, and future FIR observations will be pivotal to establish how long chemically enriched dust processing persists after star formation ends.

\begin{acknowledgements}

D.D acknowledges support from the Polish National Agency for Academic Exchange (Bekker grant BPN /BEK/2024/1/00029/DEC/1).  A.N. acknowledges support from the Narodowe Centrum Nauki (NCN), Poland, through the SONATA BIS grant UMO-2020/38/E/ST9/00077. AWSM acknowledges the support of the Natural Sciences and Engineering Research Council of Canada (NSERC) through grant reference numbers RGPIN-2021-03046 and RGPIN-2026-07024. AL has been supported by the Istituto Nazionale di Fisica Nucleare (INFN) via the specific national initiative QGSKY. DN was supported by NASA grants ATP-21-0013,  ATP-23-0002 and ADSPS-23-0007. This research was supported by the Polish National Agency for Academic Exchange under the CLEVER, Strategic Partnership program BPI/PST/2024/1/00019. This paper makes use of the following ALMA data: ADS/JAO.ALMA\#2024.1.00814.S. ALMA is a partnership of ESO (representing its member states), NSF (USA) and NINS (Japan), together with NRC (Canada), NSTC and ASIAA (Taiwan), and KASI (Republic of Korea), in cooperation with the Republic of Chile. The Joint ALMA Observatory is operated by ESO, AUI/NRAO and NAOJ. This research made use of Photutils, an Astropy package for detection and photometry of astronomical sources (Bradley et al. 2024).

\end{acknowledgements}


\bibliographystyle{aa} % style aa.bst
\bibliography{bibliography} % your references Yourfile.bib

@ARTICLE{simba2019,
       author = {{Dav{\'e}}, Romeel and {Angl{\'e}s-Alc{\'a}zar}, Daniel and {Narayanan}, Desika and {Li}, Qi and {Rafieferantsoa}, Mika H. and {Appleby}, Sarah},
        title = "{SIMBA: Cosmological simulations with black hole growth and feedback}",
      journal = {\mnras},
         year = 2019,
        month = jun,
       volume = {486},
       number = {2},
        pages = {2827-2849},
          doi = {10.1093/mnras/stz937},
archivePrefix = {arXiv},
       eprint = {1901.10203},
 primaryClass = {astro-ph.GA},
       adsurl = {https://ui.adsabs.harvard.edu/abs/2019MNRAS.486.2827D}
}

@article{dwek1998evolution,
  title={The evolution of the elemental abundances in the gas and dust phases of the galaxy},
  author={Dwek, Eli},
  journal={The Astrophysical Journal},
  volume={501},
  number={2},
  pages={643},
  year={1998},
  publisher={IOP Publishing}
}

@ARTICLE{blanquez2023gas,
       author = {{Bl{\'a}nquez-Ses{\'e}}, D. and {G{\'o}mez-Guijarro}, C. and {Magdis}, G.~E. and {Magnelli}, B. and {Gobat}, R. and {Daddi}, E. and {Franco}, M. and {Whitaker}, K. and {Valentino}, F. and {Adscheid}, S. and {Schinnerer}, E. and {Zanella}, A. and {Xiao}, M. and {Wang}, T. and {Liu}, D. and {Kokorev}, V. and {Elbaz}, D.},
        title = "{The gas mass reservoir of quiescent galaxies at cosmic noon}",
      journal = {\aap},
         year = 2023,
        month = jun,
       volume = {674},
          eid = {A166},
        pages = {A166},
          doi = {10.1051/0004-6361/202345977},
archivePrefix = {arXiv},
       eprint = {2303.12110},
 primaryClass = {astro-ph.GA},
       adsurl = {https://ui.adsabs.harvard.edu/abs/2023A&A...674A.166B}
}

@article{magdis2021interstellar,
  title={The interstellar medium of quiescent galaxies and its evolution with time},
  author={Magdis, Georgios E and Gobat, Raphael and Valentino, Francesco and Daddi, Emanuele and Zanella, Anita and Kokorev, Vasily and Toft, Sune and Jin, Shuowen and Whitaker, K},
  journal={arXiv preprint arXiv:2101.04700},
  year={2021}
}

@ARTICLE{lee2023high,
       author = {{Lee}, Minju M. and {Steidel}, Charles C. and {Brammer}, Gabriel and {F{\"o}rster-Schreiber}, Natascha and {Renzini}, Alvio and {Liu}, Daizhong and {Herrera-Camus}, Rodrigo and {Naab}, Thorsten and {Price}, Sedona H. and {{\"U}bler}, Hannah and {Arriagada-Neira}, Sebasti{\'a}n and {Magdis}, Georgios},
        title = "{High dust content of a quiescent galaxy at z   2 revealed by deep ALMA observation}",
      journal = {\mnras},
         year = 2024,
        month = feb,
       volume = {527},
       number = {4},
        pages = {9529-9547},
          doi = {10.1093/mnras/stad3718},
archivePrefix = {arXiv},
       eprint = {2311.00023},
 primaryClass = {astro-ph.GA},
       adsurl = {https://ui.adsabs.harvard.edu/abs/2024MNRAS.527.9529L}
}

@article{speagle2014highly,
  title={A highly consistent framework for the evolution of the star-forming “main sequence” from z~ 0--6},
  author={Speagle, Joshua S and Steinhardt, Charles L and Capak, Peter L and Silverman, John D},
  journal={The Astrophysical Journal Supplement Series},
  volume={214},
  number={2},
  pages={15},
  year={2014},
  publisher={IOP Publishing}
}

@ARTICLE{donevski23,
       author = {{Donevski}, D. and {Damjanov}, I. and {Nanni}, A. and {Man}, A. and {Giulietti}, M. and {Romano}, M. and {Lapi}, A. and {Narayanan}, D. and {Dav{\'e}}, R. and {Shivaei}, I. and {Sohn}, J. and {Junais} and {Pantoni}, L. and {Li}, Q.},
        title = "{In pursuit of giants. II. Evolution of dusty quiescent galaxies over the last six billion years from the hCOSMOS survey}",
      journal = {\aap},
         year = 2023,
        month = oct,
       volume = {678},
          eid = {A35},
        pages = {A35},
          doi = {10.1051/0004-6361/202346066},
archivePrefix = {arXiv},
       eprint = {2304.05842},
 primaryClass = {astro-ph.GA},
       adsurl = {https://ui.adsabs.harvard.edu/abs/2023A&A...678A..35D}
}

@ARTICLE{nanni20,
       author = {{Nanni}, A. and {Burgarella}, D. and {Theul{\'e}}, P. and
         {C{\^o}t{\'e}}, B. and {Hirashita}, H.},
        title = "{The gas, metal, and dust evolution in low-metallicity local and high-redshift galaxies}",
      journal = {\aap},
         year = 2020,
        month = sep,
       volume = {641},
          eid = {A168},
        pages = {A168},
          doi = {10.1051/0004-6361/202037833},
archivePrefix = {arXiv},
       eprint = {2006.15146},
 primaryClass = {astro-ph.GA},
       adsurl = {https://ui.adsabs.harvard.edu/abs/2020A&A...641A.168N}
}

@ARTICLE{sase23b,
       author = {{Bl{\'a}nquez-Ses{\'e}}, D. and {Magdis}, G.~E. and {G{\'o}mez-Guijarro}, C. and {Shuntov}, M. and {Kokorev}, V. and {Brammer}, G. and {Valentino}, F. and {D{\'\i}az-Santos}, T. and {Paspaliaris}, E. -D. and {Rigopoulou}, D. and {Hjorth}, J. and {Langeroodi}, D. and {Gobat}, R. and {Jin}, S. and {Sillassen}, N.~B. and {Gillman}, S. and {Greve}, T.~R. and {Lee}, M.},
        title = "{Uncovering the MIR emission of quiescent galaxies with JWST}",
      journal = {\aap},
         year = 2023,
        month = nov,
       volume = {679},
          eid = {L2},
        pages = {L2},
          doi = {10.1051/0004-6361/202347771},
archivePrefix = {arXiv},
       eprint = {2310.01601},
 primaryClass = {astro-ph.GA},
       adsurl = {https://ui.adsabs.harvard.edu/abs/2023A&A...679L...2B}
}

@ARTICLE{morishita22,
       author = {{Morishita}, T. and {Abdurro'uf} and {Hirashita}, H. and {Newman}, A.~B. and {Stiavelli}, M. and {Chiaberge}, M.},
        title = "{Compact Dust Emission in a Gravitationally Lensed Massive Quiescent Galaxy at z = 2.15 Revealed in 130 pc Resolution Observations by the Atacama Large Millimeter/submillimeter Array}",
      journal = {\apj},
         year = 2022,
        month = oct,
       volume = {938},
       number = {2},
          eid = {144},
        pages = {144},
          doi = {10.3847/1538-4357/ac9055},
archivePrefix = {arXiv},
       eprint = {2208.10525},
 primaryClass = {astro-ph.GA},
       adsurl = {https://ui.adsabs.harvard.edu/abs/2022ApJ...938..144M}
}

@ARTICLE{whitaker21a,
       author = {{Whitaker}, Katherine E. and {Narayanan}, Desika and {Williams}, Christina C. and {Li}, Qi and {Spilker}, Justin S. and {Dav{\'e}}, Romeel and {Akhshik}, Mohammad and {Akins}, Hollis B. and {Bezanson}, Rachel and {Katz}, Neal and {Leja}, Joel and {Magdis}, Georgios E. and {Mowla}, Lamiya and {Nelson}, Erica J. and {Pope}, Alexandra and {Privon}, George C. and {Toft}, Sune and {Valentino}, Francesco},
        title = "{High Molecular-gas to Dust Mass Ratios Predicted in Most Quiescent Galaxies}",
      journal = {\apjl},
         year = 2021,
        month = dec,
       volume = {922},
       number = {2},
          eid = {L30},
        pages = {L30},
          doi = {10.3847/2041-8213/ac399f},
archivePrefix = {arXiv},
       eprint = {2111.05349},
 primaryClass = {astro-ph.GA},
       adsurl = {https://ui.adsabs.harvard.edu/abs/2021ApJ...922L..30W}
}

@ARTICLE{whitaker21b,
       author = {{Whitaker}, Katherine E. and {Williams}, Christina C. and {Mowla}, Lamiya and {Spilker}, Justin S. and {Toft}, Sune and {Narayanan}, Desika and {Pope}, Alexandra and {Magdis}, Georgios E. and {van Dokkum}, Pieter G. and {Akhshik}, Mohammad and {Bezanson}, Rachel and {Brammer}, Gabriel B. and {Leja}, Joel and {Man}, Allison and {Nelson}, Erica J. and {Richard}, Johan and {Pacifici}, Camilla and {Sharon}, Keren and {Valentino}, Francesco},
        title = "{Quenching of star formation from a lack of inflowing gas to galaxies}",
      journal = {\nat},
         year = 2021,
        month = sep,
       volume = {597},
       number = {7877},
        pages = {485-488},
          doi = {10.1038/s41586-021-03806-7},
archivePrefix = {arXiv},
       eprint = {2109.10384},
 primaryClass = {astro-ph.GA},
       adsurl = {https://ui.adsabs.harvard.edu/abs/2021Natur.597..485W}
}

@ARTICLE{williams21,
       author = {{Williams}, Christina C. and {Spilker}, Justin S. and {Whitaker}, Katherine E. and {Dav{\'e}}, Romeel and {Woodrum}, Charity and {Brammer}, Gabriel and {Bezanson}, Rachel and {Narayanan}, Desika and {Weiner}, Benjamin},
        title = "{ALMA Measures Rapidly Depleted Molecular Gas Reservoirs in Massive Quiescent Galaxies at z {\ensuremath{\sim}} 1.5}",
      journal = {\apj},
         year = 2021,
        month = feb,
       volume = {908},
       number = {1},
          eid = {54},
        pages = {54},
          doi = {10.3847/1538-4357/abcbf6},
archivePrefix = {arXiv},
       eprint = {2012.01433},
 primaryClass = {astro-ph.GA},
       adsurl = {https://ui.adsabs.harvard.edu/abs/2021ApJ...908...54W}
}

@ARTICLE{1995Tsai,
       author = {{Tsai}, John C. and {Mathews}, William G.},
        title = "{Interstellar Grains in Elliptical Galaxies: Grain Evolution}",
      journal = {\apj},
         year = 1995,
        month = jul,
       volume = {448},
        pages = {84},
          doi = {10.1086/175943},
archivePrefix = {arXiv},
       eprint = {astro-ph/9502053},
 primaryClass = {astro-ph},
       adsurl = {https://ui.adsabs.harvard.edu/abs/1995ApJ...448...84T}
}

@article{McKinnon2017,
    author = {McKinnon, Ryan and Torrey, Paul and Vogelsberger, Mark and Hayward, Christopher C. and Marinacci, Federico},
    title = "{Simulating the dust content of galaxies: successes and failures}",
    journal = {Monthly Notices of the Royal Astronomical Society},
    volume = {468},
    number = {2},
    pages = {1505-1521},
    year = {2017},
    month = {02},
    issn = {0035-8711},
    doi = {10.1093/mnras/stx467},
    url = {https://doi.org/10.1093/mnras/stx467},
    eprint = {https://academic.oup.com/mnras/article-pdf/468/2/1505/11142459/stx467.pdf},
}

@article{gobat2018unexpectedly,
  title={The unexpectedly large dust and gas content of quiescent galaxies at z> 1.4},
  author={Gobat, R and Daddi, Emanuele and Magdis, G and Bournaud, F and Sargent, M and Martig, M and Jin, S and Finoguenov, A and B{\'e}thermin, M and Hwang, HS and others},
  journal={Nature Astronomy},
  volume={2},
  number={3},
  pages={239--246},
  year={2018},
  publisher={Nature Publishing Group UK London}
}

@article{park2023rapid,
  title={Rapid Quenching of Galaxies at Cosmic Noon},
  author={Park, Minjung and Belli, Sirio and Conroy, Charlie and Tacchella, Sandro and Leja, Joel and Cutler, Sam E and Johnson, Benjamin D and Nelson, Erica J and Emami, Razieh},
  journal={The Astrophysical Journal},
  volume={953},
  number={1},
  pages={119},
  year={2023},
  publisher={IOP Publishing}
}

@article{asano2013dust,
  title={Dust formation history of galaxies: A critical role of metallicity* for the dust mass growth by accreting materials in the interstellar medium},
  author={Asano, Ryosuke S and Takeuchi, Tsutomu T and Hirashita, Hiroyuki and Inoue, Akio K},
  journal={Earth, Planets and Space},
  volume={65},
  number={3},
  pages={213--222},
  year={2013},
  publisher={SpringerOpen}
}

@article{casasola2022resolved,
  title={The resolved scaling relations in DustPedia: Zooming in on the local Universe},
  author={Casasola, Viviana and Bianchi, Simone and Magrini, Laura and Mosenkov, Aleksandr V and Salvestrini, Francesco and Baes, Maarten and Calura, Francesco and Cassar{\`a}, Letizia P and Clark, Christopher JR and Corbelli, Edvige and others},
  journal={Astronomy \& Astrophysics},
  volume={668},
  pages={A130},
  year={2022},
  publisher={EDP Sciences}
}

@article{gobat2022uncertain,
  title={The uncertain interstellar medium of high-redshift quiescent galaxies: Impact of methodology},
  author={Gobat, Rapha{\"e}l and D'Eugenio, Chiara and Liu, Daizhong and Caminha, Gabriel Bartosch and Daddi, Emanuele and Bl{\'a}nquez, David},
  journal={arXiv preprint arXiv:2211.14131},
  year={2022}
}

@article{appleby2020impact,
  title={The impact of quenching on galaxy profiles in the SIMBA simulation},
  author={Appleby, Sarah and Dav{\'e}, Romeel and Kraljic, Katarina and Angl{\'e}s-Alc{\'a}zar, Daniel and Narayanan, Desika},
  journal={Monthly Notices of the Royal Astronomical Society},
  volume={494},
  number={4},
  pages={6053--6071},
  year={2020},
  publisher={Oxford University Press}
}

@article{akins2022quenching,
  title={Quenching and the UVJ Diagram in the SIMBA Cosmological Simulation},
  author={Akins, Hollis B and Narayanan, Desika and Whitaker, Katherine E and Dav{\'e}, Romeel and Lower, Sidney and Bezanson, Rachel and Feldmann, Robert and Kriek, Mariska},
  journal={The Astrophysical Journal},
  volume={929},
  number={1},
  pages={94},
  year={2022},
  publisher={IOP Publishing}
}

@article{hamadouche2023connection,
  title={The connection between stellar mass, age, and quenching time-scale in massive quiescent galaxies at z≃ 1},
  author={Hamadouche, ML and Carnall, AC and McLure, RJ and Dunlop, JS and Begley, R and Cullen, F and McLeod, DJ and Donnan, CT and Stanton, TM},
  journal={Monthly Notices of the Royal Astronomical Society},
  volume={521},
  number={4},
  pages={5400--5409},
  year={2023},
  publisher={Oxford University Press}
}

@article{magdis2012evolving,
  title={The evolving interstellar medium of star-forming galaxies since z= 2 as probed by their infrared spectral energy distributions},
  author={Magdis, Georgios E and Daddi, Emanuele and B{\'e}thermin, M and Sargent, Mark and Elbaz, David and Pannella, M and Dickinson, Mark and Dannerbauer, Helmut and da Cunha, Elisabete and Walter, F and others},
  journal={The Astrophysical Journal},
  volume={760},
  number={1},
  pages={6},
  year={2012},
  publisher={IOP Publishing}
}

@ARTICLE{popping17,
       author = {{Popping}, Gerg{\"o} and {Somerville}, Rachel S. and {Galametz}, Maud},
        title = "{The dust content of galaxies from z = 0 to z = 9}",
      journal = {\mnras},
         year = 2017,
        month = nov,
       volume = {471},
       number = {3},
        pages = {3152-3185},
          doi = {10.1093/mnras/stx1545},
archivePrefix = {arXiv},
       eprint = {1609.08622},
 primaryClass = {astro-ph.GA},
       adsurl = {https://ui.adsabs.harvard.edu/abs/2017MNRAS.471.3152P}
}

@ARTICLE{donevski20,
       author = {{Donevski}, D. and {Lapi}, A. and {Ma{\l}ek}, K. and {Liu}, D. and {G{\'o}mez-Guijarro}, C. and {Dav{\'e}}, R. and {Kraljic}, K. and {Pantoni}, L. and {Man}, A. and {Fujimoto}, S. and {Feltre}, A. and {Pearson}, W. and {Li}, Q. and {Narayanan}, D.},
        title = "{In pursuit of giants. I. The evolution of the dust-to-stellar mass ratio in distant dusty galaxies}",
      journal = {\aap},
         year = 2020,
        month = dec,
       volume = {644},
          eid = {A144},
        pages = {A144},
          doi = {10.1051/0004-6361/202038405},
archivePrefix = {arXiv},
       eprint = {2008.09995},
 primaryClass = {astro-ph.GA},
       adsurl = {https://ui.adsabs.harvard.edu/abs/2020A&A...644A.144D}
}

@ARTICLE{li19,
       author = {{Li}, Zhihui and {French}, K. Decker and {Zabludoff}, Ann I. and {Ho}, Luis C.},
        title = "{The Evolution of the Interstellar Medium in Post-starburst Galaxies}",
      journal = {\apj},
         year = 2019,
        month = jul,
       volume = {879},
       number = {2},
          eid = {131},
        pages = {131},
          doi = {10.3847/1538-4357/ab1f68},
archivePrefix = {arXiv},
       eprint = {1906.01890},
 primaryClass = {astro-ph.GA},
       adsurl = {https://ui.adsabs.harvard.edu/abs/2019ApJ...879..131L}
}

@ARTICLE{liqi19,
       author = {{Li}, Qi and {Narayanan}, Desika and {Dav{\'e}}, Romeel},
        title = "{The dust-to-gas and dust-to-metal ratio in galaxies from z = 0 to 6}",
      journal = {\mnras},
         year = 2019,
        month = nov,
       volume = {490},
       number = {1},
        pages = {1425-1436},
          doi = {10.1093/mnras/stz2684},
archivePrefix = {arXiv},
       eprint = {1906.09277},
 primaryClass = {astro-ph.GA},
       adsurl = {https://ui.adsabs.harvard.edu/abs/2019MNRAS.490.1425L}
}

@ARTICLE{vogelsberger19,
       author = {{Vogelsberger}, Mark and {McKinnon}, Ryan and {O'Neil}, Stephanie and {Marinacci}, Federico and {Torrey}, Paul and {Kannan}, Rahul},
        title = "{Dust in and around galaxies: dust in cluster environments and its impact on gas cooling}",
      journal = {\mnras},
         year = 2019,
        month = aug,
       volume = {487},
       number = {4},
        pages = {4870-4883},
          doi = {10.1093/mnras/stz1644},
archivePrefix = {arXiv},
       eprint = {1811.05477},
 primaryClass = {astro-ph.GA},
       adsurl = {https://ui.adsabs.harvard.edu/abs/2019MNRAS.487.4870V}
}

@ARTICLE{belli21,
       author = {{Belli}, Sirio and {Contursi}, Alessandra and {Genzel}, Reinhard and {Tacconi}, Linda J. and {F{\"o}rster-Schreiber}, Natascha M. and {Lutz}, Dieter and {Combes}, Fran{\c{c}}oise and {Neri}, Roberto and {Garc{\'\i}a-Burillo}, Santiago and {Schuster}, Karl F. and {Herrera-Camus}, Rodrigo and {Tadaki}, Ken-ichi and {Davies}, Rebecca L. and {Davies}, Richard I. and {Johnson}, Benjamin D. and {Lee}, Minju M. and {Leja}, Joel and {Nelson}, Erica J. and {Price}, Sedona H. and {Shangguan}, Jinyi and {Shimizu}, T. Taro and {Tacchella}, Sandro and {{\"U}bler}, Hannah},
        title = "{The Diverse Molecular Gas Content of Massive Galaxies Undergoing Quenching at z {\ensuremath{\sim}} 1}",
      journal = {\apjl},
         year = 2021,
        month = mar,
       volume = {909},
       number = {1},
          eid = {L11},
        pages = {L11},
          doi = {10.3847/2041-8213/abe6a6},
archivePrefix = {arXiv},
       eprint = {2102.07881},
 primaryClass = {astro-ph.GA},
       adsurl = {https://ui.adsabs.harvard.edu/abs/2021ApJ...909L..11B}
}

@ARTICLE{gobat20,
       author = {{Gobat}, R. and {Magdis}, G. and {D'Eugenio}, C. and {Valentino}, F.},
        title = "{The evolution of the gas fraction of quiescent galaxies modeled as a consequence of their creation rate}",
      journal = {\aap},
         year = 2020,
        month = dec,
       volume = {644},
          eid = {L7},
        pages = {L7},
          doi = {10.1051/0004-6361/202039593},
archivePrefix = {arXiv},
       eprint = {2011.10547},
 primaryClass = {astro-ph.GA},
       adsurl = {https://ui.adsabs.harvard.edu/abs/2020A&A...644L...7G}
}

@article{tacconi2018phibss,
  title={PHIBSS: unified scaling relations of gas depletion time and molecular gas fractions},
  author={Tacconi, Linda J and Genzel, Reinhard and Saintonge, Am{\'e}lie and Combes, Fran{\c{c}}oise and Garc{\'\i}a-Burillo, Santiago and Neri, Roberto and Bolatto, Alberto and Contini, Thierry and Schreiber, NM F{\"o}rster and Lilly, Simon and others},
  journal={The Astrophysical Journal},
  volume={853},
  number={2},
  pages={179},
  year={2018},
  publisher={IOP Publishing}
}

@ARTICLE{michalowski23,
       author = {{Micha{\l}owski}, Micha{\l} J. and {Gall}, C. and {Hjorth}, J. and {Frayer}, D.~T. and {Tsai}, A. -L. and {Rowlands}, K. and {Takeuchi}, T.~T. and {Le{\'s}niewska}, A. and {Behrendt}, D. and {Bourne}, N. and {Hughes}, D.~H. and {Koprowski}, M.~P. and {Nadolny}, J. and {Ryzhov}, O. and {Solar}, M. and {Spring}, E. and {Zavala}, J. and {Bartczak}, P.},
        title = "{The Fate of the Interstellar Medium in Early-type Galaxies. III. The Mechanism of Interstellar Medium Removal and the Quenching of Star Formation}",
      journal = {\apj},
         year = 2024,
        month = apr,
       volume = {964},
       number = {2},
          eid = {129},
        pages = {129},
          doi = {10.3847/1538-4357/ad1b52},
archivePrefix = {arXiv},
       eprint = {2401.04774},
 primaryClass = {astro-ph.GA},
       adsurl = {https://ui.adsabs.harvard.edu/abs/2024ApJ...964..129M}
}

@ARTICLE{wu23,
       author = {{Wu}, Po-Feng and {Bezanson}, Rachel and {D'Eugenio}, Francesco and {Gallazzi}, Anna R. and {Greene}, Jenny E. and {Maseda}, Michael V. and {Suess}, Katherine A. and {van der Wel}, Arjen},
        title = "{Stars, Gas, and Star Formation of Distant Post-starburst Galaxies}",
      journal = {\apj},
         year = 2023,
        month = sep,
       volume = {955},
       number = {1},
          eid = {75},
        pages = {75},
          doi = {10.3847/1538-4357/acf0bd},
archivePrefix = {arXiv},
       eprint = {2308.08681},
 primaryClass = {astro-ph.GA},
       adsurl = {https://ui.adsabs.harvard.edu/abs/2023ApJ...955...75W}
}

@ARTICLE{nanni13,
       author = {{Nanni}, Ambra and {Bressan}, Alessandro and {Marigo}, Paola and {Girardi}, L{\'e}o},
        title = "{Evolution of thermally pulsing asymptotic giant branch stars - II. Dust production at varying metallicity}",
      journal = {\mnras},
         year = 2013,
        month = sep,
       volume = {434},
       number = {3},
        pages = {2390-2417},
          doi = {10.1093/mnras/stt1175},
archivePrefix = {arXiv},
       eprint = {1306.6183},
 primaryClass = {astro-ph.SR},
       adsurl = {https://ui.adsabs.harvard.edu/abs/2013MNRAS.434.2390N}
}

@ARTICLE{nanni14,
       author = {{Nanni}, Ambra and {Bressan}, Alessandro and {Marigo}, Paola and {Girardi}, L{\'e}o},
        title = "{Evolution of thermally pulsing asymptotic giant branch stars - III. Dust production at supersolar metallicities}",
      journal = {\mnras},
         year = 2014,
        month = mar,
       volume = {438},
       number = {3},
        pages = {2328-2340},
          doi = {10.1093/mnras/stt2348},
archivePrefix = {arXiv},
       eprint = {1312.0875},
 primaryClass = {astro-ph.SR},
       adsurl = {https://ui.adsabs.harvard.edu/abs/2014MNRAS.438.2328N}
}

@ARTICLE{ventura20,
       author = {{Ventura}, P. and {Dell'Agli}, F. and {Lugaro}, M. and {Romano}, D. and {Tailo}, M. and {Yag{\"u}e}, A.},
        title = "{Gas and dust from metal-rich AGB stars}",
      journal = {\aap},
         year = 2020,
        month = sep,
       volume = {641},
          eid = {A103},
        pages = {A103},
          doi = {10.1051/0004-6361/202038289},
archivePrefix = {arXiv},
       eprint = {2007.02120},
 primaryClass = {astro-ph.SR},
       adsurl = {https://ui.adsabs.harvard.edu/abs/2020A&A...641A.103V}
}

@ARTICLE{greene20,
       author = {{Greene}, Jenny E. and {Setton}, David and {Bezanson}, Rachel and {Suess}, Katherine A. and {Kriek}, Mariska and {Spilker}, Justin S. and {Goulding}, Andy D. and {Feldmann}, Robert},
        title = "{The Role of Active Galactic Nuclei in the Quenching of Massive Galaxies in the SQuIGG {\v{e}}c\{L\} E Survey}",
      journal = {\apjl},
         year = 2020,
        month = aug,
       volume = {899},
       number = {1},
          eid = {L9},
        pages = {L9},
          doi = {10.3847/2041-8213/aba534},
archivePrefix = {arXiv},
       eprint = {2007.02967},
 primaryClass = {astro-ph.GA},
       adsurl = {https://ui.adsabs.harvard.edu/abs/2020ApJ...899L...9G}
}

@article{french2023state,
  title={The State of the Molecular Gas in Post-starburst Galaxies},
  author={French, K Decker and Smercina, Adam and Rowlands, Kate and Tripathi, Akshat and Zabludoff, Ann I and Smith, John-David T and Narayanan, Desika and Yang, Yujin and Shirley, Yancy and Alatalo, Katey},
  journal={The Astrophysical Journal},
  volume={942},
  number={1},
  pages={25},
  year={2023},
  publisher={IOP Publishing}
}

@ARTICLE{man16,
       author = {{Man}, Allison W.~S. and {Greve}, Thomas R. and {Toft}, Sune and {Magnelli}, Benjamin and {Karim}, Alexander and {Ilbert}, Olivier and {Salvato}, Mara and {Le Floc'h}, Emeric and {Bertoldi}, Frank and {Casey}, Caitlin M. and {Lee}, Nicholas and {Li}, Yanxia and {Navarrete}, Felipe and {Sheth}, Kartik and {Smol{\v{c}}i{\'c}}, Vernesa and {Sanders}, David B. and {Schinnerer}, Eva and {Zirm}, Andrew W.},
        title = "{Confirming the Existence of a Quiescent Galaxy Population out to z=3: A Stacking Analysis of Mid-, Far-Infrared and Radio Data}",
      journal = {\apj},
         year = 2016,
        month = mar,
       volume = {820},
       number = {1},
          eid = {11},
        pages = {11},
          doi = {10.3847/0004-637X/820/1/11},
archivePrefix = {arXiv},
       eprint = {1411.2870},
 primaryClass = {astro-ph.GA},
       adsurl = {https://ui.adsabs.harvard.edu/abs/2016ApJ...820...11M}
}

@ARTICLE{liu19b,
       author = {{Liu}, Daizhong and {Schinnerer}, E. and {Groves}, B. and {Magnelli}, B. and {Lang}, P. and {Leslie}, S. and {Jim{\'e}nez-Andrade}, E. and {Riechers}, D.~A. and {Popping}, G. and {Magdis}, Georgios E. and {Daddi}, E. and {Sargent}, M. and {Gao}, Yu and {Fudamoto}, Y. and {Oesch}, P.~A. and {Bertoldi}, F.},
        title = "{Automated Mining of the ALMA Archive in the COSMOS Field (A$^{3}$COSMOS). II. Cold Molecular Gas Evolution out to Redshift 6}",
      journal = {\apj},
         year = 2019,
        month = dec,
       volume = {887},
       number = {2},
          eid = {235},
        pages = {235},
          doi = {10.3847/1538-4357/ab578d},
archivePrefix = {arXiv},
       eprint = {1910.12883},
 primaryClass = {astro-ph.GA},
       adsurl = {https://ui.adsabs.harvard.edu/abs/2019ApJ...887..235L}
}

@ARTICLE{setton24,
       author = {{Setton}, David J. and {Khullar}, Gourav and {Miller}, Tim B. and {Bezanson}, Rachel and {Greene}, Jenny E. and {Suess}, Katherine A. and {Whitaker}, Katherine E. and {Antwi-Danso}, Jacqueline and {Atek}, Hakim and {Brammer}, Gabriel and {Cutler}, Sam E. and {Dayal}, Pratika and {Feldmann}, Robert and {Fujimoto}, Seiji and {Furtak}, Lukas J. and {Glazebrook}, Karl and {Goulding}, Andy D. and {Kokorev}, Vasily and {Labbe}, Ivo and {Leja}, Joel and {Ma}, Yilun and {Marchesini}, Danilo and {Nanayakkara}, Themiya and {Pan}, Richard and {Price}, Sedona H. and {Siegel}, Jared C. and {Shipley}, Heath and {Weaver}, John R. and {van Dokkum}, Pieter and {Wang}, Bingjie and {Williams}, Christina C.},
        title = "{UNCOVER NIRSpec/PRISM Spectroscopy Unveils Evidence of Early Core Formation in a Massive, Centrally Dusty Quiescent Galaxy at z $_{spec}$ = 3.97}",
      journal = {\apj},
         year = 2024,
        month = oct,
       volume = {974},
       number = {1},
          eid = {145},
        pages = {145},
          doi = {10.3847/1538-4357/ad6a18},
archivePrefix = {arXiv},
       eprint = {2402.05664},
 primaryClass = {astro-ph.GA},
       adsurl = {https://ui.adsabs.harvard.edu/abs/2024ApJ...974..145S}
}

@article{richie2024dust,
  title={Dust Survival in Galactic Winds},
  author={Richie, Helena M and Schneider, Evan E and Abruzzo, Matthew W and Torrey, Paul},
  journal={arXiv preprint arXiv:2403.03711},
  year={2024}
}

@ARTICLE{lorenzon25a,
       author = {{Lorenzon}, G. and {Donevski}, D. and {Lisiecki}, K. and {Lovell}, C. and {Romano}, M. and {Narayanan}, D. and {Dav{\'e}}, R. and {Man}, A. and {Whitaker}, K.~E. and {Nanni}, A. and {Long}, A. and {Lee}, M.~M. and {Junais} and {Ma{\l}ek}, K. and {Rodighiero}, G. and {Li}, Q.},
        title = "{Tracing the evolutionary pathways of dust and cold gas in high-z quiescent galaxies with SIMBA}",
      journal = {\aap},
         year = 2025,
        month = jan,
       volume = {693},
          eid = {A118},
        pages = {A118},
          doi = {10.1051/0004-6361/202450393},
archivePrefix = {arXiv},
       eprint = {2404.10568},
 primaryClass = {astro-ph.GA},
       adsurl = {https://ui.adsabs.harvard.edu/abs/2025A&A...693A.118L}
}

@ARTICLE{lorenzon25b,
       author = {{Lorenzon}, G. and {Donevski}, D. and {Man}, A.~W.~S. and {Romano}, M. and {Whitaker}, K.~E. and {Belli}, S. and {Liu}, D. and {Lee}, M.~M. and {Narayanan}, D. and {Long}, A. and {Shivaei}, I. and {Nanni}, A. and {Lisiecki}, K. and {Sawant}, P. and {Rodighiero}, G. and {Damjanov}, I. and {Junais} and {Dav{\'e}}, R. and {Pappalardo}, C. and {Lovell}, C. and {Hamed}, M.},
        title = "{ALMA Reveals Diverse Dust-to-gas Mass Ratios and Quenching Modes in Old Quiescent Galaxies}",
      journal = {\apjl},
         year = 2025,
        month = dec,
       volume = {995},
       number = {2},
          eid = {L63},
        pages = {L63},
          doi = {10.3847/2041-8213/ae226c},
archivePrefix = {arXiv},
       eprint = {2509.10079},
 primaryClass = {astro-ph.GA},
       adsurl = {https://ui.adsabs.harvard.edu/abs/2025ApJ...995L..63L}
}

@ARTICLE{dl07,
       author = {{Draine}, B.~T. and {Li}, Aigen},
        title = "{Infrared Emission from Interstellar Dust. IV. The Silicate-Graphite-PAH Model in the Post-Spitzer Era}",
      journal = {\apj},
         year = "2007",
        month = "Mar",
       volume = {657},
       number = {2},
        pages = {810-837},
          doi = {10.1086/511055},
archivePrefix = {arXiv},
       eprint = {astro-ph/0608003},
 primaryClass = {astro-ph},
       adsurl = {https://ui.adsabs.harvard.edu/abs/2007ApJ...657..810D}
}

@ARTICLE{smercina22,
       author = {{Smercina}, Adam and {Smith}, John-David T. and {French}, K. Decker and {Bell}, Eric F. and {Dale}, Daniel A. and {Medling}, Anne M. and {Nyland}, Kristina and {Privon}, George C. and {Rowlands}, Kate and {Walter}, Fabian and {Zabludoff}, Ann I.},
        title = "{After The Fall: Resolving the Molecular Gas in Post-starburst Galaxies}",
      journal = {\apj},
         year = 2022,
        month = apr,
       volume = {929},
       number = {2},
          eid = {154},
        pages = {154},
          doi = {10.3847/1538-4357/ac5d5f},
archivePrefix = {arXiv},
       eprint = {2108.03231},
 primaryClass = {astro-ph.GA},
       adsurl = {https://ui.adsabs.harvard.edu/abs/2022ApJ...929..154S}
}

@article{Umehata_2025,
            doi = {10.3847/2041-8213/add1d4},
            url = {https://dx.doi.org/10.3847/2041-8213/add1d4},
            year = {2025},
            month = {may},
            publisher = {The American Astronomical Society},
            volume = {985},
            number = {1},
            pages = {L8},
            author = {Umehata, Hideki and Kubo, Mariko and Nakanishi, Kouichiro},
            title = {ADF22-WEB: Detection of a Molecular Gas Reservoir in a Massive Quiescent Galaxy Located in a z ≈ 3 Protocluster Core},
            journal = {The Astrophysical Journal Letters}
            }

@ARTICLE{bezanson21,
       author = {{Bezanson}, Rachel and {Spilker}, Justin S. and {Suess}, Katherine A. and {Setton}, David J. and {Feldmann}, Robert and {Greene}, Jenny E. and {Kriek}, Mariska and {Narayanan}, Desika and {Verrico}, Margaret},
        title = "{Now You See It, Now You Don't: Star Formation Truncation Precedes the Loss of Molecular Gas by 100 Myr in Massive Poststarburst Galaxies at z   0.6}",
      journal = {\apj},
         year = 2022,
        month = feb,
       volume = {925},
       number = {2},
          eid = {153},
        pages = {153},
          doi = {10.3847/1538-4357/ac3dfa},
archivePrefix = {arXiv},
       eprint = {2111.14877},
 primaryClass = {astro-ph.GA},
       adsurl = {https://ui.adsabs.harvard.edu/abs/2022ApJ...925..153B}
}

@ARTICLE{weibel25,
       author = {{Weibel}, Andrea and {de Graaff}, Anna and {Setton}, David J. and {Miller}, Tim B. and {Oesch}, Pascal A. and {Brammer}, Gabriel and {Lagos}, Claudia D.~P. and {Whitaker}, Katherine E. and {Williams}, Christina C. and {Baggen}, Josephine F.~W. and {Bezanson}, Rachel and {Boogaard}, Leindert A. and {Cleri}, Nikko J. and {Greene}, Jenny E. and {Hirschmann}, Michaela and {Hviding}, Raphael E. and {Kuruvanthodi}, Adarsh and {Labb{\'e}}, Ivo and {Leja}, Joel and {Maseda}, Michael V. and {Matthee}, Jorryt and {McConachie}, Ian and {Naidu}, Rohan P. and {Roberts-Borsani}, Guido and {Schaerer}, Daniel and {Suess}, Katherine A. and {Valentino}, Francesco and {van Dokkum}, Pieter and {Wang}, Bingjie},
        title = "{RUBIES Reveals a Massive Quiescent Galaxy at z = 7.3}",
      journal = {\apj},
         year = 2025,
        month = apr,
       volume = {983},
       number = {1},
          eid = {11},
        pages = {11},
          doi = {10.3847/1538-4357/adab7a},
archivePrefix = {arXiv},
       eprint = {2409.03829},
 primaryClass = {astro-ph.GA},
       adsurl = {https://ui.adsabs.harvard.edu/abs/2025ApJ...983...11W}
}

@ARTICLE{zanella23,
       author = {{Zanella}, A. and {Valentino}, F. and {Gallazzi}, A. and {Belli}, S. and {Magdis}, G. and {Bolamperti}, A.},
        title = "{The large molecular gas fraction of post-starburst galaxies at z > 1}",
      journal = {\mnras},
         year = 2023,
        month = sep,
       volume = {524},
       number = {1},
        pages = {923-939},
          doi = {10.1093/mnras/stad1821},
archivePrefix = {arXiv},
       eprint = {2306.08120},
 primaryClass = {astro-ph.GA},
       adsurl = {https://ui.adsabs.harvard.edu/abs/2023MNRAS.524..923Z}
}

@ARTICLE{spilker25,
       author = {{Spilker}, Justin S. and {Whitaker}, Katherine E. and {Narayanan}, Desika and {Bezanson}, Rachel and {Bodansky}, Sarah and {D'Onofrio}, Vincenzo R. and {Feldmann}, Robert and {Goulding}, Andy D. and {Greene}, Jenny E. and {Kriek}, Mariska and {Luo}, Yuanze and {Setton}, David J. and {Suess}, Katherine A. and {van der Wel}, Arjen and {Verrico}, Margaret E. and {Williams}, Christina C. and {Woodrum}, Charity and {Wu}, Po-Feng},
        title = "{Unusually High Gas-to-Dust Ratios Observed in High-Redshift Quiescent Galaxies}",
      journal = {arXiv e-prints},
         year = 2025,
        month = jul,
          eid = {arXiv:2507.16914},
        pages = {arXiv:2507.16914},
archivePrefix = {arXiv},
       eprint = {2507.16914},
 primaryClass = {astro-ph.GA},
       adsurl = {https://ui.adsabs.harvard.edu/abs/2025arXiv250716914S}
}

@ARTICLE{hayashi18,
       author = {{Hayashi}, Masao and {Tadaki}, Ken-ichi and {Kodama}, Tadayuki and {Kohno}, Kotaro and {Yamaguchi}, Yuki and {Hatsukade}, Bunyo and {Koyama}, Yusei and {Shimakawa}, Rhythm and {Tamura}, Yoichi and {Suzuki}, Tomoko L.},
        title = "{Molecular Gas Reservoirs in Cluster Galaxies at z = 1.46}",
      journal = {\apj},
         year = 2018,
        month = apr,
       volume = {856},
       number = {2},
          eid = {118},
        pages = {118},
          doi = {10.3847/1538-4357/aab3e7},
archivePrefix = {arXiv},
       eprint = {1803.00298},
 primaryClass = {astro-ph.GA},
       adsurl = {https://ui.adsabs.harvard.edu/abs/2018ApJ...856..118H}
}

@ARTICLE{siegel2025,
       author = {{Siegel}, Jared C. and {Setton}, David J. and {Greene}, Jenny E. and {Suess}, Katherine A. and {Whitaker}, Katherine E. and {Bezanson}, Rachel and {Leja}, Joel and {Furtak}, Lukas J. and {Cutler}, Sam E. and {de Graaff}, Anna and {Feldmann}, Robert and {Khullar}, Gourav and {Labbe}, Ivo and {Marchesini}, Danilo and {Miller}, Tim B. and {Nanayakkara}, Themiya and {Pan}, Richard and {Price}, Sedona H. and {Treiber}, Helena P. and {van Dokkum}, Pieter and {Wang}, Bingjie and {Weaver}, John R.},
        title = "{UNCOVER: Significant Reddening in Cosmic Noon Quiescent Galaxies}",
      journal = {\apj},
         year = 2025,
        month = may,
       volume = {985},
       number = {1},
          eid = {125},
        pages = {125},
          doi = {10.3847/1538-4357/adc7b7},
archivePrefix = {arXiv},
       eprint = {2409.11457},
 primaryClass = {astro-ph.GA},
       adsurl = {https://ui.adsabs.harvard.edu/abs/2025ApJ...985..125S}
}

@ARTICLE{slavin15,
       author = {{Slavin}, Jonathan D. and {Dwek}, Eli and {Jones}, Anthony P.},
        title = "{Destruction of Interstellar Dust in Evolving Supernova Remnant Shock Waves}",
      journal = {\apj},
         year = 2015,
        month = apr,
       volume = {803},
       number = {1},
          eid = {7},
        pages = {7},
          doi = {10.1088/0004-637X/803/1/7},
archivePrefix = {arXiv},
       eprint = {1502.00929},
 primaryClass = {astro-ph.GA},
       adsurl = {https://ui.adsabs.harvard.edu/abs/2015ApJ...803....7S}
}

@ARTICLE{zhukovska16,
       author = {{Zhukovska}, Svitlana and {Dobbs}, Clare and {Jenkins}, Edward B. and {Klessen}, Ralf S.},
        title = "{Modeling Dust Evolution in Galaxies with a Multiphase, Inhomogeneous ISM}",
      journal = {\apj},
         year = 2016,
        month = nov,
       volume = {831},
       number = {2},
          eid = {147},
        pages = {147},
          doi = {10.3847/0004-637X/831/2/147},
archivePrefix = {arXiv},
       eprint = {1608.04781},
 primaryClass = {astro-ph.GA},
       adsurl = {https://ui.adsabs.harvard.edu/abs/2016ApJ...831..147Z}
}

@ARTICLE{chandro26,
       author = {{Chandro-G{\'o}mez}, {\'A}ngel and {Lagos}, Claudia del P. and {Power}, Chris and {Baker}, Willian M. and {Ben{\'\i}tez-Llambay}, Alejandro and {Chaikin}, Evgenii and {Chittenden}, Harry G. and {Correa}, Camila and {Frenk}, Carlos S. and {Hu{\v{s}}ko}, Filip and {McGibbon}, Robert J. and {Nanayakkara}, Themiya and {Ploeckinger}, Sylvia and {Richings}, Alexander J. and {Schaller}, Matthieu and {Schaye}, Joop and {Trayford}, James W.},
        title = "{Unveiling the properties and origin of massive quenched galaxies at $z\ge2$ in the COLIBRE hydrodynamical simulations}",
      journal = {arXiv e-prints},
         year = 2025,
        month = dec,
          eid = {arXiv:2512.16208},
        pages = {arXiv:2512.16208},
          doi = {10.48550/arXiv.2512.16208},
archivePrefix = {arXiv},
       eprint = {2512.16208},
 primaryClass = {astro-ph.GA},
       adsurl = {https://ui.adsabs.harvard.edu/abs/2025arXiv251216208C}
}

@ARTICLE{narayanan23,
       author = {{Narayanan}, Desika and {Smith}, J.-D.~T. and {Hensley}, Brandon S. and {Li}, Qi and {Hu}, Chia-Yu and {Sandstrom}, Karin and {Torrey}, Paul and {Vogelsberger}, Mark and {Marinacci}, Federico and {Sales}, Laura V.},
        title = "{A Framework for Modeling Polycyclic Aromatic Hydrocarbon Emission in Galaxy Evolution Simulations}",
      journal = {\apj},
         year = 2023,
        month = jul,
       volume = {951},
       number = {2},
          eid = {100},
        pages = {100},
          doi = {10.3847/1538-4357/accf8d},
archivePrefix = {arXiv},
       eprint = {2301.07136},
 primaryClass = {astro-ph.GA},
       adsurl = {https://ui.adsabs.harvard.edu/abs/2023ApJ...951..100N}
}

@ARTICLE{dwek07,
       author = {{Dwek}, Eli and {Galliano}, Fr{\'e}d{\'e}ric and {Jones}, Anthony P.},
        title = "{The Evolution of Dust in the Early Universe with Applications to the Galaxy SDSS J1148+5251}",
      journal = {\apj},
         year = 2007,
        month = jun,
       volume = {662},
       number = {2},
        pages = {927-939},
          doi = {10.1086/518430},
archivePrefix = {arXiv},
       eprint = {0705.3799},
 primaryClass = {astro-ph},
       adsurl = {https://ui.adsabs.harvard.edu/abs/2007ApJ...662..927D}
}

@ARTICLE{hou19,
       author = {{Hou}, Kuan-Chou and {Aoyama}, Shohei and {Hirashita}, Hiroyuki and {Nagamine}, Kentaro and {Shimizu}, Ikkoh},
        title = "{Dust scaling relations in a cosmological simulation}",
      journal = {\mnras},
         year = 2019,
        month = may,
       volume = {485},
       number = {2},
        pages = {1727-1744},
          doi = {10.1093/mnras/stz121},
archivePrefix = {arXiv},
       eprint = {1901.02886},
 primaryClass = {astro-ph.GA},
       adsurl = {https://ui.adsabs.harvard.edu/abs/2019MNRAS.485.1727H}
}

@ARTICLE{hirashita17,
       author = {{Hirashita}, Hiroyuki and {Nozawa}, Takaya},
        title = "{Dust evolution with active galactic nucleus feedback in elliptical galaxies}",
      journal = {\planss},
         year = 2017,
        month = dec,
       volume = {149},
        pages = {45-55},
          doi = {10.1016/j.pss.2017.01.009},
archivePrefix = {arXiv},
       eprint = {1701.07200},
 primaryClass = {astro-ph.GA},
       adsurl = {https://ui.adsabs.harvard.edu/abs/2017P&SS..149...45H}
}

@ARTICLE{hirashita15a,
       author = {{Hirashita}, Hiroyuki},
        title = "{Two-size approximation: a simple way of treating the evolution of grain size distribution in galaxies}",
      journal = {\mnras},
         year = 2015,
        month = mar,
       volume = {447},
       number = {3},
        pages = {2937-2950},
          doi = {10.1093/mnras/stu2617},
archivePrefix = {arXiv},
       eprint = {1412.3866},
 primaryClass = {astro-ph.GA},
       adsurl = {https://ui.adsabs.harvard.edu/abs/2015MNRAS.447.2937H}
}

@ARTICLE{hirashita15b,
       author = {{Hirashita}, Hiroyuki and {Nozawa}, Takaya and {Villaume}, Alexa and {Srinivasan}, Sundar},
        title = "{Dust processing in elliptical galaxies}",
      journal = {\mnras},
         year = 2015,
        month = dec,
       volume = {454},
       number = {2},
        pages = {1620-1633},
          doi = {10.1093/mnras/stv2095},
archivePrefix = {arXiv},
       eprint = {1509.03978},
 primaryClass = {astro-ph.GA},
       adsurl = {https://ui.adsabs.harvard.edu/abs/2015MNRAS.454.1620H}
}

@ARTICLE{hirashita21,
       author = {{Hirashita}, Hiroyuki and {Lan}, Ting-Wen},
        title = "{Shattering as a source of small grains in the circum-galactic medium}",
      journal = {\mnras},
         year = 2021,
        month = aug,
       volume = {505},
       number = {2},
        pages = {1794-1805},
          doi = {10.1093/mnras/stab1416},
archivePrefix = {arXiv},
       eprint = {2105.06611},
 primaryClass = {astro-ph.GA},
       adsurl = {https://ui.adsabs.harvard.edu/abs/2021MNRAS.505.1794H}
}

@ARTICLE{delooze20,
       author = {{De Looze}, I. and {Lamperti}, I. and {Saintonge}, A. and {Rela{\~n}o}, M. and {Smith}, M.~W.~L. and {Clark}, C.~J.~R. and {Wilson}, C.~D. and {Decleir}, M. and {Jones}, A.~P. and {Kennicutt}, R.~C. and {Accurso}, G. and {Brinks}, E. and {Bureau}, M. and {Cigan}, P. and {Clements}, D.~L. and {De Vis}, P. and {Fanciullo}, L. and {Gao}, Y. and {Gear}, W.~K. and {Ho}, L.~C. and {Hwang}, H.~S. and {Micha{\l}owski}, M.~J. and {Lee}, J.~C. and {Li}, C. and {Lin}, L. and {Liu}, T. and {Lomaeva}, M. and {Pan}, H.-A. and {Sargent}, M. and {Williams}, T. and {Xiao}, T. and {Zhu}, M.},
        title = "{JINGLE - IV. Dust, H I gas, and metal scaling laws in the local Universe}",
      journal = {\mnras},
         year = 2020,
        month = aug,
       volume = {496},
       number = {3},
        pages = {3668-3687},
          doi = {10.1093/mnras/staa1496},
archivePrefix = {arXiv},
       eprint = {2006.01856},
 primaryClass = {astro-ph.GA},
       adsurl = {https://ui.adsabs.harvard.edu/abs/2020MNRAS.496.3668D}
}

@ARTICLE{ormel09,
       author = {{Ormel}, C.~W. and {Paszun}, D. and {Dominik}, C. and {Tielens}, A.~G.~G.~M.},
        title = "{Dust coagulation and fragmentation in molecular clouds. I. How collisions between dust aggregates alter the dust size distribution}",
      journal = {\aap},
         year = 2009,
        month = aug,
       volume = {502},
       number = {3},
        pages = {845-869},
          doi = {10.1051/0004-6361/200811158},
archivePrefix = {arXiv},
       eprint = {0906.1770},
 primaryClass = {astro-ph.SR},
       adsurl = {https://ui.adsabs.harvard.edu/abs/2009A&A...502..845O}
}

@ARTICLE{galliano08,
       author = {{Galliano}, Fr{\'e}d{\'e}ric and {Dwek}, Eli and {Chanial}, Pierre},
        title = "{Stellar Evolutionary Effects on the Abundances of Polycyclic Aromatic Hydrocarbons and Supernova-Condensed Dust in Galaxies}",
      journal = {\apj},
         year = 2008,
        month = jan,
       volume = {672},
       number = {1},
        pages = {214-243},
          doi = {10.1086/523621},
archivePrefix = {arXiv},
       eprint = {0708.0790},
 primaryClass = {astro-ph},
       adsurl = {https://ui.adsabs.harvard.edu/abs/2008ApJ...672..214G}
}

@ARTICLE{vega10,
       author = {{Vega}, O. and {Bressan}, A. and {Panuzzo}, P. and {Rampazzo}, R. and {Clemens}, M. and {Granato}, G.~L. and {Buson}, L. and {Silva}, L. and {Zeilinger}, W.~W.},
        title = "{Unusual PAH Emission in Nearby Early-type Galaxies: A Signature of an Intermediate-age Stellar Population?}",
      journal = {\apj},
         year = 2010,
        month = oct,
       volume = {721},
       number = {2},
        pages = {1090-1104},
          doi = {10.1088/0004-637X/721/2/1090},
archivePrefix = {arXiv},
       eprint = {1008.0009},
 primaryClass = {astro-ph.CO},
       adsurl = {https://ui.adsabs.harvard.edu/abs/2010ApJ...721.1090V}
}

@ARTICLE{perez25,
       author = {{Smith-Perez}, Charlotte and {Hembruff}, Aidan and {Peeters}, Els and {Tielens}, Alexander G.~G.~M. and {Ricca}, Alessandra},
        title = "{Polycyclic aromatic hydrocarbon spectral diversity in NGC 7027 and the evolution of aromatic carriers}",
      journal = {\aap},
         year = 2026,
        month = mar,
       volume = {707},
          eid = {A201},
        pages = {A201},
          doi = {10.1051/0004-6361/202557653},
archivePrefix = {arXiv},
       eprint = {2510.09972},
 primaryClass = {astro-ph.SR},
       adsurl = {https://ui.adsabs.harvard.edu/abs/2026A&A...707A.201S}
}

@ARTICLE{montillaud13,
       author = {{Montillaud}, J. and {Joblin}, C. and {Toublanc}, D.},
        title = "{Evolution of polycyclic aromatic hydrocarbons in photodissociation regions. Hydrogenation and charge states}",
      journal = {\aap},
         year = 2013,
        month = apr,
       volume = {552},
          eid = {A15},
        pages = {A15},
          doi = {10.1051/0004-6361/201220757},
archivePrefix = {arXiv},
       eprint = {1301.6507},
 primaryClass = {astro-ph.GA},
       adsurl = {https://ui.adsabs.harvard.edu/abs/2013A&A...552A..15M}
}

@ARTICLE{aoyama20,
       author = {{Aoyama}, Shohei and {Hirashita}, Hiroyuki and {Nagamine}, Kentaro},
        title = "{Galaxy simulation with the evolution of grain size distribution}",
      journal = {\mnras},
         year = 2020,
        month = jan,
       volume = {491},
       number = {3},
        pages = {3844-3859},
          doi = {10.1093/mnras/stz3253},
archivePrefix = {arXiv},
       eprint = {1906.01917},
 primaryClass = {astro-ph.GA},
       adsurl = {https://ui.adsabs.harvard.edu/abs/2020MNRAS.491.3844A}
}

@ARTICLE{baron22,
       author = {{Baron}, Dalya and {Netzer}, Hagai and {Lutz}, Dieter and {Prochaska}, J. Xavier and {Davies}, Ric I.},
        title = "{Multiphase outflows in post-starburst E+A galaxies - I. General sample properties and the prevalence of obscured starbursts}",
      journal = {\mnras},
         year = 2022,
        month = jan,
       volume = {509},
       number = {3},
        pages = {4457-4479},
          doi = {10.1093/mnras/stab3232},
archivePrefix = {arXiv},
       eprint = {2105.08071},
 primaryClass = {astro-ph.GA},
       adsurl = {https://ui.adsabs.harvard.edu/abs/2022MNRAS.509.4457B}
}

@ARTICLE{fluetsch19,
       author = {{Fluetsch}, A. and {Maiolino}, R. and {Carniani}, S. and {Marconi}, A. and {Cicone}, C. and {Bourne}, M.~A. and {Costa}, T. and {Fabian}, A.~C. and {Ishibashi}, W. and {Venturi}, G.},
        title = "{Cold molecular outflows in the local Universe and their feedback effect on galaxies}",
      journal = {\mnras},
         year = 2019,
        month = mar,
       volume = {483},
       number = {4},
        pages = {4586-4614},
          doi = {10.1093/mnras/sty3449},
archivePrefix = {arXiv},
       eprint = {1805.05352},
 primaryClass = {astro-ph.GA},
       adsurl = {https://ui.adsabs.harvard.edu/abs/2019MNRAS.483.4586F}
}

@ARTICLE{valentino25,
       author = {{Valentino}, F. and {Heintz}, K.~E. and {Brammer}, G. and {Ito}, K. and {Kokorev}, V. and {Whitaker}, K.~E. and {Gallazzi}, A. and {de Graaff}, A. and {Weibel}, A. and {Frye}, B.~L. and {Kamieneski}, P.~S. and {Jin}, S. and {Ceverino}, D. and {Faisst}, A. and {Farcy}, M. and {Fujimoto}, S. and {Gillman}, S. and {Gottumukkala}, R. and {Hamadouche}, M. and {Harrington}, K.~C. and {Hirschmann}, M. and {Jespersen}, C.~K. and {Kakimoto}, T. and {Kubo}, M. and {Lagos}, C. d. P. and {Lee}, M. and {Magdis}, G.~E. and {Man}, A.~W.~S. and {Onodera}, M. and {Rizzo}, F. and {Shimakawa}, R. and {Setton}, D.~J. and {Tanaka}, M. and {Toft}, S. and {Wu}, P.-F. and {Zhu}, P.},
        title = "{Gas outflows in two recently quenched galaxies at z = 4 and 7}",
      journal = {\aap},
         year = 2025,
        month = jul,
       volume = {699},
          eid = {A358},
        pages = {A358},
          doi = {10.1051/0004-6361/202553908},
archivePrefix = {arXiv},
       eprint = {2503.01990},
 primaryClass = {astro-ph.GA},
       adsurl = {https://ui.adsabs.harvard.edu/abs/2025A&A...699A.358V}
}

@ARTICLE{davies24,
       author = {{Davies}, Rebecca L. and {Belli}, Sirio and {Park}, Minjung and {Mendel}, J. Trevor and {Johnson}, Benjamin D. and {Conroy}, Charlie and {Benton}, Chlo{\"e} and {Bugiani}, Letizia and {Emami}, Razieh and {Leja}, Joel and {Li}, Yijia and {Maheson}, Gabriel and {Mathews}, Elijah P. and {Naidu}, Rohan P. and {Nelson}, Erica J. and {Tacchella}, Sandro and {Terrazas}, Bryan A. and {Weinberger}, Rainer},
        title = "{JWST reveals widespread AGN-driven neutral gas outflows in massive z   2 galaxies}",
      journal = {\mnras},
         year = 2024,
        month = mar,
       volume = {528},
       number = {3},
        pages = {4976-4992},
          doi = {10.1093/mnras/stae327},
archivePrefix = {arXiv},
       eprint = {2310.17939},
 primaryClass = {astro-ph.GA},
       adsurl = {https://ui.adsabs.harvard.edu/abs/2024MNRAS.528.4976D}
}

@ARTICLE{cronin26,
       author = {{Cronin}, Serena A. and {Bolatto}, Alberto D. and {Richie}, Helena M. and {Donnelly}, Grant P. and {Levy}, Rebecca C. and {Gordon}, Karl D. and {Tarantino}, Elizabeth and {Boyer}, Martha L. and {Armus}, Lee and {Arens}, Patricia A. and {Boogaard}, Leindert A. and {Dale}, Daniel A. and {Donaghue}, Keaton and {Draine}, Bruce T. and {Duval}, Sara E. and {Emig}, Kimberly and {Fisher}, Deanne B. and {Glover}, Simon C.~O. and {Hensley}, Brandon S. and {Herrera-Camus}, Rodrigo and {Klessen}, Ralf S. and {Lai}, Thomas S.-Y. and {Lenki{\'c}}, Laura and {Leroy}, Adam K. and {Lieber}, Ashley E. and {De Looze}, Ilse and {Lopez}, Sebastian and {Meier}, David S. and {Mills}, Elisabeth A.~C. and {Sandstrom}, Karin M. and {Schneider}, Evan and {Sheriff}, Kaitlyn E. and {Siwakoti}, Utsav and {Skillman}, Evan D. and {Smith}, J.~D.~T. and {Teng}, Yu-Hsuan and {Thompson}, Todd A. and {Tielens}, Alexander G.~G.~M. and {Veilleux}, Sylvain and {Villanueva}, Vicente and {Walter}, Fabian and {van der Werf}, Paul P.},
        title = "{JWST Observations of Starbursts: Dust Processing in the M82 Superwind}",
      journal = {\apj},
         year = 2026,
        month = may,
       volume = {1002},
       number = {2},
          eid = {217},
        pages = {217},
          doi = {10.3847/1538-4357/ae5f66},
archivePrefix = {arXiv},
       eprint = {2604.11873},
 primaryClass = {astro-ph.GA},
       adsurl = {https://ui.adsabs.harvard.edu/abs/2026ApJ..1002..217C}
}

@ARTICLE{veilleux25,
       author = {{Veilleux}, Sylvain and {Shockley}, Steven D. and {Mel{\'e}ndez}, Marcio and {Rupke}, David S.~N. and {Coil}, Alison L. and {Diamond-Stanic}, Aleksandar M. and {Geach}, James E. and {Hickox}, Ryan C. and {Moustakas}, John and {Rudnick}, Gregory H. and {Sell}, Paul H. and {Tremonti}, Christy A. and {Cha}, Hojoon},
        title = "{JWST Discovery of Warm Dust in the Circumgalactic Medium of the Makani Galaxy}",
      journal = {\apj},
         year = 2025,
        month = sep,
       volume = {990},
       number = {1},
          eid = {57},
        pages = {57},
          doi = {10.3847/1538-4357/adee91},
archivePrefix = {arXiv},
       eprint = {2507.08098},
 primaryClass = {astro-ph.GA},
       adsurl = {https://ui.adsabs.harvard.edu/abs/2025ApJ...990...57V}
}

@ARTICLE{richie26,
       author = {{Richie}, Helena M. and {Schneider}, Evan E.},
        title = "{Dust Evolution in Simulated Multiphase Galactic Outflows}",
      journal = {\apj},
         year = 2026,
        month = jan,
       volume = {996},
       number = {1},
          eid = {17},
        pages = {17},
          doi = {10.3847/1538-4357/ae12e8},
archivePrefix = {arXiv},
       eprint = {2505.11734},
 primaryClass = {astro-ph.GA},
       adsurl = {https://ui.adsabs.harvard.edu/abs/2026ApJ...996...17R}
}

@ARTICLE{micelota13,
       author = {{Micelotta}, E.~R. and {Jones}, A.~P. and {Tielens}, A.~G.~G.~M.},
        title = "{Polycyclic aromatic hydrocarbon processing in interstellar shocks}",
      journal = {\aap},
         year = 2010,
        month = feb,
       volume = {510},
          eid = {A36},
        pages = {A36},
          doi = {10.1051/0004-6361/200911682},
archivePrefix = {arXiv},
       eprint = {0910.2461},
 primaryClass = {astro-ph.GA},
       adsurl = {https://ui.adsabs.harvard.edu/abs/2010A&A...510A..36M}
}

@ARTICLE{li_SNIa_20,
       author = {{Li}, Miao and {Li}, Yuan and {Bryan}, Greg L. and {Ostriker}, Eve C. and {Quataert}, Eliot},
        title = "{The Impact of Type Ia Supernovae in Quiescent Galaxies. II. Energetics and Turbulence}",
      journal = {\apj},
         year = 2020,
        month = jul,
       volume = {898},
       number = {1},
          eid = {23},
        pages = {23},
          doi = {10.3847/1538-4357/ab9c22},
archivePrefix = {arXiv},
       eprint = {1909.04204},
 primaryClass = {astro-ph.GA},
       adsurl = {https://ui.adsabs.harvard.edu/abs/2020ApJ...898...23L}
}

@ARTICLE{viaene20,
       author = {{Viaene}, S. and {Nersesian}, A. and {Fritz}, J. and {Verstocken}, S. and {Baes}, M. and {Bianchi}, S. and {Casasola}, V. and {Cassar{\`a}}, L. and {Clark}, C. and {Davies}, J. and {De Looze}, I. and {De Vis}, P. and {Dobbels}, W. and {Galametz}, M. and {Galliano}, F. and {Jones}, A. and {Madden}, S. and {Mosenkov}, A. and {Trcka}, A. and {Xilouris}, E.~M. and {Ysard}, N.},
        title = "{High-resolution, 3D radiative transfer modelling. IV. AGN-powered dust heating in NGC 1068}",
      journal = {\aap},
         year = 2020,
        month = jun,
       volume = {638},
          eid = {A150},
        pages = {A150},
          doi = {10.1051/0004-6361/202037476},
archivePrefix = {arXiv},
       eprint = {2005.01720},
 primaryClass = {astro-ph.GA},
       adsurl = {https://ui.adsabs.harvard.edu/abs/2020A&A...638A.150V}
}

@ARTICLE{gaspari12,
       author = {{Gaspari}, M. and {Brighenti}, F. and {Temi}, P.},
        title = "{Mechanical AGN feedback: controlling the thermodynamical evolution of elliptical galaxies}",
      journal = {\mnras},
         year = 2012,
        month = jul,
       volume = {424},
       number = {1},
        pages = {190-209},
          doi = {10.1111/j.1365-2966.2012.21183.x},
archivePrefix = {arXiv},
       eprint = {1202.6054},
 primaryClass = {astro-ph.CO},
       adsurl = {https://ui.adsabs.harvard.edu/abs/2012MNRAS.424..190G}
}

@ARTICLE{haidar26,
       author = {{Haidar}, Houda and {Rosario}, David J. and {Garc{\'\i}a-Bernete}, Ismael and {Alonso-Herrero}, Almudena and {Audibert}, Anelise and {Campbell}, Steph and {Harrison}, Chris M. and {Costa}, Tiago and {Mu{\~n}oz}, Laura Hermosa and {Combes}, Fran{\c{c}}oise and {Rigopoulou}, Dimitra and {Ricci}, Claudio and {Almeida}, Cristina Ramos and {Bellocchi}, Enrica and {Boorman}, Peter and {Bunker}, Andrew and {Davies}, Richard and {Delaney}, Daniel and {Santos}, Tanio D{\'\i}az and {Esposito}, Federico and {Fawcett}, Victoria A. and {Gandhi}, Poshak and {Garc{\'\i}a-Burillo}, Santiago and {Gonz{\'a}lez-Mart{\'\i}n}, Omaira and {Hicks}, Erin K.~S. and {H{\"o}nig}, Sebastian F. and {Labiano}, Alvaro and {Levenson}, Nancy A. and {Lopez-Rodriguez}, Enrique and {Packham}, Chris and {Pereira-Santaella}, Miguel and {Riffel}, Rogemar A. and {Rodr{\'\i}guez Ardila}, Alberto and {Schneider}, John and {Shimizu}, T. Taro and {Stalevski}, Marko and {Mart{\'\i}n}, Montserrat Villar and {Ward}, Martin and {Zhang}, Lulu and {Leeds}, Gillian and {Donnan}, Fergus R.},
        title = "{GATOS ─ XI. Excess dust heating in the narrow-line regions of nearby AGN revealed with JWST/MIRI}",
      journal = {\mnras},
         year = 2026,
        month = mar,
       volume = {546},
       number = {4},
          eid = {stag069},
        pages = {stag069},
          doi = {10.1093/mnras/stag069},
archivePrefix = {arXiv},
       eprint = {2601.02865},
 primaryClass = {astro-ph.GA},
       adsurl = {https://ui.adsabs.harvard.edu/abs/2026MNRAS.546ag069H}
}

@ARTICLE{bernete24,
       author = {{Garc{\'\i}a-Bernete}, I. and {Rigopoulou}, D. and {Donnan}, F.~R. and {Alonso-Herrero}, A. and {Pereira-Santaella}, M. and {Shimizu}, T. and {Davies}, R. and {Roche}, P.~F. and {Garc{\'\i}a-Burillo}, S. and {Labiano}, A. and {Hermosa Mu{\~n}oz}, L. and {Zhang}, L. and {Audibert}, A. and {Bellocchi}, E. and {Bunker}, A. and {Combes}, F. and {Delaney}, D. and {Esparza-Arredondo}, D. and {Gandhi}, P. and {Gonz{\'a}lez-Mart{\'\i}n}, O. and {H{\"o}nig}, S.~F. and {Imanishi}, M. and {Hicks}, E.~K.~S. and {Fuller}, L. and {Leist}, M. and {Levenson}, N.~A. and {Lopez-Rodriguez}, E. and {Packham}, C. and {Ramos Almeida}, C. and {Ricci}, C. and {Stalevski}, M. and {Villar Mart{\'\i}n}, M. and {Ward}, M.~J.},
        title = "{The Galaxy Activity, Torus, and Outflow Survey (GATOS): V. Unveiling PAH survival and resilience in the circumnuclear regions of AGNs with JWST}",
      journal = {\aap},
         year = 2024,
        month = nov,
       volume = {691},
          eid = {A162},
        pages = {A162},
          doi = {10.1051/0004-6361/202450086},
archivePrefix = {arXiv},
       eprint = {2409.05686},
 primaryClass = {astro-ph.GA},
       adsurl = {https://ui.adsabs.harvard.edu/abs/2024A&A...691A.162G}
}

@ARTICLE{aoyama18,
       author = {{Aoyama}, Shohei and {Hou}, Kuan-Chou and {Hirashita}, Hiroyuki and {Nagamine}, Kentaro and {Shimizu}, Ikkoh},
        title = "{Cosmological simulation with dust formation and destruction}",
      journal = {\mnras},
         year = 2018,
        month = aug,
       volume = {478},
       number = {4},
        pages = {4905-4921},
          doi = {10.1093/mnras/sty1431},
archivePrefix = {arXiv},
       eprint = {1802.04027},
 primaryClass = {astro-ph.GA},
       adsurl = {https://ui.adsabs.harvard.edu/abs/2018MNRAS.478.4905A}
}

@ARTICLE{hirashita19,
       author = {{Hirashita}, Hiroyuki and {Aoyama}, Shohei},
        title = "{Remodelling the evolution of grain size distribution in galaxies}",
      journal = {\mnras},
         year = 2019,
        month = jan,
       volume = {482},
       number = {2},
        pages = {2555-2572},
          doi = {10.1093/mnras/sty2838},
archivePrefix = {arXiv},
       eprint = {1810.07962},
 primaryClass = {astro-ph.GA},
       adsurl = {https://ui.adsabs.harvard.edu/abs/2019MNRAS.482.2555H}
}

@ARTICLE{romano22,
       author = {{Romano}, Leonard E.~C. and {Nagamine}, Kentaro and {Hirashita}, Hiroyuki},
        title = "{Dust diffusion in SPH simulations of an isolated galaxy}",
      journal = {\mnras},
         year = 2022,
        month = jul,
       volume = {514},
       number = {1},
        pages = {1441-1460},
          doi = {10.1093/mnras/stac1385},
archivePrefix = {arXiv},
       eprint = {2202.05243},
 primaryClass = {astro-ph.GA},
       adsurl = {https://ui.adsabs.harvard.edu/abs/2022MNRAS.514.1441R}
}

@ARTICLE{biscaro16,
       author = {{Biscaro}, Chiara and {Cherchneff}, Isabelle},
        title = "{Molecules and dust in Cassiopeia A. II. Dust sputtering and diagnosis of supernova dust survival in remnants}",
      journal = {\aap},
         year = 2016,
        month = may,
       volume = {589},
          eid = {A132},
        pages = {A132},
          doi = {10.1051/0004-6361/201527769},
archivePrefix = {arXiv},
       eprint = {1511.05487},
 primaryClass = {astro-ph.SR},
       adsurl = {https://ui.adsabs.harvard.edu/abs/2016A&A...589A.132B}
}

@ARTICLE{slavin20,
       author = {{Slavin}, Jonathan D. and {Dwek}, Eli and {Mac Low}, Mordecai-Mark and {Hill}, Alex S.},
        title = "{The Dynamics, Destruction, and Survival of Supernova-formed Dust Grains}",
      journal = {\apj},
         year = 2020,
        month = oct,
       volume = {902},
       number = {2},
          eid = {135},
        pages = {135},
          doi = {10.3847/1538-4357/abb5a4},
archivePrefix = {arXiv},
       eprint = {2009.01895},
 primaryClass = {astro-ph.HE},
       adsurl = {https://ui.adsabs.harvard.edu/abs/2020ApJ...902..135S}
}

@ARTICLE{adscheid25,
       author = {{Adscheid}, Sylvia and {Magnelli}, Benjamin and {Ciesla}, Laure and {Liu}, Daizhong and {Schinnerer}, Eva and {Bertoldi}, Frank},
        title = "{A$^{3}$COSMOS: The dust content of massive quiescent galaxies and its evolution with cosmic time}",
      journal = {\aap},
         year = 2025,
        month = oct,
       volume = {702},
          eid = {A186},
        pages = {A186},
          doi = {10.1051/0004-6361/202554400},
archivePrefix = {arXiv},
       eprint = {2508.18097},
 primaryClass = {astro-ph.GA},
       adsurl = {https://ui.adsabs.harvard.edu/abs/2025A&A...702A.186A}
}

@ARTICLE{deugenio26,
       author = {{D'Eugenio}, C. and {Daddi}, E. and {Gobat}, R. and {Jin}, S. and {Liu}, D. and {Sun}, H. and {Gentile}, F. and {Bruckmann}, F. and {Liu}, Z. and {Delvecchio}, I. and {Vallini}, L. and {Magnelli}, B. and {Zanella}, A.},
        title = "{A first [CII] view of high-z quiescent galaxies}",
      journal = {arXiv e-prints},
         year = 2026,
        month = apr,
          eid = {arXiv:2604.09347},
        pages = {arXiv:2604.09347},
          doi = {10.48550/arXiv.2604.09347},
archivePrefix = {arXiv},
       eprint = {2604.09347},
 primaryClass = {astro-ph.GA},
       adsurl = {https://ui.adsabs.harvard.edu/abs/2026arXiv260409347D}
}

@ARTICLE{kriek19,
       author = {{Kriek}, Mariska and {Price}, Sedona H. and {Conroy}, Charlie and {Suess}, Katherine A. and {Mowla}, Lamiya and {Pasha}, Imad and {Bezanson}, Rachel and {van Dokkum}, Pieter and {Barro}, Guillermo},
        title = "{Stellar Metallicities and Elemental Abundance Ratios of z {\ensuremath{\sim}} 1.4 Massive Quiescent Galaxies}",
      journal = {\apjl},
         year = 2019,
        month = aug,
       volume = {880},
       number = {2},
          eid = {L31},
        pages = {L31},
          doi = {10.3847/2041-8213/ab2e75},
archivePrefix = {arXiv},
       eprint = {1907.04327},
 primaryClass = {astro-ph.GA},
       adsurl = {https://ui.adsabs.harvard.edu/abs/2019ApJ...880L..31K}
}

@ARTICLE{osman25,
       author = {{Osman}, Omima and {De Lucia}, Gabriella and {Fontanot}, Fabio and {Xie}, Lizhi and {Hirschmann}, Michaela},
        title = "{Dust evolution across cosmic times as seen through DUSTY-GAEA}",
      journal = {arXiv e-prints},
         year = 2025,
        month = dec,
          eid = {arXiv:2512.15902},
        pages = {arXiv:2512.15902},
          doi = {10.48550/arXiv.2512.15902},
archivePrefix = {arXiv},
       eprint = {2512.15902},
 primaryClass = {astro-ph.GA},
       adsurl = {https://ui.adsabs.harvard.edu/abs/2025arXiv251215902O}
}

@ARTICLE{whitaker26,
       author = {{Whitaker}, Katherine E. and {Bezanson}, Rachel},
        title = "{Quenching of Star Formation in Massive Galaxies}",
      journal = {arXiv e-prints},
         year = 2026,
        month = jun,
          eid = {arXiv:2606.12156},
        pages = {arXiv:2606.12156},
          doi = {10.48550/arXiv.2606.12156},
archivePrefix = {arXiv},
       eprint = {2606.12156},
 primaryClass = {astro-ph.GA},
       adsurl = {https://ui.adsabs.harvard.edu/abs/2026arXiv260612156W}
}

@ARTICLE{hirashita20,
       author = {{Hirashita}, Hiroyuki and {Deng}, Weining and {Murga}, Maria S.},
        title = "{Spectral energy distributions of dust and PAHs based on the evolution of grain size distribution in galaxies}",
      journal = {\mnras},
         year = 2020,
        month = dec,
       volume = {499},
       number = {2},
        pages = {3046-3060},
          doi = {10.1093/mnras/staa3101},
archivePrefix = {arXiv},
       eprint = {2010.00922},
 primaryClass = {astro-ph.GA},
       adsurl = {https://ui.adsabs.harvard.edu/abs/2020MNRAS.499.3046H}
}

@ARTICLE{matsumoto24,
       author = {{Matsumoto}, Kosei and {Hirashita}, Hiroyuki and {Nagamine}, Kentaro and {van der Giessen}, Stefan and {Romano}, Leonard E.~C. and {Rela{\~n}o}, Monica and {De Looze}, Ilse and {Baes}, Maarten and {Nersesian}, Angelos and {Camps}, Peter and {Hou}, Kuan-chou and {Oku}, Yuri},
        title = "{Observational signatures of the dust size evolution in isolated galaxy simulations}",
      journal = {\aap},
         year = 2024,
        month = sep,
       volume = {689},
          eid = {A79},
        pages = {A79},
          doi = {10.1051/0004-6361/202449454},
archivePrefix = {arXiv},
       eprint = {2402.02659},
 primaryClass = {astro-ph.GA},
       adsurl = {https://ui.adsabs.harvard.edu/abs/2024A&A...689A..79M}
}

@ARTICLE{lu25,
       author = {{Lu}, Shiying and {Daddi}, Emanuele and {Maraston}, Claudia and {Dickinson}, Mark and {Haro}, Pablo Arrabal and {Gobat}, Raphael and {Renzini}, Alvio and {Giavalisco}, Mauro and {Bagley}, Micaela B. and {Calabr{\`o}}, Antonello and {Cheng}, Yingjie and {de la Vega}, Alexander and {D'Eugenio}, Chiara and {Elbaz}, David and {Finkelstein}, Steven L. and {G{\'o}mez-Guijarro}, Carlos and {Gu}, Qiusheng and {Hathi}, Nimish P. and {Huertas-Company}, Marc and {Kartaltepe}, Jeyhan S. and {Koekemoer}, Anton M. and {Henry}, Aur{\'e}lien and {Lyu}, Yipeng and {Magnelli}, Benjamin and {Mobasher}, Bahram and {Papovich}, Casey and {Pirzkal}, Nor and {Rich}, R. Michael and {Tacchella}, Sandro and {Yung}, L.~Y. Aaron},
        title = "{Strong spectral features from asymptotic giant branch stars in distant quiescent galaxies}",
      journal = {Nature Astronomy},
         year = 2025,
        month = jan,
       volume = {9},
        pages = {128-140},
          doi = {10.1038/s41550-024-02391-9},
archivePrefix = {arXiv},
       eprint = {2403.07414},
 primaryClass = {astro-ph.GA},
       adsurl = {https://ui.adsabs.harvard.edu/abs/2025NatAs...9..128L}
}

@ARTICLE{bevacqua25,
       author = {{Bevacqua}, Davide and {Saracco}, Paolo and {La Barbera}, Francesco and {De Marchi}, Guido and {De Propris}, Roberto and {Ditrani}, Fabio R. and {Gallazzi}, Anna R. and {Giardino}, Giovanna and {Marchesini}, Danilo and {Pasquali}, Anna and {Rawle}, Tim D. and {Spiniello}, Chiara and {Vazdekis}, Alexandre and {Zibetti}, Stefano},
        title = "{TP-AGB stars and stellar population properties of a post-starburst galaxy at z {\ensuremath{\sim}} 2 through optical and near-infrared spectroscopy with JWST}",
      journal = {\aap},
         year = 2025,
        month = jul,
       volume = {699},
          eid = {A203},
        pages = {A203},
          doi = {10.1051/0004-6361/202553736},
archivePrefix = {arXiv},
       eprint = {2501.07291},
 primaryClass = {astro-ph.GA},
       adsurl = {https://ui.adsabs.harvard.edu/abs/2025A&A...699A.203B}
}

@ARTICLE{lisiecki26,
       author = {{Lisiecki}, K. and {Donevski}, D. and {Man}, A.~W.~S. and {Damjanov}, I. and {Romano}, M. and {Belli}, S. and {Long}, A. and {Lorenzon}, G. and {Ma{\l}ek}, K. and {Junais} and {Lovell}, C.~C. and {Nanni}, A. and {Bertemes}, C. and {Pearson}, W.~J. and {Ryzhov}, O. and {Koprowski}, M. and {Pollo}, A. and {Dey}, S. and {Thuruthipilly}, H.},
        title = "{Impact of stochastic star formation histories and dust information on selecting quiescent galaxies with JWST photometry}",
      journal = {\aap},
         year = 2026,
        month = apr,
       volume = {708},
          eid = {A235},
        pages = {A235},
          doi = {10.1051/0004-6361/202557118},
archivePrefix = {arXiv},
       eprint = {2509.10117},
 primaryClass = {astro-ph.GA},
       adsurl = {https://ui.adsabs.harvard.edu/abs/2026A&A...708A.235L}
}

@ARTICLE{kelson10,
       author = {{Kelson}, Daniel D. and {Holden}, Bradford P.},
        title = "{The Mid-infrared Luminosities of Normal Galaxies Over Cosmic Time}",
      journal = {\apjl},
         year = 2010,
        month = apr,
       volume = {713},
       number = {1},
        pages = {L28-L32},
          doi = {10.1088/2041-8205/713/1/L28},
archivePrefix = {arXiv},
       eprint = {1003.1420},
 primaryClass = {astro-ph.CO},
       adsurl = {https://ui.adsabs.harvard.edu/abs/2010ApJ...713L..28K}
}

@ARTICLE{barrufet25,
       author = {{Barrufet}, L. and {Oesch}, P.~A. and {Marques-Chaves}, R. and {Arellano-Cordova}, K. and {Baggen}, J.~F.~W. and {Carnall}, A.~C. and {Cullen}, F. and {Dunlop}, J.~S. and {Gottumukkala}, R. and {Fudamoto}, Y. and {Illingworth}, G.~D. and {Magee}, D. and {McLure}, R.~J. and {McLeod}, D.~J. and {Micha{\l}owski}, M.~J. and {Stefanon}, M. and {van Dokkum}, P.~G. and {Weibel}, A.},
        title = "{Quiescent or dusty? Unveiling the nature of extremely red galaxies at z > 3}",
      journal = {\mnras},
         year = 2025,
        month = mar,
       volume = {537},
       number = {4},
        pages = {3453-3469},
          doi = {10.1093/mnras/staf013},
archivePrefix = {arXiv},
       eprint = {2404.08052},
 primaryClass = {astro-ph.GA},
       adsurl = {https://ui.adsabs.harvard.edu/abs/2025MNRAS.537.3453B}
}

@ARTICLE{zhu26,
       author = {{Zhu}, Pengpei and {Ito}, Kei and {Valentino}, Francesco and {Hamadouche}, Massissilia and {Scarpe}, Gianluca and {Whitaker}, Katherine E. and {Kakimoto}, Takumi and {Baker}, William M. and {Gallazzi}, Anna R. and {Gillman}, Steven and {Gottumukkala}, Rashmi and {Jespersen}, Christian Kragh and {Lee}, Minju and {Man}, Allison W.~S. and {Magdis}, Georgios and {Onodera}, Masato and {Shimakawa}, Rhythm and {Vijayan}, Aswin and {Wu}, Po-Feng},
        title = "{There and back again? Neutral outflows in z\raisebox{-0.5ex}\textasciitilde3.5 quiescent galaxies}",
      journal = {arXiv e-prints},
         year = 2026,
        month = feb,
          eid = {arXiv:2602.17767},
        pages = {arXiv:2602.17767},
          doi = {10.48550/arXiv.2602.17767},
archivePrefix = {arXiv},
       eprint = {2602.17767},
 primaryClass = {astro-ph.GA},
       adsurl = {https://ui.adsabs.harvard.edu/abs/2026arXiv260217767Z}
}

@ARTICLE{shivaei24,
       author = {{Shivaei}, Irene and {Alberts}, Stacey and {Florian}, Michael and {Rieke}, George and {Wuyts}, Stijn and {Bodansky}, Sarah and {Bunker}, Andrew J. and {Cameron}, Alex J. and {Curti}, Mirko and {D'Eugenio}, Francesco and {Dudzevi{\v{c}}i{\={u}}t{\.{e}}}, Ugn{\.{e}} and {Ji}, Zhiyuan and {Johnson}, Benjamin D. and {Kramarenko}, Ivan and {Lyu}, Jianwei and {Matthee}, Jorryt and {Morrison}, Jane and {Naidu}, Rohan and {P{\'e}rez-Gonz{\'a}lez}, Pablo G. and {Reddy}, Naveen and {Robertson}, Brant and {Sun}, Yang and {Tacchella}, Sandro and {Whitaker}, Katherine and {Williams}, Christina C. and {Willmer}, Christopher N.~A. and {Witstok}, Joris and {Xiao}, Mengyuan and {Zhu}, Yongda},
        title = "{A new census of dust and polycyclic aromatic hydrocarbons at z = 0.7─2 with JWST MIRI}",
      journal = {\aap},
         year = 2024,
        month = oct,
       volume = {690},
          eid = {A89},
        pages = {A89},
          doi = {10.1051/0004-6361/202449579},
archivePrefix = {arXiv},
       eprint = {2402.07989},
 primaryClass = {astro-ph.GA},
       adsurl = {https://ui.adsabs.harvard.edu/abs/2024A&A...690A..89S}
}

@ARTICLE{chestenet25,
       author = {{Chastenet}, J{\'e}r{\'e}my and {Sandstrom}, Karin and {Leroy}, Adam K. and {Bot}, Caroline and {Chiang}, I-Da and {Chown}, Ryan and {Gordon}, Karl D. and {Koch}, Eric W. and {Roussel}, H{\'e}l{\`e}ne and {Sutter}, Jessica and {Williams}, Thomas G.},
        title = "{The Resolved Behavior of Dust Mass, Polycyclic Aromatic Hydrocarbon Fraction, and Radiation Field in {\ensuremath{\sim}}800 Nearby Galaxies}",
      journal = {\apjs},
         year = 2025,
        month = jan,
       volume = {276},
       number = {1},
          eid = {2},
        pages = {2},
          doi = {10.3847/1538-4365/ad8a5c},
archivePrefix = {arXiv},
       eprint = {2410.03835},
 primaryClass = {astro-ph.GA},
       adsurl = {https://ui.adsabs.harvard.edu/abs/2025ApJS..276....2C}
}

@ARTICLE{setton25,
       author = {{Setton}, David J. and {Spilker}, Justin S. and {Bezanson}, Rachel and {Suess}, Katherine A. and {Greene}, Jenny E. and {Goulding}, Andy D. and {Cenci}, Elia and {D'Onofrio}, Vincenzo R. and {Feldmann}, Robert and {Kriek}, Mariska and {Kumar}, Anika and {Luo}, Yuanze and {Narayanan}, Desika and {Verrico}, Margaret E. and {Zhu}, Pengpei},
        title = "{SQuIGGL{\textrightarrow}E: Buried Star Formation Cannot Explain the Rapidly Fading CO(2─1) Luminosity in Massive, z {\ensuremath{\sim}} 0.7 Post-starburst Galaxies}",
      journal = {\aj},
         year = 2025,
        month = dec,
       volume = {170},
       number = {6},
          eid = {351},
        pages = {351},
          doi = {10.3847/1538-3881/ae1607},
archivePrefix = {arXiv},
       eprint = {2509.00148},
 primaryClass = {astro-ph.GA},
       adsurl = {https://ui.adsabs.harvard.edu/abs/2025AJ....170..351S}
}

@ARTICLE{woodrum22,
       author = {{Woodrum}, Charity and {Williams}, Christina C. and {Rieke}, Marcia and {Leja}, Joel and {Johnson}, Benjamin D. and {Bezanson}, Rachel and {Kennicutt}, Robert and {Spilker}, Justin and {Tacchella}, Sandro},
        title = "{Molecular Gas Reservoirs in Massive Quiescent Galaxies at z   0.7 Linked to Late-time Star Formation}",
      journal = {\apj},
         year = 2022,
        month = nov,
       volume = {940},
       number = {1},
          eid = {39},
        pages = {39},
          doi = {10.3847/1538-4357/ac9af7},
archivePrefix = {arXiv},
       eprint = {2210.03832},
 primaryClass = {astro-ph.GA},
       adsurl = {https://ui.adsabs.harvard.edu/abs/2022ApJ...940...39W}
}

@ARTICLE{wild16,
       author = {{Wild}, Vivienne and {Almaini}, Omar and {Dunlop}, Jim and {Simpson}, Chris and {Rowlands}, Kate and {Bowler}, Rebecca and {Maltby}, David and {McLure}, Ross},
        title = "{The evolution of post-starburst galaxies from z=2 to 0.5}",
      journal = {\mnras},
         year = 2016,
        month = nov,
       volume = {463},
       number = {1},
        pages = {832-844},
          doi = {10.1093/mnras/stw1996},
archivePrefix = {arXiv},
       eprint = {1608.00588},
 primaryClass = {astro-ph.GA},
       adsurl = {https://ui.adsabs.harvard.edu/abs/2016MNRAS.463..832W}
}

@ARTICLE{chang26,
       author = {{Chang}, Wenjun and {Wilson}, Gillian and {Forrest}, Ben and {McConachie}, Ian and {Noble}, Allison and {Muzzin}, Adam and {Marchesini}, Danilo and {Cooper}, Michael C. and {Webb}, Tracy and {Canalizo}, Gabriela and {Gomez}, Percy L. and {Zhu}, Yongda and {Edward}, Adit and {Lei}, Han and {Henry}, Aur{\'e}lien and {Urbano Stawinski}, Stephanie M. and {Wisz}, Marie E.},
        title = "{MAGAZ3NE: Dust Deficiency in Ultramassive Quiescent Galaxies at $3<z<4$ with ALMA Observations}",
      journal = {arXiv e-prints},
         year = 2026,
        month = jan,
          eid = {arXiv:2601.22844},
        pages = {arXiv:2601.22844},
          doi = {10.48550/arXiv.2601.22844},
archivePrefix = {arXiv},
       eprint = {2601.22844},
 primaryClass = {astro-ph.GA},
       adsurl = {https://ui.adsabs.harvard.edu/abs/2026arXiv260122844C}
}

@ARTICLE{yasuda12,
       author = {{Yasuda}, Yuki and {Kozasa}, Takashi},
        title = "{Formation of SiC Grains in Pulsation-enhanced Dust-driven Wind around Carbon-rich Asymptotic Giant Branch Stars}",
      journal = {\apj},
         year = 2012,
        month = feb,
       volume = {745},
       number = {2},
          eid = {159},
        pages = {159},
          doi = {10.1088/0004-637X/745/2/159},
archivePrefix = {arXiv},
       eprint = {1109.6386},
 primaryClass = {astro-ph.SR},
       adsurl = {https://ui.adsabs.harvard.edu/abs/2012ApJ...745..159Y}
}

@ARTICLE{nanni18,
       author = {{Nanni}, Ambra and {Marigo}, Paola and {Girardi}, L{\'e}o and {Rubele}, Stefano and {Bressan}, Alessandro and {Groenewegen}, Martin A.~T. and {Pastorelli}, Giada and {Aringer}, Bernhard},
        title = "{Estimating the dust production rate of carbon stars in the Small Magellanic Cloud}",
      journal = {\mnras},
         year = 2018,
        month = feb,
       volume = {473},
       number = {4},
        pages = {5492-5513},
          doi = {10.1093/mnras/stx2641},
archivePrefix = {arXiv},
       eprint = {1710.02591},
 primaryClass = {astro-ph.SR},
       adsurl = {https://ui.adsabs.harvard.edu/abs/2018MNRAS.473.5492N}
}

@ARTICLE{relano22,
       author = {{Rela{\~n}o}, M. and {De Looze}, I. and {Saintonge}, A. and {Hou}, K.-C. and {Romano}, L.~E.~C. and {Nagamine}, K. and {Hirashita}, H. and {Aoyama}, S. and {Lamperti}, I. and {Lisenfeld}, U. and {Smith}, M.~W.~L. and {Chastenet}, J. and {Xiao}, T. and {Gao}, Y. and {Sargent}, M. and {van der Giessen}, S.~A.},
        title = "{Dust grain size evolution in local galaxies: a comparison between observations and simulations}",
      journal = {\mnras},
         year = 2022,
        month = oct,
       volume = {515},
       number = {4},
        pages = {5306-5334},
          doi = {10.1093/mnras/stac2108},
archivePrefix = {arXiv},
       eprint = {2207.13196},
 primaryClass = {astro-ph.GA},
       adsurl = {https://ui.adsabs.harvard.edu/abs/2022MNRAS.515.5306R}
}

@ARTICLE{alberts24,
       author = {{Alberts}, Stacey and {Lyu}, Jianwei and {Shivaei}, Irene and {Rieke}, George H. and {P{\'e}rez-Gonz{\'a}lez}, Pablo G. and {Bonaventura}, Nina and {Zhu}, Yongda and {Helton}, Jakob M. and {Ji}, Zhiyuan and {Morrison}, Jane and {Robertson}, Brant E. and {Stone}, Meredith A. and {Sun}, Yang and {Williams}, Christina C. and {Willmer}, Christopher N.~A.},
        title = "{SMILES Initial Data Release: Unveiling the Obscured Universe with MIRI Multiband Imaging}",
      journal = {\apj},
         year = 2024,
        month = dec,
       volume = {976},
       number = {2},
          eid = {224},
        pages = {224},
          doi = {10.3847/1538-4357/ad7396},
archivePrefix = {arXiv},
       eprint = {2405.15972},
 primaryClass = {astro-ph.GA},
       adsurl = {https://ui.adsabs.harvard.edu/abs/2024ApJ...976..224A}
}

@ARTICLE{letter91,
       author = {{Latter}, William B.},
        title = "{Large Molecule Production by Mass-losing Carbon Stars: The Primary Source of Interstellar Polycyclic Aromatic Hydrocarbons?}",
      journal = {\apj},
         year = 1991,
        month = aug,
       volume = {377},
        pages = {187},
          doi = {10.1086/170346},
       adsurl = {https://ui.adsabs.harvard.edu/abs/1991ApJ...377..187L}
}

@ARTICLE{jones96,
       author = {{Jones}, A.~P. and {Tielens}, A.~G.~G.~M. and {Hollenbach}, D.~J.},
        title = "{Grain Shattering in Shocks: The Interstellar Grain Size Distribution}",
      journal = {\apj},
         year = 1996,
        month = oct,
       volume = {469},
        pages = {740},
          doi = {10.1086/177823},
       adsurl = {https://ui.adsabs.harvard.edu/abs/1996ApJ...469..740J}
}

@ARTICLE{neumann23,
       author = {{Neumann}, Lukas and {Gallagher}, Molly J. and {Bigiel}, Frank and {Leroy}, Adam K. and {Barnes}, Ashley T. and {Usero}, Antonio and {den Brok}, Jakob S. and {Belfiore}, Francesco and {Be{\v{s}}li{\'c}}, Ivana and {Cao}, Yixian and {Chevance}, M{\'e}lanie and {Dale}, Daniel A. and {Eibensteiner}, Cosima and {Glover}, Simon C.~O. and {Grasha}, Kathryn and {Henshaw}, Jonathan D. and {Jim{\'e}nez-Donaire}, Mar{\'\i}a J. and {Klessen}, Ralf S. and {Kruijssen}, J.~M. Diederik and {Liu}, Daizhong and {Meidt}, Sharon and {Pety}, J{\'e}r{\^o}me and {Puschnig}, Johannes and {Querejeta}, Miguel and {Rosolowsky}, Erik and {Schinnerer}, Eva and {Schruba}, Andreas and {Sormani}, Mattia C. and {Sun}, Jiayi and {Teng}, Yu-Hsuan and {Williams}, Thomas G.},
        title = "{The ALMOND survey: molecular cloud properties and gas density tracers across 25 nearby spiral galaxies with ALMA}",
      journal = {\mnras},
         year = 2023,
        month = may,
       volume = {521},
       number = {3},
        pages = {3348-3383},
          doi = {10.1093/mnras/stad424},
archivePrefix = {arXiv},
       eprint = {2302.03042},
 primaryClass = {astro-ph.GA},
       adsurl = {https://ui.adsabs.harvard.edu/abs/2023MNRAS.521.3348N}
}

@ARTICLE{lin24,
       author = {{Lin}, Lihwai and {Pan}, Hsi-An and {Ellison}, Sara L. and {Harada}, Nanase and {Jim{\'e}nez-Donaire}, Mar{\'\i}a J. and {French}, K. Decker and {Baker}, William M. and {Hsieh}, Bau-Ching and {Koyama}, Yusei and {L{\'o}pez-Cob{\'a}}, Carlos and {Michiyama}, Tomonari and {Rowlands}, Kate and {S{\'a}nchez}, Sebasti{\'a}n F. and {Thorp}, Mallory D.},
        title = "{The ALMaQUEST Survey. XII. Dense Molecular Gas as Traced by HCN and HCO$^{+}$ in Green Valley Galaxies}",
      journal = {\apj},
         year = 2024,
        month = mar,
       volume = {963},
       number = {2},
          eid = {115},
        pages = {115},
          doi = {10.3847/1538-4357/ad18b9},
archivePrefix = {arXiv},
       eprint = {2401.05976},
 primaryClass = {astro-ph.GA},
       adsurl = {https://ui.adsabs.harvard.edu/abs/2024ApJ...963..115L}
}

@ARTICLE{draine01,
       author = {{Draine}, B.~T. and {Li}, Aigen},
        title = "{Infrared Emission from Interstellar Dust. I. Stochastic Heating of Small Grains}",
      journal = {\apj},
         year = 2001,
        month = apr,
       volume = {551},
       number = {2},
        pages = {807-824},
          doi = {10.1086/320227},
archivePrefix = {arXiv},
       eprint = {astro-ph/0011318},
 primaryClass = {astro-ph},
       adsurl = {https://ui.adsabs.harvard.edu/abs/2001ApJ...551..807D}
}

@ARTICLE{hirashita23,
       author = {{Hirashita}, Hiroyuki},
        title = "{Evolution of grain size distribution with enhanced abundance of small carbonaceous grains in galactic environments}",
      journal = {\mnras},
         year = 2023,
        month = jan,
       volume = {518},
       number = {3},
        pages = {3827-3837},
          doi = {10.1093/mnras/stac3394},
archivePrefix = {arXiv},
       eprint = {2211.09982},
 primaryClass = {astro-ph.GA},
       adsurl = {https://ui.adsabs.harvard.edu/abs/2023MNRAS.518.3827H}
}

@ARTICLE{faisst17,
       author = {{Faisst}, A.~L. and {Carollo}, C.~M. and {Capak}, P.~L. and {Tacchella}, S. and {Renzini}, A. and {Ilbert}, O. and {McCracken}, H.~J. and {Scoville}, N.~Z.},
        title = "{Constraints on Quenching of Z {\ensuremath{\lesssim}} 2 Massive Galaxies from the Evolution of the Average Sizes of Star-forming and Quenched Populations in COSMOS}",
      journal = {\apj},
         year = 2017,
        month = apr,
       volume = {839},
       number = {2},
          eid = {71},
        pages = {71},
          doi = {10.3847/1538-4357/aa697a},
archivePrefix = {arXiv},
       eprint = {1703.09234},
 primaryClass = {astro-ph.GA},
       adsurl = {https://ui.adsabs.harvard.edu/abs/2017ApJ...839...71F}
}

@ARTICLE{backhaus25,
       author = {{Backhaus}, Bren E. and {Kirkpatrick}, Allison and {Yang}, Guang and {Troiani}, Gregory and {Hamblin}, Kurt and {Kartaltepe}, Jeyhan S. and {Kocevski}, Dale D. and {Koekemoer}, Anton M. and {Lambrides}, Erini and {Papovich}, Casey and {Ronayne}, Kaila},
        title = "{MEGA Mass Assembly with JWST: The MIRI EGS Galaxy and Active Galactic Nucleus Survey}",
      journal = {\aj},
         year = 2025,
        month = dec,
       volume = {170},
       number = {6},
          eid = {300},
        pages = {300},
          doi = {10.3847/1538-3881/ae0cc4},
archivePrefix = {arXiv},
       eprint = {2503.19078},
 primaryClass = {astro-ph.GA},
       adsurl = {https://ui.adsabs.harvard.edu/abs/2025AJ....170..300B}
}

@ARTICLE{pantoni26,
       author = {{Pantoni}, L. and {Baes}, M. and {Decin}, L. and {Guillard}, P. and {Alonso Herrero}, A. and {Hermosa Mu{\~n}oz}, L. and {Evangelista}, L. and {Garc{\'\i}a-Bernete}, I. and {Donnan}, F. and {Buiten}, V. and {Garcia-Burillo}, S. and {Wright}, G. and {Colina}, L. and {B{\"o}ker}, T. and {{\"O}stlin}, G. and {Dicken}, D. and {Labiano}, A. and {Rouan}, D. and {van der Werf}, P. and {Eckart}, A. and {Garc{\'\i}a-Mar{\'\i}n}, M. and {G{\"u}del}, M. and {Henning}, Th. and {Lagage}, P.-O. and {Walter}, F. and {Ward}, M.~J.},
        title = "{MICONIC: JWST/MIRI-MRS reveals heavily reprocessed polycyclic aromatic hydrocarbons in the circumnuclear disc of Centaurus A}",
      journal = {\aap},
         year = 2026,
        month = may,
       volume = {709},
          eid = {A237},
        pages = {A237},
          doi = {10.1051/0004-6361/202558839},
archivePrefix = {arXiv},
       eprint = {2603.23674},
 primaryClass = {astro-ph.GA},
       adsurl = {https://ui.adsabs.harvard.edu/abs/2026A&A...709A.237P}
}

@ARTICLE{romano24,
       author = {{Romano}, M. and {Donevski}, D. and {Junais} and {Nanni}, A. and {Ginolfi}, M. and {Jones}, G.~C. and {Shivaei}, I. and {Lorenzon}, G. and {Hamed}, M. and {Salak}, D. and {Sawant}, P.},
        title = "{Evidence of extended [CII] and dust emission in local dwarf galaxies}",
      journal = {\aap},
         year = 2024,
        month = mar,
       volume = {683},
          eid = {L9},
        pages = {L9},
          doi = {10.1051/0004-6361/202349111},
archivePrefix = {arXiv},
       eprint = {2402.17662},
 primaryClass = {astro-ph.GA},
       adsurl = {https://ui.adsabs.harvard.edu/abs/2024A&A...683L...9R}
}

@ARTICLE{fogarty19,
       author = {{Fogarty}, Kevin and {Postman}, Marc and {Li}, Yuan and {Dannerbauer}, Helmut and {Liu}, Hauyu Baobab and {Donahue}, Megan and {Ziegler}, Bodo and {Koekemoer}, Anton and {Frye}, Brenda},
        title = "{The Dust and Molecular Gas in the Brightest Cluster Galaxy in MACS 1931.8-2635}",
      journal = {\apj},
         year = 2019,
        month = jul,
       volume = {879},
       number = {2},
          eid = {103},
        pages = {103},
          doi = {10.3847/1538-4357/ab22a4},
archivePrefix = {arXiv},
       eprint = {1905.01377},
 primaryClass = {astro-ph.GA},
       adsurl = {https://ui.adsabs.harvard.edu/abs/2019ApJ...879..103F}
}

@ARTICLE{hu19,
       author = {{Hu}, Chia-Yu and {Zhukovska}, Svitlana and {Somerville}, Rachel S. and {Naab}, Thorsten},
        title = "{Thermal and non-thermal dust sputtering in hydrodynamical simulations of the multiphase interstellar medium}",
      journal = {\mnras},
         year = 2019,
        month = aug,
       volume = {487},
       number = {3},
        pages = {3252-3269},
          doi = {10.1093/mnras/stz1481},
archivePrefix = {arXiv},
       eprint = {1902.01368},
 primaryClass = {astro-ph.GA},
       adsurl = {https://ui.adsabs.harvard.edu/abs/2019MNRAS.487.3252H}
}

@INCOLLECTION{long26,
       author = {{Long}, Arianna and {McKinney}, Jed and {Lambrides}, Erini and {Cooper}, Olivia and {Manning}, Sinclaire and {Donevski}, Darko and {Kubo}, Mariko and {Clements}, Dave},
        title = "{Dust in High Redshift Quiescent Galaxies}",
    booktitle = {PRIMA General Observer Science Book Volume 2},
         year = 2025,
       editor = {{Moullet}, A. and {Burgarella}, D. and {Kataria}, T. and {Beuther}, H. and {Battersby}, C. and {Cheng}, M. and {Essinger-Hileman}, T. and {Inami}, H. and {Mills}, E. and {Nagao}, T. and {Unwin}, S.},
       volume = {2},
        pages = {191-195},
       adsurl = {https://ui.adsabs.harvard.edu/abs/2025prim.book..191L}
}

@ARTICLE{peissker26,
       author = {{Pei{\ss}ker}, F. and {Garc{\'\i}a Mar{\'\i}n}, M. and {Eckart}, A. and {Wright}, G. and {Jones}, O.~C. and {Dicken}, D. and {Alonso Herrero}, A. and {Rouan}, D. and {Law}, D. and {B{\"o}ker}, T. and {Henning}, T. and {Baes}, M. and {Labiano}, A. and {Pantoni}, L. and {Hermosa Mu{\~n}oz}, L. and {Lagage}, P.~O. and {van der Werf}, P. and {{\"O}stlin}, G. and {Blommaert}, J.~A.~D.~L. and {G{\"u}del}, M. and {Guillard}, P.},
        title = "{Dust production in the harsh environment of Sgr A* - MIRI/JWST observation of the O-rich asymptotic giant branch star IRS\raisebox{-0.5ex}\textasciitilde3}",
      journal = {arXiv e-prints},
         year = 2026,
        month = aug,
          eid = {arXiv:2608.09511},
        pages = {arXiv:2608.09511},
archivePrefix = {arXiv},
       eprint = {2608.09511},
 primaryClass = {astro-ph.GA},
       adsurl = {https://ui.adsabs.harvard.edu/abs/2026arXiv260809511P}
}

@ARTICLE{prima25,
       author = {{Glenn}, Jason and {Meixner}, Margaret and {Bradford}, Charles M. and {Pontoppidan}, Klaus and {Pope}, Alexandra and {Kataria}, Tiffany and {Rocca}, Jennifer and {Luthman}, Elizabeth and {Armus}, Lee and {Baselmans}, Jochem and {Battersby}, Cara and {Bollato}, Alberto and {Burgarella}, Denis and {Chen}, Weibo and {Ciesla}, Laure and {Day}, Peter and {Di Giorgio}, Anna and {Dipirro}, Michael and {Dowell}, Charles Darren and {Echternach}, Pierre and {Essinger-Hileman}, Thomas and {Foote}, Marc and {Gruppioni}, Carlotta and {Hensley}, Brandon and {Henning}, Thomas and {Jellema}, Willem and {Johnson}, Matthew and {Kogut}, Alan and {Krause}, Oliver and {McGuire}, James and {Mills}, Elisabeth and {Moullet}, Arielle and {Rodgers}, Michael and {Sauvage}, Marc and {Smith}, John D. and {Somerville}, Rachel and {Staguhn}, Johannes and {Stevenson}, Thomas and {Tucker}, Carole and {Unwin}, Stephen and {Ziemer}, John and {Cannella}, Matthew and {Dissly}, Richard},
        title = "{PRIMA mission concept}",
      journal = {Journal of Astronomical Telescopes, Instruments, and Systems},
         year = 2025,
        month = jul,
       volume = {11},
          eid = {031628},
        pages = {031628},
          doi = {10.1117/1.JATIS.11.3.031628},
       adsurl = {https://ui.adsabs.harvard.edu/abs/2025JATIS..11c1628G}
}

@ARTICLE{rau19,
       author = {{Rau}, Shiau-Jie and {Hirashita}, Hiroyuki and {Murga}, Maria},
        title = "{Modelling the evolution of PAH abundance in galaxies}",
      journal = {\mnras},
         year = 2019,
        month = nov,
       volume = {489},
       number = {4},
        pages = {5218-5224},
          doi = {10.1093/mnras/stz2532},
archivePrefix = {arXiv},
       eprint = {1909.02725},
 primaryClass = {astro-ph.GA},
       adsurl = {https://ui.adsabs.harvard.edu/abs/2019MNRAS.489.5218R}
}

@ARTICLE{carnall19,
       author = {{Carnall}, A.~C. and {McLure}, R.~J. and {Dunlop}, J.~S. and {Cullen}, F. and {McLeod}, D.~J. and {Wild}, V. and {Johnson}, B.~D. and {Appleby}, S. and {Dav{\'e}}, R. and {Amorin}, R. and {Bolzonella}, M. and {Castellano}, M. and {Cimatti}, A. and {Cucciati}, O. and {Gargiulo}, A. and {Garilli}, B. and {Marchi}, F. and {Pentericci}, L. and {Pozzetti}, L. and {Schreiber}, C. and {Talia}, M. and {Zamorani}, G.},
        title = "{The VANDELS survey: the star-formation histories of massive quiescent galaxies at 1.0 < z < 1.3}",
      journal = {\mnras},
         year = 2019,
        month = nov,
       volume = {490},
       number = {1},
        pages = {417-439},
          doi = {10.1093/mnras/stz2544},
archivePrefix = {arXiv},
       eprint = {1903.11082},
 primaryClass = {astro-ph.GA},
       adsurl = {https://ui.adsabs.harvard.edu/abs/2019MNRAS.490..417C}
}

@ARTICLE{beverage24,
       author = {{Beverage}, Aliza G. and {Kriek}, Mariska and {Suess}, Katherine A. and {Conroy}, Charlie and {Price}, Sedona H. and {Barro}, Guillermo and {Bezanson}, Rachel and {Franx}, Marijn and {Lorenz}, Brian and {Ma}, Yilun and {Mowla}, Lamiya A. and {Pasha}, Imad and {van Dokkum}, Pieter and {Weisz}, Daniel R.},
        title = "{The Heavy Metal Survey: The Evolution of Stellar Metallicities, Abundance Ratios, and Ages of Massive Quiescent Galaxies since z {\ensuremath{\sim}} 2}",
      journal = {\apj},
         year = 2024,
        month = may,
       volume = {966},
       number = {2},
          eid = {234},
        pages = {234},
          doi = {10.3847/1538-4357/ad372d},
archivePrefix = {arXiv},
       eprint = {2312.05307},
 primaryClass = {astro-ph.GA},
       adsurl = {https://ui.adsabs.harvard.edu/abs/2024ApJ...966..234B}
}

@ARTICLE{toyouchi26,
       author = {{Toyouchi}, Daisuke and {Ferrara}, Andrea and {Nakazato}, Yurina and {Matsumoto}, Kosei and {Schneider}, Raffaella and {Otaki}, Koki},
        title = "{Grain-size evolution and rapid dust growth in high-redshift galaxies}",
      journal = {arXiv e-prints},
         year = 2026,
        month = jun,
          eid = {arXiv:2606.06108},
        pages = {arXiv:2606.06108},
          doi = {10.48550/arXiv.2606.06108},
archivePrefix = {arXiv},
       eprint = {2606.06108},
 primaryClass = {astro-ph.GA},
       adsurl = {https://ui.adsabs.harvard.edu/abs/2026arXiv260606108T}
}

@ARTICLE{donnelly25,
       author = {{Donnelly}, Grant P. and {Lai}, Thomas S.-Y. and {Armus}, Lee and {D{\'\i}az-Santos}, Tanio and {Larson}, Kirsten L. and {Barcos-Mu{\~n}oz}, Loreto and {Bianchin}, Marina and {Bohn}, Thomas and {B{\"o}ker}, Torsten and {Buiten}, Victorine A. and {Charmandaris}, Vassilis and {Evans}, Aaron S. and {Howell}, Justin and {Inami}, Hanae and {Kakkad}, Darshan and {Lenki{\'c}}, Laura and {Linden}, Sean T. and {Lofaro}, Cristina M. and {Malkan}, Matthew A. and {Medling}, Anne M. and {Privon}, George C. and {Ricci}, Claudio and {Smith}, J.~D.~T. and {Song}, Yiqing and {Stierwalt}, Sabrina and {van der Werf}, Paul P. and {U}, Vivian},
        title = "{A Spectroscopically Calibrated Prescription for Extracting Polycyclic Aromatic Hydrocarbon Flux from JWST MIRI Imaging}",
      journal = {\apj},
         year = 2025,
        month = apr,
       volume = {983},
       number = {1},
          eid = {79},
        pages = {79},
          doi = {10.3847/1538-4357/adb97f},
archivePrefix = {arXiv},
       eprint = {2501.19397},
 primaryClass = {astro-ph.GA},
       adsurl = {https://ui.adsabs.harvard.edu/abs/2025ApJ...983...79D}
}

@ARTICLE{vijayan26,
       author = {{Vijayan}, Aswin P. and {Trayford}, James W. and {Schaye}, Joop and {Ploeckinger}, Sylvia and {Gebek}, Andrea and {Andreadis}, Nick and {Baes}, Maarten and {Ben{\'\i}tez-Llambay}, Alejandro and {Chaikin}, Evgenii and {Frenk}, Carlos S. and {Hu{\v{s}}ko}, Filip and {McGibbon}, Robert J. and {Richings}, Alexander J. and {Schaller}, Matthieu},
        title = "{The evolution of galaxy dust scaling relations in the COLIBRE simulations}",
      journal = {arXiv e-prints},
         year = 2026,
        month = jul,
          eid = {arXiv:2607.26058},
        pages = {arXiv:2607.26058},
archivePrefix = {arXiv},
       eprint = {2607.26058},
 primaryClass = {astro-ph.GA},
       adsurl = {https://ui.adsabs.harvard.edu/abs/2026arXiv260726058V}
}

@ARTICLE{gangula26,
       author = {{Gangula}, Sai and {Newman}, Andrew B. and {Gu}, Meng and {Belli}, Sirio and {Whitaker}, Katherine E. and {Barone}, Tania M. and {Beverage}, Aliza and {Bolamperti}, Andrea and {Bugiani}, Letizia and {Ellis}, Richard S. and {Kriek}, Mariska and {Matthews}, Allison and {Nanayakkara}, Themiya},
        title = "{Resolved Maps of Gas and Dust in a Massive Quiescent Galaxy at z=2 from INQUEST-JWST: Evidence of Accretion and Rejuvenation}",
      journal = {arXiv e-prints},
         year = 2026,
        month = apr,
          eid = {arXiv:2604.26195},
        pages = {arXiv:2604.26195},
          doi = {10.48550/arXiv.2604.26195},
archivePrefix = {arXiv},
       eprint = {2604.26195},
 primaryClass = {astro-ph.GA},
       adsurl = {https://ui.adsabs.harvard.edu/abs/2026arXiv260426195G}
}

@ARTICLE{valentino26,
       author = {{Valentino}, F. and {Pensabene}, A. and {Weibel}, A. and {de Graaff}, A. and {Setton}, D.~J. and {Oesch}, P. and {Brammer}, G. and {Baker}, W.~M. and {Bezanson}, R. and {Greene}, J.~E. and {Heintz}, K.~E. and {Ito}, K. and {Lee}, M. and {Leja}, J. and {Matthee}, J. and {Wang}, B. and {Whitaker}, K.~E. and {Williams}, C.~C. and {Zhu}, P.},
        title = "{Extended [CII] gas emission in and around a massive quiescent galaxy at z=7.3}",
      journal = {arXiv e-prints},
         year = 2026,
        month = jun,
          eid = {arXiv:2606.21361},
        pages = {arXiv:2606.21361},
archivePrefix = {arXiv},
       eprint = {2606.21361},
 primaryClass = {astro-ph.GA},
       adsurl = {https://ui.adsabs.harvard.edu/abs/2026arXiv260621361V}
}

@ARTICLE{wang25,
       author = {{Wang}, Bingjie and {de Graaff}, Anna and {Davies}, Rebecca L. and {Greene}, Jenny E. and {Leja}, Joel and {Brammer}, Gabriel B. and {Goulding}, Andy D. and {Miller}, Tim B. and {Suess}, Katherine A. and {Weibel}, Andrea and {Williams}, Christina C. and {Bezanson}, Rachel and {Boogaard}, Leindert A. and {Cleri}, Nikko J. and {Hirschmann}, Michaela and {Katz}, Harley and {Labb{\'e}}, Ivo and {Maseda}, Michael V. and {Matthee}, Jorryt and {McConachie}, Ian and {Naidu}, Rohan P. and {Oesch}, Pascal A. and {Rix}, Hans-Walter and {Setton}, David J. and {Whitaker}, Katherine E.},
        title = "{RUBIES: JWST/NIRSpec Confirmation of an Infrared-luminous, Broad-line Little Red Dot with an Ionized Outflow}",
      journal = {\apj},
         year = 2025,
        month = may,
       volume = {984},
       number = {2},
          eid = {121},
        pages = {121},
          doi = {10.3847/1538-4357/adc1ca},
archivePrefix = {arXiv},
       eprint = {2403.02304},
 primaryClass = {astro-ph.GA},
       adsurl = {https://ui.adsabs.harvard.edu/abs/2025ApJ...984..121W}
}

@ARTICLE{casey26,
       author = {{Casey}, Caitlin M. and {Akins}, Hollis B. and {Battisti}, Andrew J. and {McKinney}, Jed and {Treister}, Ezequiel and {Zavala}, Jorge A. and {Algera}, Hiddo and {Aravena}, Manuel and {Cheng}, Yingjie and {Drakos}, Nicole E. and {Faisst}, Andreas L. and {Franco}, Maximilien and {Fujimoto}, Seiji and {Gozaliasl}, Ghassem and {Hadi}, Ali and {Harish}, Santosh and {Hirschmann}, Michaela and {Ilbert}, Olivier and {Inayoshi}, Kohei and {Kartaltepe}, Jeyhan S. and {Koekemoer}, Anton M. and {Lagos}, Claudia del P. and {Lambrides}, Erini and {Laishram}, Ronaldo and {Liu}, Daizhong and {Long}, Arianna S. and {Magdis}, Georgios E. and {Manning}, Sinclaire M. and {Martin}, Crystal L. and {Martinez}, III, Felix and {Massey}, Richard and {McCleary}, Jacqueline E. and {McCracken}, Henry Joy and {Moscardini}, Lauro and {Narayanan}, Desika and {Paquereau}, Louise and {Rhodes}, Jason and {Robertson}, Brant E. and {Samir}, Rasha M. and {Scarlata}, Claudia and {Shuntov}, Marko and {Sommovigo}, Laura and {Vijayan}, Aswin P. and {Wang}, Wuji and {Xu}, Can and {Zimmerman}, Dhruv},
        title = "{Dust in the Average Galaxy: Attenuation, Emission, and Opacity from 0<z<7}",
      journal = {arXiv e-prints},
         year = 2026,
        month = jun,
          eid = {arXiv:2606.17270},
        pages = {arXiv:2606.17270},
          doi = {10.48550/arXiv.2606.17270},
archivePrefix = {arXiv},
       eprint = {2606.17270},
 primaryClass = {astro-ph.GA},
       adsurl = {https://ui.adsabs.harvard.edu/abs/2026arXiv260617270C}
}

@ARTICLE{narayanan26,
       author = {{Narayanan}, Desika and {Torrey}, Paul and {Parente}, Massimiliano and {Donnelly}, Grant and {Zimmerman}, Dhruv and {Garcia}, Alex and {Richie}, Helena and {Smith}, J.-D.~T. and {Hensley}, Brandon and {Marinacci}, Federico and {McKinney}, Jed and {Pope}, Alexandra and {Popping}, Gergo and {Sales}, Laura and {Sandstrom}, Karin and {Savitch}, Ethan and {Shivaei}, Irene and {Spilker}, Justin and {Whitcomb}, Corey},
        title = "{The Lifecycle and Emission Properties of PAHs in Cosmological Hydrodynamic Galaxy Formation Simulations}",
      journal = {arXiv e-prints},
         year = 2026,
        month = jun,
          eid = {arXiv:2606.20809},
        pages = {arXiv:2606.20809},
          doi = {10.48550/arXiv.2606.20809},
archivePrefix = {arXiv},
       eprint = {2606.20809},
 primaryClass = {astro-ph.GA},
       adsurl = {https://ui.adsabs.harvard.edu/abs/2026arXiv260620809N}
}

\twocolumn

\begin{appendix} %First appendix

% In your preamble:
% \usepackage{tikz}
% \usetikzlibrary{arrows.meta,positioning,calc,fit,shapes.misc,shapes.geometric}

%=====================================================================
\section{Effect of varying AGN gas outflow strengths}
\label{AppendixA}
%=====================================================================

Our fiducial AGN-regulated models adopt a characteristic outflow strength close to unity. However, observations of molecular-gas removal in massive galaxies indicate a broad range of outflow rates across cosmic time \citep[e.g.,][]{fluetsch19, belli21, davies24}. To test how this uncertainty affects our conclusions, we repeat the model runs with higher effective AGN mass-loading factors, $\eta_{\rm out}=2$, 5, 10, and 15, while keeping all other initial conditions fixed.

\begin{figure*}
\centering
\includegraphics[width=0.64\textwidth]{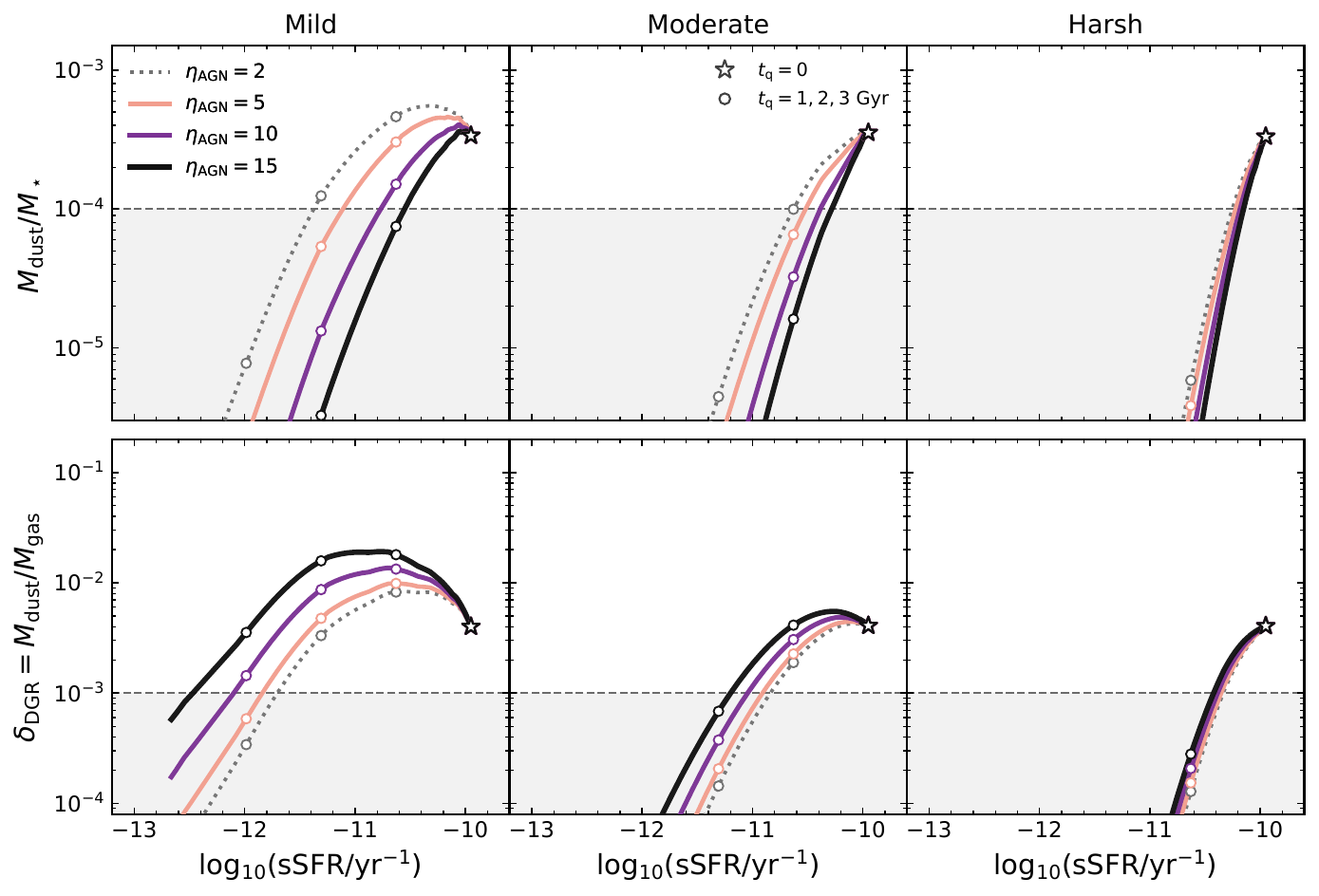}
\includegraphics[width=0.64\textwidth]{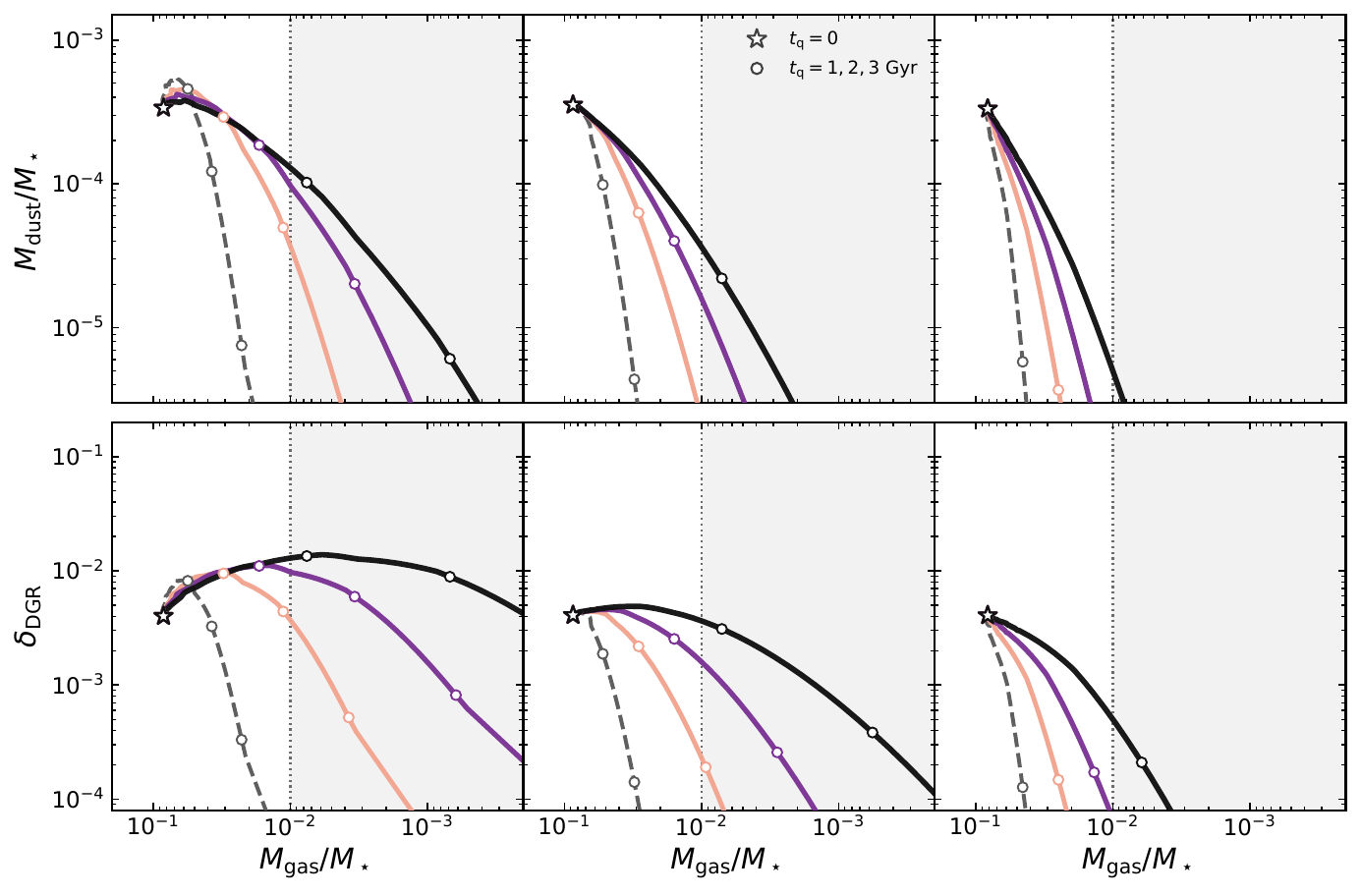}
\caption{
Effect of varying the effective AGN outflow strength on the evolution of dust and gas in quenched galaxies. Columns show the mild, moderate, and harsh destruction regimes. The upper set of panels shows $M_{\rm dust}/M_\star$ and $\delta_{\rm DGR}$ as a function of sSFR, while the lower set shows the same quantities as a function of $M_{\rm gas}/M_\star$. Coloured tracks correspond to different values of the effective AGN mass-loading factor, $\eta_{\rm out}$, while symbols mark the onset of quenching and successive post-quenching times, with values indicated in the legend. The grey dashed curve shows the extreme case in which strong gas removal is combined with fully unshielded dust grains. Larger $\eta_{\rm out}$ values mainly accelerate gas removal, but the dust response remains governed by the competition between destruction, delayed stellar replenishment, and residual grain growth. As a result, temporarily elevated $\delta_{\rm DGR}$ values can occur even when the absolute dust fraction is rapidly declining, demonstrating that dust-to-gas ratio alone is not a unique tracer of dust regrowth.
}
\label{fig:Fig3app}
\end{figure*}

As evident from Fig.~\ref{fig:Fig3app}, increasing $\eta_{\rm out}$ primarily accelerates the decline of $f_{\rm gas}\propto M_{\rm H_2}/M_\star$, shifting the tracks toward lower $f_{\rm gas}$ at fixed sSFR. The dust response is less obvious, due to a complex interplay of remnant-dust survival, delayed TP-AGB dust injection, ability for efficient ISM grain growth, and grain destruction. As a result, stronger outflows do not simply map onto proportionally lower $f_{\rm gas}$, instead they create a stronger decoupling between gas and dust evolution. This reinforces our main conclusion that small differences in the cold-gas reservoir can be amplified into large variations in $f_{\rm dust}$ and/or $\delta_{\rm DGR}$.

A particularly important feature appears in the $\delta_{\rm DGR}$ tracks. Strong outflows can temporarily bring $\delta_{\rm DGR}$ close to nominal SF values, $\delta_{\rm DGR}\sim1/100$, which can wrongly be interpreted as evidence for enhanced dust regrowth. In these cases, both gas and dust are decreasing in tandem, and the apparently high $\delta_{\rm DGR}$ reflects a short phase in which their depletion timescales are temporarily comparable. This behaviour highlights why considering only a single diagnostic (e.g. dust or gas fraction) to describe post-quenching ISM can be misleading.

These higher-$\eta_{\rm out}$ experiments are relevant for recently reported strong outflows in quiescent or rapidly quenching systems \citep[e.g.,][]{park2023rapid,zhu26,valentino26}, and may also help interpret QGs with inferred dust temperatures higher than expected from simple modified-blackbody modelling \citep[e.g.,][]{deugenio26}. In our framework, such systems could correspond to short-lived phases in which AGN-regulated gas removal and heating strongly modify the cold ISM while a residual dust reservoir is still present. At the same time, the figure shows that similar locations in $f_{\rm dust}$ or $\delta_{\rm DGR}$ can be reached by different combinations of outflow strength, shielding loss, destruction, and replenishment. Breaking these degeneracies requires simultaneous constraints on $f_{\rm dust}$, $\delta_{\rm DGR}$, $f_{\rm H_2}$, sSFR, and $t_{\rm q}$.

The shaded regions in Fig.~\ref{fig:Fig3app} illustrate when systems enter regimes that are difficult to access with current individual ALMA observations. The horizontal grey region marks the regime currently beyond the sensitivity of individual cold-dust continuum observations, corresponding to $M_{\rm dust}/M_\star\lesssim10^{-4}$ (\citealt{lorenzon25b}), while the vertical grey region in the gas-fraction plane marks the similarly ALMA-faint molecular-gas regime, $M_{\rm gas}/M_\star\lesssim10^{-2}$. We see that stronger AGN outflows move QGs into the gas-faint regime earlier, but not always simultaneously into the dust-faint regime. This produces a short-lived phase in which $M_{\rm gas}/M_\star$ has already dropped below current detectability limits, while a residual dust reservoir remains measurable or close to the ALMA continuum threshold. Conversely, in harsh destruction models $f_{\rm dust}$ can fall below $10^{-4}$ while the gas reservoir is not yet fully depleted. These offsets between the gas- and dust-detectability windows highlight why non-detections in only one tracer cannot uniquely determine whether a QG is genuinely ISM-poor or instead caught during a rapid, feedback-regulated transition. Such offsets may be further complicated by dust entrainment, as outflows can redistribute dust into extended reservoirs rather than remove it entirely, as observed in local outflowing galaxies \citep[e.g.,][]{fogarty19, romano24}; this spatial transport is not captured by our one-zone treatment. Finally, it is also interesting to note that in the harsh-destruction regime, the limited divergence of the tracks with $\eta_{\rm out}$ suggests that strong AGN radiative processing dominates over variations in the outflow loading factor, thereby driving a more direct coupling between the ISM evolution and sSFR.

%=====================================================================
\section{Effect of gas-phase metallicity on modeled dust and PAH evolution}
\label{AppendixB}
%=====================================================================

We test whether the main \texttt{UNDUST} trends depend strongly on the assumed post-quenching gas metallicity. Our fiducial calculations adopt a solar value, $Z_{\rm gas}=0.02$. Here we compare this reference case to a super-solar ($Z_{\rm gas}=0.04$) model while keeping the same initial gas fraction, grain populations, and quenching history. This isolates the impact of metallicity on ISM grain growth and TP-AGB dust injection, without changing the underlying gas-evolution pathway.

\begin{figure}[h]
	\centering
	\includegraphics[width=0.35\textwidth]{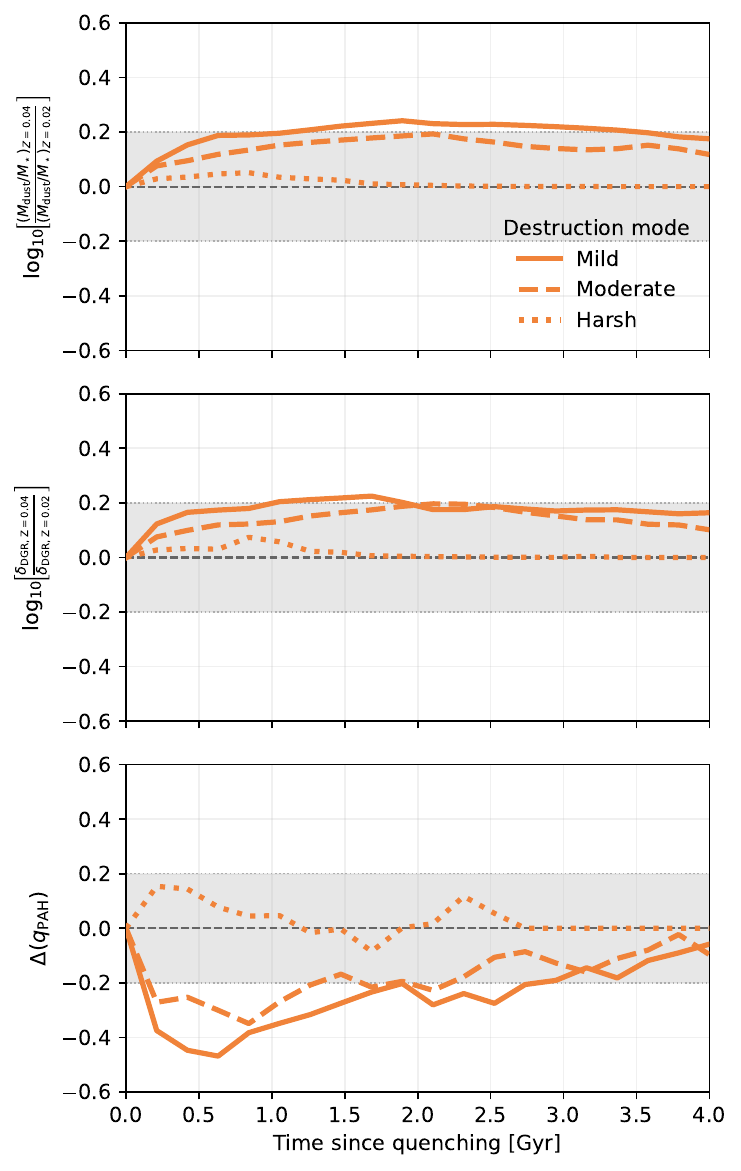}
	\caption{
		Effect of increasing the reference gas-phase metallicity on the predicted dust and PAH evolution. The panels show logarithmic offsets between the super-solar and fiducial solar metallicity runs as a function of $t_{\rm q}$ for the mild, moderate, and harsh destruction regimes. From top to bottom, we show the change in $M_{\rm dust}/M_\star$, $\delta_{\rm DGR}$, and $q_{\rm PAH}$. The grey band marks arbitrary reference $\pm0.2$ dex interval. Higher $Z_{\rm gas}$ produces a modest increase in the bulk dust fraction and dust-to-gas ratio, typically $\sim0.1$--$0.2$ dex in the mild and moderate modes, but shows opposite effect in $q_{\rm PAH}$ where solar values tend to return higher $q_{\rm PAH}$.
	}
	\label{fig:Fig2app}
\end{figure}

Figure~\ref{fig:Fig2app} shows that increasing $Z_{\rm gas}$ produces only a modest enhancement in the bulk dust reservoir. In the mild and moderate destruction regimes, $f_{\rm dust}$ and $\delta_{\rm DGR}$ are typically higher by $\sim0.1$--$0.25$ dex over most of the post-quenching phase. The harsh model remains almost insensitive to metallicity, because rapid destruction and inefficient growth dominate over the additional metal supply. Thus, higher metallicity can slightly raise the normalisation of the cold-dust reservoir, but it does not alter the qualitative ordering between destruction modes or the main evolutionary pathways.

The PAH response is different. We see that a larger $Z_{\rm gas}$ does not produce a proportional increase in $q_{\rm PAH}$. In the mild and moderate modes the PAH fraction is even slightly lower, by up to $\sim0.2$--$0.3$ dex at some epochs. This occurs because higher metallicity mainly supports the growth and survival of the bulk dust reservoir, including $\mathrm{Sil-}$rich grains, while the PAH reservoir is tied specifically to the availability of carbonaceous material. In our adopted AGB yields, super-solar metallicities do not strongly boost carbon-dust production; instead, TP-AGB injection becomes increasingly dominated by O-rich, $\mathrm{Sil-}$producing phases. As a result, the total $M_{rm dust}$ can increase while the $q_{\rm PAH}$ remains unchanged or decreases, partly because the denominator of $q_{\rm PAH}=M_{\rm PAH}/M_{\rm dust}$ grows faster than the PAH-bearing carbonaceous component. The only clear positive PAH offset appears briefly in the harsh model, likely due to a slightly longer survival of large C-grains that can temporarily feed shattering into PAH-sized material before destruction dominates.

The behaviour presented in Figure~\ref{fig:Fig2app} generally differs from the usual picture in SFGs, where dust and PAH enrichment are often described in terms of a metallicity-regulated transition above a critical metallicity for efficient grain growth \citep[e.g.,][]{asano2013dust, popping17, liqi19, donevski20,shivaei24}. In QGs, metallicity alone is not the controlling parameter. As we discussed in Section~\ref{sec:results}, grain growth is limited by the shrinking cold dense gas reservoir and AGN-regulated heating. Similarly, PAH evolution emerges from the complex competition between carbon-grain injection, shattering, coagulation, and destruction (see also \cite{narayanan23}). Therefore, the weak metallicity dependence shown here supports our main conclusion that post-quenching dust and PAH evolution is governed primarily by gas survival and grain-processing physics, while variations in $Z_{\rm gas}$ introduce mainly a second-order effect.

%=====================================================================
\section{TP-AGB injection rates of C- and $\mathrm{Sil-}$rich dust with stellar population age}
\label{AppendixC}
%=====================================================================

The delayed dust contribution from TP-AGB stars depends on both the stellar age at quenching and the metallicity-dependent composition of the TP-AGB ejecta. At fixed final stellar mass, a younger quenched system must have formed a larger fraction of its stars shortly before quenching. It therefore contains more intermediate-mass AGB progenitors close to the quenching epoch, leading to higher instantaneous dust injection rates at $t_{\rm q}=0$. Older systems, by contrast, assembled their stellar mass over a longer period and their TP-AGB contribution at quenching is dominated by lower-mass, older progenitors with weaker and more delayed dust yields.

\begin{figure}[ht!]
\centering
\includegraphics[width=0.44\textwidth]{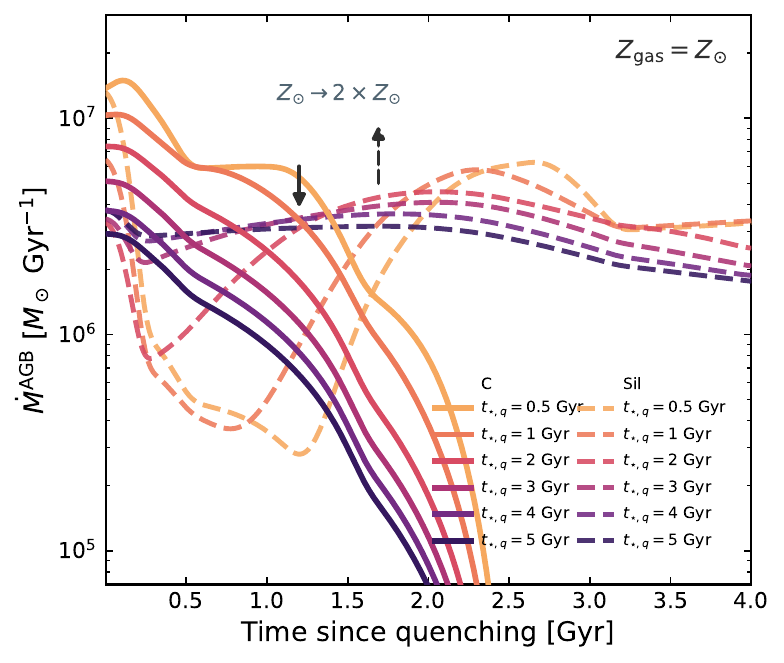}
\caption{
TP-AGB dust injection rates as a function of time since quenching for different stellar population ages at the onset of quenching ($t_{\star,q}$). Solid lines show carbonaceous dust injection, while dashed lines show silicate dust injection; colours indicate the assumed stellar population age at quenching. All curves are normalized to the same final stellar mass, so younger systems have higher instantaneous injection rates at $t_{\rm q}=0$ because a larger fraction of their intermediate-mass TP-AGB progenitors are still entering the TP-AGB phase. The arrows indicate the direction and approximate magnitude of the change expected when increasing the metallicity from solar to twice solar: carbonaceous dust injection is suppressed, while silicate-rich injection is enhanced. The figure illustrates why C-rich TP-AGB enrichment, and therefore PAH replenishment, is most efficient in younger post-quenching populations, whereas older or more metal-rich systems are expected to become increasingly dominated by $\mathrm{Sil-}$rich dust.
}
\label{fig:Fig4app}
\end{figure}

Figure~\ref{fig:Fig4app} illustrates this effect by showing the post-quenching injection rates of carbonaceous and silicate dust for different stellar population ages at quenching. The normalization varies strongly with $t_{\star,q}$: younger populations can reach instantaneous TP-AGB dust injection rates of order $\sim10^7\,M_\odot\,{\rm Gyr}^{-1}$, while older populations typically remain closer to a few $\times10^6\,M_\odot\,{\rm Gyr}^{-1}$. The carbonaceous contribution declines rapidly after quenching, especially for young systems, because the C-rich TP-AGB phase is restricted to a finite range of progenitor masses and lifetimes. In contrast, the silicate component evolves more gradually and can remain important for several Gyr, sometimes showing a delayed enhancement when lower-mass O-rich AGB stars enter the TP-AGB phase.

The arrows in Fig.~\ref{fig:Fig4app} indicate the direction and approximate magnitude of the change expected when moving from $Z_{\rm gas}=Z_{\odot}$ to $Z_{\rm gas}=2\times Z_{\odot}$. Super-solar metallicities tend to suppress the carbonaceous TP-AGB contribution while enhancing the silicate-rich component. This reflects the metallicity dependence of TP-AGB dust chemistry: at super-solar values, the larger oxygen abundance and the reduced efficiency of carbon-star formation favour oxygen-rich, $\mathrm{Sil-}$producing phases, while the C-rich dust channel becomes more limited. This behaviour of our adopted TP-AGB model by \citet{nanni14} is qualitatively consistent with the metal-rich TP-AGB dust models of \citet{ventura20}, where silicate production fully dominates at $Z_{\rm gas}\gtrsim0.02-0.03$.

These trends are important for interpreting the general PAH evolution. Since PAHs in \texttt{UNDUST} are tied to the small carbonaceous reservoir, younger high-redshift QGs are expected to experience a longer or stronger phase of C-rich replenishment, while older or more metal-rich QGs transition more rapidly toward $\mathrm{Sil-}$dominated TP-AGB injection. Therefore, the window in which delayed TP-AGB enrichment can support PAH production is not universal. Instead, it depends on the stellar population age at quenching and on the $Z_{\rm gas}$-dependent balance between C- and $\mathrm{Sil-}$rich TP-AGB dust. We do not interpret the exact timing or amplitude of individual bumps as robust predictions, since they depend on the adopted yield tables, interpolation method and lifetime mapping. Instead, Fig.~\ref{fig:Fig4app} illustrates the physical origin of the stellar age and metallicity sensitivity that enters the delayed dust-replenishment channel.

\end{appendix}

\end{document}